\documentclass[
superscriptaddress,
preprint,
 amsmath,amssymb,
 aps,
]{revtex4-2}

\usepackage{graphicx}
\usepackage{dcolumn}
\usepackage{bm}
\usepackage{amssymb}
\usepackage{xcolor} 
\usepackage{physics}
\usepackage{multirow}
\usepackage{booktabs}
\usepackage{subfigure}
\usepackage{hyperref}
\usepackage{caption} 
\usepackage{soul}

\begin{document}

\title{Structural transformations driven by local disorder at interfaces}

\author{Yanyan Liang}
\email{yanyan.liang@rub.de}
\affiliation{ICAMS, Ruhr-Universit{\"a}t Bochum, 44801 Bochum, Germany}
\author{Grisell D\'{i}az Leines}%
\affiliation{European Bioinformatics Institute (EMBL- EBI), Wellcome Genome Campus, Hinxton, Cambridge, United Kingdom} %
\author{Ralf Drautz}
\affiliation{ICAMS, Ruhr-Universit{\"a}t Bochum, 44801 Bochum, Germany}
\author{Jutta Rogal}
\affiliation{Department of Chemistry, New York University, New York, NY 10003, USA}
\affiliation{Fachbereich Physik, Freie Universit{\"a}t Berlin, 14195 Berlin, Germany}

\date{\today}

\begin{abstract}
Despite the fundamental importance of solid–solid transformations in many technologies, the microscopic mechanisms remain poorly understood. Here, we explore the atomistic mechanisms at the migrating interface during solid-solid phase transformations between the topologically closed-packed A15 and body-centred cubic phase in tungsten.  The high energy barriers and slow dynamics associated with this transformation require the application of enhanced molecular sampling approaches.  To this end, we performed metadynamics simulations in combination with a path collective variable derived from a machine learning classification of local structural environments, which allows the system to freely sample the complex interface structure.
A disordered region of varying width forming at the migrating interface is identified as a key physical descriptor of the transformation mechanisms, facilitating the atomic shuffling and rearrangement necessary for structural transformations. Furthermore, this can directly be linked to the differences in interface mobility for distinct orientation relationships as well as the formation of interfacial ledges during the migration along low-mobility directions. 
\end{abstract}

\maketitle

\section{Introduction}
\label{sec:Intro}
Solid-solid phase transformations are ubiquitous in nature~\cite{kirbyMantlePhaseChanges1991}, and are of particular significance in the realm of materials science, where they have practical implications in metallurgy~\cite{porter2009phase}, ceramics~\cite{smith1986principles}, and colloidal matter~\cite{manoharan2015colloidal}, among others. 
These transformations are primarily governed by the nucleation and growth of new phases at internal interfaces, where structural transformations are continuously driven by the migration of the interface. Over the years, there have been numerous experimental and computational studies focusing on structural transformations at grain~\cite{molodovMagneticallyForcedMotion1999,straumal2004grain,meiners2020observations,niuAtomisticInsightsShearcoupled2016,olmsted2007grain,olmstedSurveyComputedGrain2009,rahman2014comprehensive,race2014role,hoytAtomisticSimulationsGrain2014,korneva2020atomistic,wei2021direct,wang2022tracking} and phase boundaries~\cite{pitsch1959martensite,furuhara1990interphase,bos2006molecular,fukino2008situ,song2012molecular,wang2013molecular,yang2014dissecting,duncan2016collective,ouMolecularDynamicsSimulations2017,meijer2017observation,fei2018observable,immink2020crystal,chen2021molecular,kuznetsov2001effect,li2021revealing,fu2022atomic,zhang2022defect,jiang2023dynamics,yang2015mechanism} in metals and alloys. 
Still, the microscopic processes associated with interface migration remain, to a large extent, poorly understood due to limitations in accessible timescales and the complexity of the microscopic dynamics.

Internal interfaces are intrinsic planar defects between distinct phases directly linking the parent and product phase in solid-solid transformations~\cite{howeInterfacesMaterialsAtomic1997}. 
The properties of these interfaces hold a vital role in the phase transformation kinetics and mechanisms, impacting the microstructural evolution and thereby shaping the mechanical or functional properties of materials. Since interface migration often involves atomic shuffling and rearrangement, interface velocities are markedly influenced by various interface characteristics, including lattice mismatch, orientation relationship, density, impurity contents, and so on.
Indeed, experimental studies of colloidal systems evidence that interface velocities strongly depend on the interface structure, resulting in largely different velocities for distinct orientation relationships~\cite{peng2015two,peng2017diffusive}.
More specifically, a strong correlation was found between interface coherency and mobility, with incoherent interfaces migrating much faster than semicoherent and coherent interfaces. At incoherent interfaces, disordered interface layers play a prominent role in facilitating the rearrangement of particles at the interface, thereby enhancing its mobility~\cite{peng2015two,peng2017diffusive}. Similarly, the formation of intermediate amorphous phases enabling solid-solid transformations has been reported in perovskites~\cite{levitas2014solid,song2022planar} and metals~\cite{wang1998thermodynamic,luo2005segregation,levitasCrystalcrystalPhaseTransformation2012,su2021solid}, but little is known for systems including more complex crystallographic phases. 

In the present study, we focus on the importance of disordered interface regions during structural transformations between the complex topological-closed-packed (TCP)~\cite{watson1984transition,long2018minimum,seiser2011tcp,long2020structural} A15 and body-centred cubic (bcc) phase in tungsten.
In recent years, these two phases have gained significant attention due to their relevance in practical applications, where bcc-W with its high melting temperature and relatively short electron mean free path is a promising candidate for microelectronics~\cite{choi_PhaseGrain_2011a,choi_ElectronMean_2012}, while A15-W exhibits a giant spin Hall effect~\cite{paiSpinTransferTorque2012} desired in spintronic devices. The presence of both bcc-W and A15-W is often reported in W thin-films~\cite{okeefePhaseTransformationSputter1996,narasimhamFabrication520Nm2014,paiSpinTransferTorque2012,vullers2015alpha,liuTopologicallyClosepackedPhases2016,zhu2018phase,barmak_TransformationTopologically_2017a,wang2019demand} exhibiting different orientational geometries with semicoherent or incoherent interfaces~\cite{vullers2015alpha,wang2019demand,zhu2018phase}, most notably with (210), (200), and (110) planes. 
To investigate the atomic-scale processes and find the relevant physical descriptors during structural transformations at these heterophase interfaces with distinct orientation relationships, we perform extensive enhanced sampling simulations following the moving interface over extended timescales.  A particular challenge is the accurate characterization and meaningful representation of the inherently complex A15 phase as well as the bcc-A15 interface.  Here, we combine the enhanced sampling with a machine-learning based classification of local structural environments.
Following the approach of Rogal {\it et al.}~\cite{rogal2019neural}, we describe the phase transformation as a path in the global classifier space~\cite{rogal2019neural,rogal2021pathways} of A15-W and bcc-W, and combine the resulting path collective variable with metadynamics~\cite{laioEscapingFreeenergyMinima2002,laioAssessingAccuracyMetadynamics2005} to drive the transformation between different structures. 
Similar to the findings in colloids~\cite{peng2015two,peng2017diffusive,peng2023situ}, pervoskites~\cite{levitas2014solid,song2022planar}, and metals~\cite{wang1998thermodynamic,luo2005segregation,levitasCrystalcrystalPhaseTransformation2012,su2021solid}, a disordered region is formed at the moving interface.  This disordered interface region constitutes a key physical descriptor of the transformation mechanism that determines the change in interface mobility for different orientation relationships. Furthermore, our enhanced sampling simulations reveal the formation of interfacial ledges resulting from the intrinsic difference in interface mobilities along different directions that is not captured by straightforward high-temperature molecular dynamics simulations. This study takes a step forward in deepening our understanding of phase transformations between A15 and bcc phases by identifying key microscopic descriptors that could be used to guide and interpret future experiments, towards the design of materials with targeted properties.

\section{Computational approach}
\subsection{Machine learning classification of local structural environments}

Characterizing local structural environments~\cite{behler2007generalized,Bartok2013,drautz2019} is an essential first step when investigating interface migration processes. 
In our study, we apply a classification neural network (NN) based on Geiger and Dellago's work~\cite{geiger2013neural}, that has been demonstrated to provide accurate structural fingerprints for amorphous and complex crystalline phases in water~\cite{geiger2013neural} and metallic systems~\cite{rogal2019neural}.  
We adapt the same NN architecture as in our previous work~\cite{rogal2019neural}, with 14 features in the input layer, two hidden layers with 25 nodes each, and an output layer with five different classes.  
The input features comprise 11 Behler-Parrinello symmetry functions~\cite{behler2007generalized,behler2011atom} and three Steinhardt bond order parameters~\cite{steinhardt1983bond,lechnerAccurateDeterminationCrystal2008}.
Details of these descriptors are given in the supplementary information. 
The output layer of the NN provides a  vector $\bm{y}_i$ that measures the similarity of local structural environments of atom $i$ with the phase of interest $j$ with $y_i^j\in[0,1]$. A value close to 1 indicates a large similarity to the target phase. Here, the five considered classes include crystalline bcc, A15, fcc, hcp, and an amorphous/disordered phase. 
   
For our study of the W system, we do not need to train a new classification NN but can utilize the network that  was originally trained for Mo~\cite{rogal2019neural}.  The dataset for Mo included local environments of all relevant crystalline bulk phases, amorphous phases, and interface configurations at various temperatures and pressures. 
Since interatomic distances in Mo and W are quite similar, with lattice constants $a_{0}(\text{Mo})\approx3.15~\mathrm{\AA}$ and $a_{0}(\text{W})\approx3.17~\mathrm{\AA}$, respectively, the classification network showed excellent transferability and accuracy when applied to our dataset for the various phases in W. Additional details can be found in the supplementary information.

\subsection{Path collective variable in classifier space}
\label{subsec:pathcv}
\begin{figure}[htbp]
    \centering
    \includegraphics[scale=0.65]{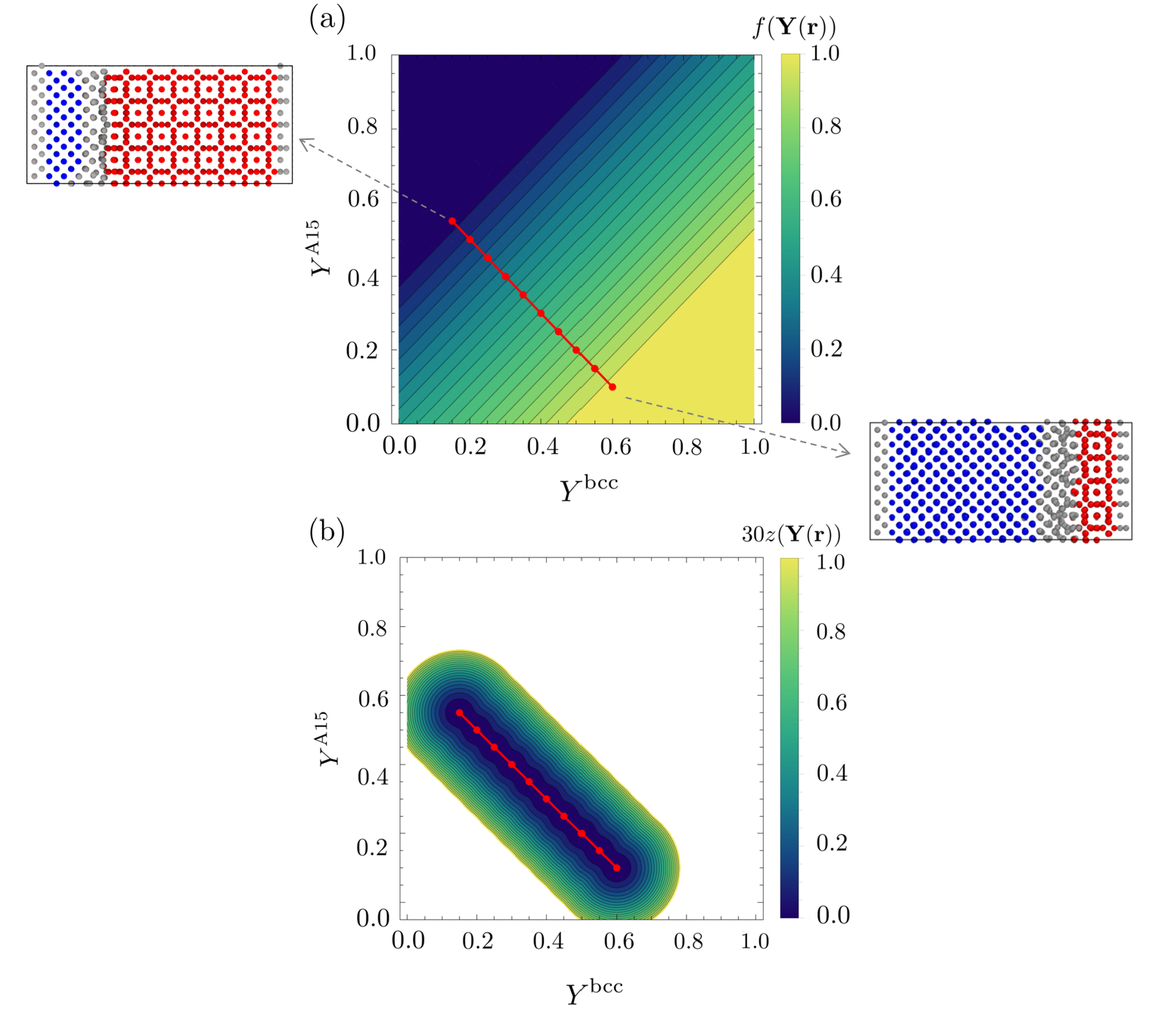}
    \caption{Illustration of (a) the path collective variable $f(\mathbf{Y}(\mathbf{r}))$ and (b) the distance function $z(\mathbf{Y}(\mathbf{r}))$ (multiplied by 30) together with the path (red line) in the $Y^\text{bcc}$-$Y^\text{A15}$ space. 
    There are 10 equidistant points along the path, starting at $\mathbf{Y}_1=\{Y^\text{bcc}:0.15,Y^\text{A15}:0.55\}$ and ending at $\mathbf{Y}_{10}=\{Y^\text{bcc}:0.60,Y^\text{A15}:0.10\}$. 
    Representative configurations are shown for the two end points of the path. In these configurations, bcc atoms are shown in blue, A15 in red, and disorder in gray.}
    \label{fig:pathCV}
\end{figure}

Based on the information about the local structural environment of each atom $i$, $y_i^j$, we define global classifiers that effectively measure the respective phase fractions in a given configuration,  
\begin{equation}
    Y^{j}=\frac{1}{N}\sum_{i}^{N}y_i^{j}\quad,
    \label{eqn:globalCV}
\end{equation}
where $N$ denotes the total number of atoms.  
For each configuration, we compute five phase fractions, 
$Y^\text{bcc}$,  $Y^\text{A15}$, $Y^\text{fcc}$, $Y^\text{hcp}$, and $Y^\text{dis}$, where `dis' stands  for disordered/amorphous phase, respectively. 
Using the global classifiers of the considered phases separately, for example using only $Y^\text{bcc}$ as a one-dimensional collective variable (CV) in the enhanced sampling, is insufficient to drive the phase transformation between bcc and A15, since decreasing  $Y^\text{bcc}$ does not promote the growth of A15.
Furthermore, if both $Y^\text{bcc}$ and $Y^\text{A15}$ are used as separate CVs in 2D enhanced sampling, the simulations do not converge due to the mutual dependence of these CVs: as $Y^\text{bcc}$ increases, $Y^\text{A15}$ needs to decrease and vice versa.  
Instead, the global classifiers are combined in a non-linear way by defining a path collective variable~\cite{branduardi2007b} in the space of global classifiers~\cite{rogal2019neural,rogal2021pathways}.
Since we are interested in the structural transformation between the A15 and bcc phase, we construct a linear path in the $Y^\text{bcc}-Y^\text{A15}$ space, keeping the sum of the two phase fraction constant (red line in Fig.~\ref{fig:pathCV}(a)). 
The corresponding  path CV is given by
\begin{equation}
f(\mathbf{Y}(\mathbf{r}))=\frac{1}{K-1} \frac{\sum_{k=1}^{K}(k-1) \exp \left[-\lambda\left|\mathbf{Y}(\mathbf{r})-\mathbf{Y}_{k}\right|^{2}\right]}{\sum_{k=1}^{K} \exp \left[-\lambda\left|\mathbf{Y}(\mathbf{r})-\mathbf{Y}_{k}\right|^{2}\right]}  \quad,
\label{eqn:pathCVeq}
\end{equation}
where $\mathbf{Y}(\mathbf{r})=\{Y^\text{bcc},Y^\text{A15}\}$ represents the position of a configuration $\mathbf{r}$ in the $Y^\text{bcc}$-$Y^\text{A15}$ space, $\mathbf{Y}_k$ are the $K$ points, $k:\{1,...,K\}$, defining the path, and $|\mathbf{Y}(\mathbf{r})-\mathbf{Y}_k|^2$ marks the squared distance of a configuration from a point $k$ 
in the $Y^\text{bcc}-Y^\text{A15}$ space. The parameter $\lambda$ is derived from the inverse of the distance between subsequent points along the path. 
The value of the path CV, shown in Fig.~\ref{fig:pathCV}(a), increases smoothly from 0 to 1 along the path and is constant perpendicular to the path.   
When the path CV is used in enhanced sampling, the system is pushed along the path while it can explore all degrees of freedom normal to the path.
In addition to the path CV in Eq.~\eqref{eqn:pathCVeq}, a function $z(\mathbf{Y}(\mathbf{r}))$, shown in Fig.~\ref{fig:pathCV}(b), can be defined measuring the distance from the path in the $Y^\text{bcc}$-$Y^\text{A15}$ space, thus forming a tube around the path~\cite{branduardi2007b}
\begin{equation} 
z(\mathbf{Y}(\mathbf{r}))=-\frac{1}{\lambda} \ln \left(\sum_{k=1}^{K} \exp \left[-\lambda\left(\mathbf{Y}(\mathbf{r})-\mathbf{Y}_{k}\right)^{2}\right]\right)
\quad.
\label{eqn:zfunc}
\end{equation}
This distance function can be used as an additional CV in enhanced sampling or as a restraint on the sampling of $f(\mathbf{Y}(\mathbf{r}))$. Utilizing $z(\mathbf{Y}(\mathbf{r}))$ together with $f(\mathbf{Y}(\mathbf{r}))$ as CVs is efficient if the proposed path is not a suitable sampling coordinate and an exploration of phase space away from the predefined path is needed. 
Conversely, if  $z(\mathbf{Y}(\mathbf{r}))$ is used to add a restraining potential to $f(\mathbf{Y}(\mathbf{r}))$, it imposes sampling close to the proposed path~\cite{cuendet2018endpoint} and limits the sampling of configuration space orthogonal to the path.
In this work, we utilized the distance function in a restraining potential to sample the configuration space deviating only little from the predefined path in the $Y^\text{bcc}$-$Y^\text{A15}$ space. 
The restraining potential~\cite{cuendet2018endpoint} takes the form
\begin{equation}
    V^\text{res}(z(\mathbf{Y}(\mathbf{r})))=\kappa\left(\frac{z(\mathbf{Y}(\mathbf{r}))-z_0}{\epsilon}\right)^n\quad,
    \label{eqn:res_V}
\end{equation}
where $z_0$ sets the center from which the distance is measured, $n$ is the exponent number, $\kappa$ is the prefactor, and $\epsilon$ defines the width of the restraining potential.

\subsection{Metadynamics}
\label{subsec:metady}
To drive the phase transformation between A15 and bcc  and sample the corresponding free energy profiles, we use  metadynamics together with the path CV defined in section~\ref{subsec:pathcv}. In metadynamics~\cite{laioEscapingFreeenergyMinima2002,laioAssessingAccuracyMetadynamics2005}, 
the exploration of the phase space is accelerated by adding a time-dependent bias potential along a predefined CV. 
Here, we only summarize the essential equations of metadynamics combined with the path CV and refer to the original papers and reviews~\cite{laioEscapingFreeenergyMinima2002,iannuzzi2003efficient,laioAssessingAccuracyMetadynamics2005,laioMetadynamicsMethodSimulate2008,
valsson2016enhancing,bussi2020using} for more details on the methodology. 
The time-dependent  bias potential deposited along the path CV is given by a sum of Gaussians
\begin{equation}
   V_\text{bias}(s,t)=h\sum^{t_{i}<t}_{t_i=\tau_G,2\tau_G,...}
   \exp \bigg(-\frac{(S({t_i})-s)^2}{2\sigma^2}\bigg)\quad,
   \label{eqn:metady}
\end{equation}
where $h$ and $\sigma$ are the height and width of the Gaussians, respectively. $S({t_i})=f(\mathbf{Y}(\mathbf{r}(t_i)))$ represents the value of the path CV at time $t_i$. The free energy profile along $s$ is approximated by the negative of the accumulated bias with
\begin{equation}
    F(s)=-\lim_{t\to \infty} V_\text{bias}(s,t)+\text{const}\quad.
    \label{eqn:mtdfe}
\end{equation}
Metadynamics can be used to sample the transformation close to the proposed path by adding the restraining potential on the distance function, Eq.~\eqref{eqn:res_V}. The unbiased free energy $F(s)$ can be derived from the simulations employing a restraining potential by~\cite{cuendet2018endpoint}
\begin{equation}
    F(s)=F_{r}(s)-k_{B}T\ln\langle e^{\beta V^\text{res}(z)}\rangle_{r,s}\quad,
    \label{eqn:fe_unrestraint}
\end{equation}
where  $F_{r}(s)$ is the free energy in the restrained system, $k_B$ is the Boltzmann constant, and $\langle \dots \rangle$ denotes the ensemble average.

\subsection{Computational details}
\label{subsec:comp_detail}
\begin{figure}
    \centering
    \includegraphics[width=0.95\textwidth]{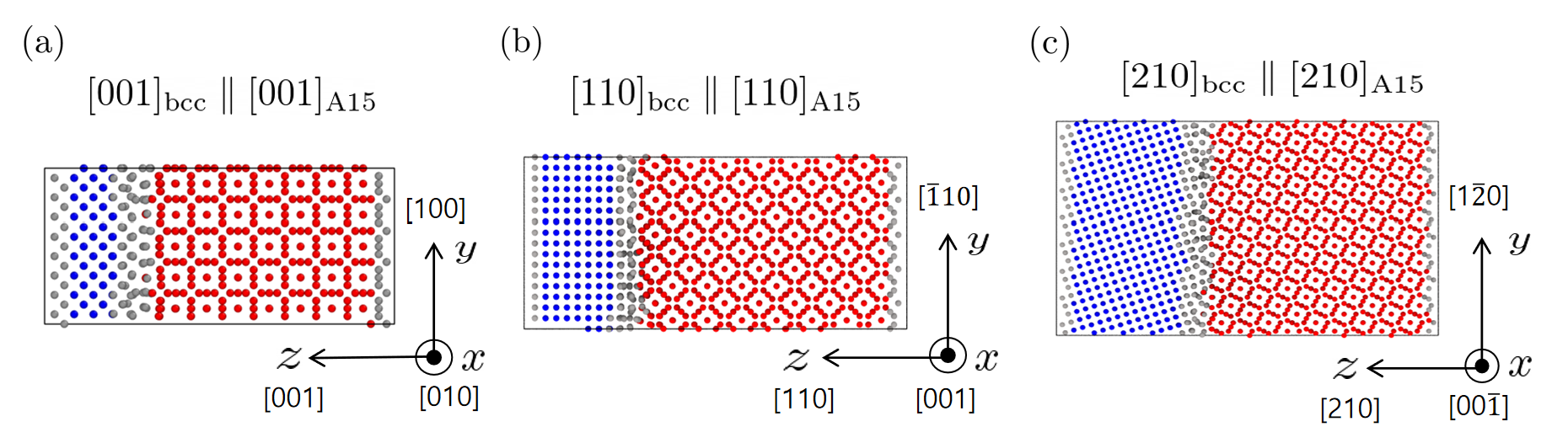}
    \caption{Illustration of supercells with interfaces for different orientation relationships. (a) $[001]_{\text{bcc}}\parallel$~$[001]_{\text{A15}}$, (b) $[110]_{\text{bcc}}\parallel$~$[110]_{\text{A15}}$, and (c) $[210]_{\text{bcc}}\parallel$~$[210]_{\text{A15}}$. 
    Gray atoms represent the interface, where the left- and right-end gray atoms are fixed so that only one interface (the middle one) is mobile in simulations. A15 is colored in red and bcc in blue.}
    \label{fig:config_difforient}
\end{figure}

Semicoherent interfaces between bcc-W and A15-W  are constructed by matching different numbers of A15 and bcc unit cells in the interface plane. By pairing both phases along the [001], [110], and [210] direction normal to the interface, we obtain supercells of three distinct orientations relevant to the experimental observations in W-thin films, $[001]_{\text{bcc}}\parallel$~$[001]_{\text{A15}}$, $[110]_{\text{bcc}}\parallel$~$[110]_{\text{A15}}$, and $[210]_{\text{bcc}}\parallel$~$[210]_{\text{A15}}$. The dimensions of the supercells parallel to the interface are optimized for the bcc lattice constant ($T=0$~K), with 128 atoms in a bcc layer, resulting in a slight compression of 0.57\% of A15. The initial setup of the interfaces is optimized by relaxing atomic positions in $xyz$ and cell dimensions in $z$ direction normal to the interface. Periodic boundary conditions are applied in all dimensions. To track the motion of a single interface,  the atoms in one of the two interfaces in the supercell are kept fixed. Supercells of various orientations are visualized in Fig,~\ref{fig:config_difforient} with more details provided in Tab.~\ref{tab:difforient}.
\begingroup
\setlength{\tabcolsep}{7pt} 
\renewcommand{\arraystretch}{1.1} 
\begin{table}[htbp]
\caption{Supercells with different orientation relationships with  bcc and A15 paired along $z$ direction. The corresponding atomic configurations are shown in Fig.~\ref{fig:config_difforient}. All three systems have 128 atoms in a bcc layer and same mismatches between bcc and A15, resulting in a 0.57\% compression of A15. The A15 phase has two planar densities as it has two inequivalent sublattices that are 12- and 14-fold coordinated resulting in two types of layers.}
\begin{tabular}{lccccccc}
\toprule
\multirow{2}{*}{Paired orientation} &
\multirow{2}{*}{\begin{tabular}[c]{@{}c@{}}Cell length (\AA)\\  $L_{x}\times L_{y}\times L_{z}$\end{tabular}} & 
\multicolumn{2}{c}{\begin{tabular}[c]{@{}c@{}}$N_\text{unit cell}$  in \\ $x\times y$ direction \end{tabular}} & 
\multirow{2}{*}{$N_\text{atoms}$} & \multicolumn{3}{c}{\begin{tabular}[c]{@{}c@{}}Planar densities\\  (atom/nm$^2$)\end{tabular}} \\ \cmidrule{3-4} \cmidrule{6-8}  &  & bcc & A15   &    & bcc  & \multicolumn{2}{c}{A15}\\ 
\midrule 
$[001]_\text{bcc} \parallel [001]_\text{A15}$ &  $25.34\times 50.69\times57.11$ &   $8\times16$   &  $5\times10$ &    4352   &  9.96 &  11.68   &  3.89  \\ 
$[110]_\text{bcc} \parallel [110]_\text{A15}$  &   $35.84\times25.34\times79.92$  &  $8\times8$  &  $5\times5$  &     4302   &  14.09  &      11.01 &   5.50     \\ 
$[210]_\text{bcc} \parallel [210]_\text{A15}$ &      $ 56.67\times50.69\times96.62$   &   $8\times16$     &    $5\times10$    &  15966   &    4.46        &    6.96     &   3.48        \\ \bottomrule
\end{tabular}
\label{tab:difforient}
\end{table}
\endgroup

The LAMMPS package~\cite{LAMMPS} is used for molecular dynamics (MD) simulations. To describe the interactions between W atoms, we use an embedded atom method (EAM) potential~\cite{zhou2004}, which has been shown to provide an accurate description of the fundamental properties of W. All MD simulations are performed in the canonical ensemble ($NVT$) using a Nos\'{e}-Hoover thermostat~\cite{evans1985nose,noseUnifiedFormulationConstant1984,hooverCanonicalDynamicsEquilibrium1985} with an integration time step of $\Delta t=2$~fs. The damping parameter of the thermostat is well tested and set to 0.1~ps to ensure the temperature fluctuations are appropriate. Before every simulation, the system is thermally equilibrated for $t_\text{eq}=1$~ns.

In the MD simulations of interface migration at elevated temperatures, the formation time of a bcc layer $\tau_\text{bcc}$ is recorded in each run. By monitoring the bcc layer adjacent to the migrating interface and tracking the formation of a new bcc layer, the interface velocity is calculated as $v=d_{(hkl)}/\tau_{bcc}$, where $d_\text{(hkl)}$ is the inter-plane distance normal to the interface. An average velocity is computed from a minimum of 100 values at each temperature. The formation rate of bcc layers relates to the formation time by $k=1/\bar{\tau}_\text{bcc}$. An apparent activation energy can then be derived from the Arrhenius relationship between transformation rate and temperature via $k=A \exp(-\frac{E^\text{act}}{k_{B}T})$, where $A$ is a prefactor.

Metadynamics simulations are performed with an in-house developed \textit{fix} package that is added to the LAMMPS~\cite{LAMMPS} code. For the $[110]_\text{bcc}\parallel[110]_\text{A15}$ system, we employed the system size listed in Tab.~\ref{tab:difforient}, while for $[001]_\text{bcc}\parallel[001]_\text{A15}$, we used a smaller simulation cell with $N=2176$ atoms consisting of an $8\times8$ bcc unit cells paired with $5\times5$ A15 unit cells along the $x$ and $y$ directions. 
All metadynamics parameters have been well-tuned for each run. The bias potential is deposited with a frequency of $\nu=0.4$~ps. The width of the deposited Gaussians is set to $\sigma=2\times10^{-3}$ for the $[001]_\text{bcc}\parallel[001]_\text{A15}$ and $\sigma=1\times10^{-3}$ for the $[110]_\text{bcc}\parallel[110]_\text{A15}$ system. 
In the sampling of the A15$\to$bcc transformation, the height of the Gaussians is $h=10^{-3}$~eV for $[001]_\text{bcc}\parallel[001]_\text{A15}$ and $h=4\times10^{-3}$~eV for $[110]_\text{bcc}\parallel[110]_\text{A15}$. In order to sample both transformation directions A15$\leftrightarrow$bcc, the height of the Gaussians needed to be increased to $h=10^{-2}$~eV due to the large energy difference between the two phases. 
All metadynamics simulations are performed at  $T=1300$~K.
The deposited bias potential was recorded on a grid with 1401 points for $s\in[-0.2,1.2]$ in all runs. 

\section{Results and discussion}
\subsection{Importance of disorder during interface migration }
\label{subsec:disorderinterface}
To explore the interface migration process between A15 and bcc in the $[001]_\text{bcc}\parallel[001]_\text{A15}$ orientation, 
we perform metadynamics simulations sampling the A15$\to$ bcc transformation along the path CV shown in Fig.~\ref{fig:pathCV}(a) in the $Y^\text{bcc}$-$Y^\text{A15}$ space. The path consists of ten equidistant nodes between the two endpoints at $\{Y^\text{bcc}:0.15, Y^\text{A15}:0.55\}$ and $\{Y^\text{bcc}:0.60, Y^\text{A15}:0.10\}$.  
Metadynamics runs are terminated when the value of the path CV reaches $s=1$, just before recrossing, when a single one-way A15$\to$bcc transformation is fully sampled, and the accumulated bias potential is recorded  at this point.
As discussed below, a continued sampling of both the A15$\to$bcc and bcc$\to$A15 transformations requires the addition of a restraining potential.  In this initial sampling, we focus on the A15$\to$bcc transformation only allowing us to analyse the atomistic mechanisms without any restrains.
In total, 60 independent metadynamics runs were performed.

\begin{figure}[htbp]
    \centering
    \includegraphics[width=0.99\textwidth]{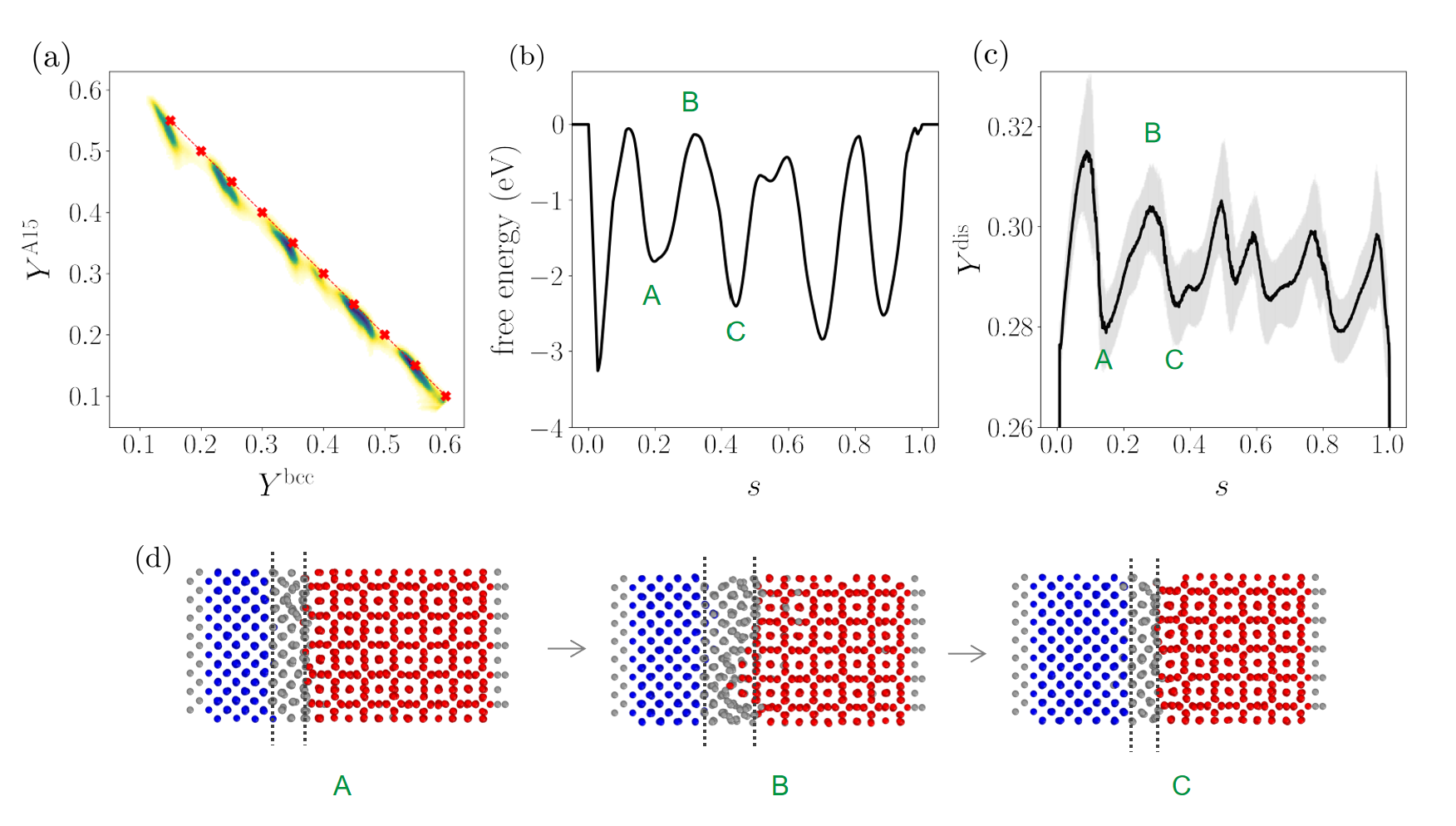}
    \caption{(a) Average path density in the $Y^\text{bcc}$-$Y^\text{A15}$ space
 together with the predefined path (red). The color gradient corresponds to an increase in density from light to dark. (b) Average free energy profile along the path collective variable. (c) Average amount of disordered interface projected on the path collective variable. The shaded area indicates the standard deviation in $Y^\text{dis}$. (d) Representative configurations at energy minima A and C, and transition state B. A15 phase is colored in red, bcc in blue, and disorder in gray, respectively. The dashed lines mark the boundary of the disordered interface between A15 and bcc. When the system is in the free energy minima, it has well defined crystalline layers with minimum amount of disorder at the  interface; at the transition state, the width of the disordered region increases  to facilitate the migration. From  A to C, approximately 4 layers of A15 are transformed into bcc along the [001] direction.}
\label{fig:forward_metady}
\end{figure}

In all runs, a complete A15$\to$bcc transformations is captured within 70~ns, showing a stepwise decrease in A15 and corresponding increase in bcc which indicates a layer-by-layer growth (see Supplementary Fig. S1). In addition, there is a noticeable amount of disorder at the migrating interface. The average path density in the $Y^\text{bcc}$-$Y^\text{A15}$ 
space extracted from the metadynamics runs is shown in Fig.~\ref{fig:forward_metady}(a).  
Since there is no restraining potential enforcing the A15$\to$bcc transformation to closely follow the predefined path in the $Y^\text{bcc}$-$Y^\text{A15}$ space, the system is free to explore a wide range of interface configurations that correspond to the same value of the path CV perpendicular to the path.
Indeed, the average path density  deviates from the predefined linear path and follows a zig-zag pattern, where high density regions are coloured in  blue/green and low density regions in  yellow in Fig.~\ref{fig:forward_metady}(a).  
The highest path density regions correspond to the five energy minima in the average free energy profile shown in Fig.~\ref{fig:forward_metady}(b), and the lowest density regions 
to the transition states. 
The zig-zag pattern indicates a gradual decrease in the amount of crystalline bcc and A15 and corresponding increase in disorder as the system moves away from the linear path, followed by a transition restoring the amount of crystalline phases but with changed relative phase fractions. 
Correspondingly, each minimum in the free energy profile represents configurations with complete layers of the two crystalline phases  along the [001] direction. As the interface migrates, moving the system from one free energy minimum to the next, an energy barrier of approximately $1.7\pm0.6$~eV needs to be overcome.  
The barrier is in agreement with the activation energy of the A15$\to$bcc transformation in W thin films extracted from high-temperature MD simulations ($\Delta E = 1.7$ eV) and differential scanning calorimetry measurements ($\Delta E = 2.2\pm 0.1$ eV)~\cite{barmak_TransformationTopologically_2017a}. 
In this study, the mechanism underlying the transformation was described as collective movements of W atoms within a disordered interface layer~\cite{barmak_TransformationTopologically_2017a}.
Indeed, our results similarly show that the disordered region at the interface is closely associated with the transformation mechanism and free energy profile. This interface region between A15 and bcc spans $4-6$ layers along [001] with a width of $4-7$~\AA. Within the current supercell, this corresponds to approximately 28-33\% disorder including a constant contribution from the second, fixed interface.
As shown in Fig.~\ref{fig:forward_metady}(c), the average amount of disorder at the interface strongly correlates with the free energy profile. The minima and maxima in the amount of disorder are located at the same positions as the free energy minima and saddles. At the free energy minima, the system exhibits the least amount of disorder at the interface with well-defined crystalline layers, see  configurations A and C in Fig~\ref{fig:forward_metady}(d). As the interface migrates, disorder increases, where a 5\% increase corresponds to an expansion of the interface region by approximately 3~\AA~in the [001] direction.
As the width of the disordered region increases (configuration B in Fig~\ref{fig:forward_metady}(d)), it enables more flexibility in atomic shuffling and structural rearrangement to facilitate structural transformations. The growth of the disordered interface region dissolves antecedent A15 layers, facilitating barrier crossings, followed by the formation of bcc layers along [001]. Moving from one free energy minimum  to the next, approximately four layers of A15 are transformed into bcc through the migration of a disordered interface with varying thickness.

\begin{figure}[tp]
    \subfigure
    {\includegraphics[width=0.34\textwidth]
    {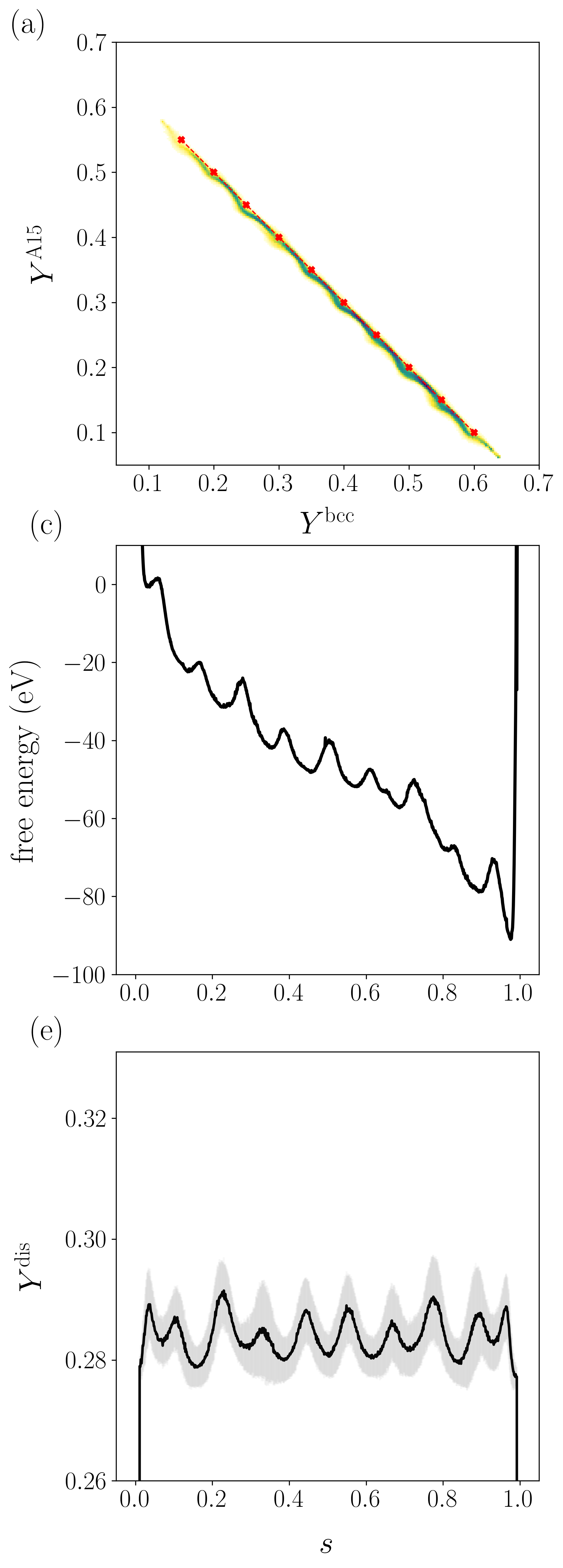}}
    \subfigure
    {\includegraphics[width=0.34\textwidth]{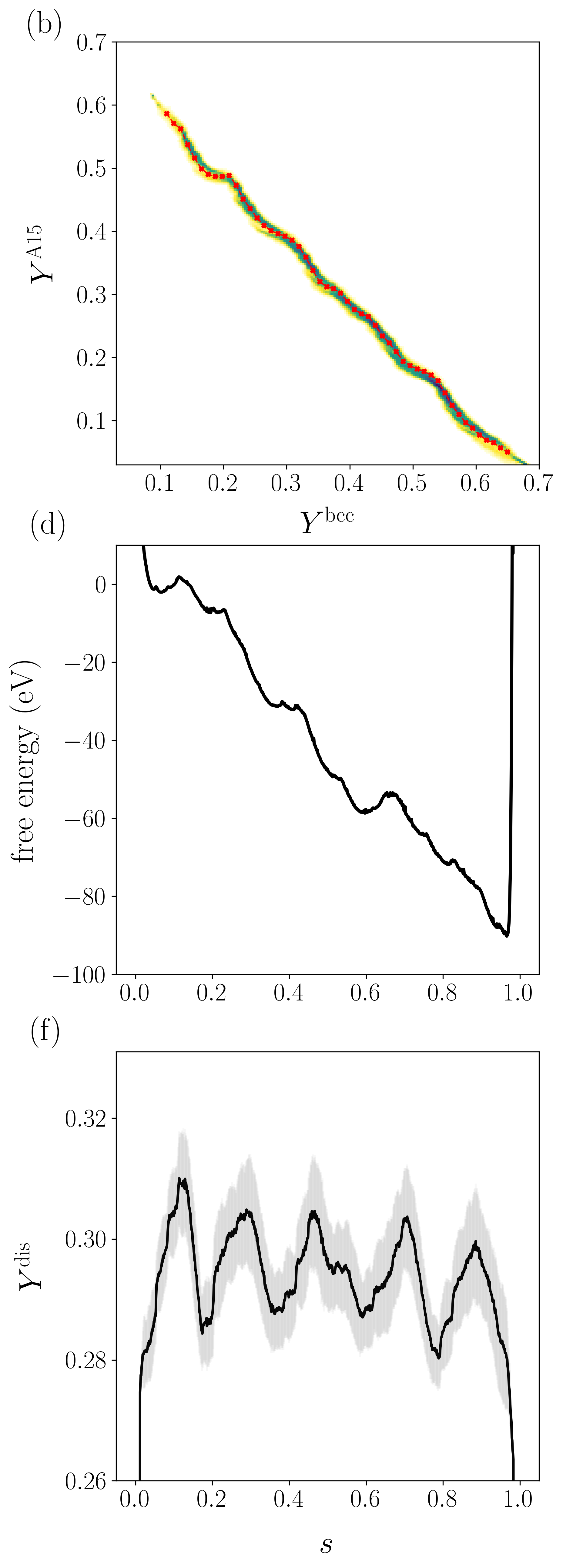}}
    \caption{Average path density with the predefined (a) linear and (b) bumpy path (red) in the $Y^\text{bcc}$-$Y^\text{A15}$ space. (c) and (d) Free energy profiles along the path collective variable for the (c) linear path after $t=781$~ns and (d) the bumpy path after $t=1554$~ns. The average amount of disorder along the path collective variable is shown for the (e) linear and (f) bumpy path. 
    }
    \label{fig:tight_bumpy_combined}
\end{figure}

The average free energy profile depicted in Fig~\ref{fig:forward_metady}(b) only reflects the one-way transformation from A15 to bcc, but does not resemble the complete free energy profile since the reverse bcc$\to$A15 transformation has not been sampled.
To be able to sample the transformation in both directions, A15$\leftrightarrow$bcc, and obtain a converged estimate of the complete free energy profile, we need to apply a tight restraining potential on the distance function of the path, Eq.~\eqref{eqn:res_V}.
This enforces the transformations to proceed closely along the predefined path in the $Y^\text{bcc}$-$Y^\text{A15}$ space and avoids that the system is being pushed into unphysical high-energy regions of configuration space. 
The parameters of the restraining potential are set to $\kappa=30$~eV, $n=2, \epsilon=0.05$, and $z_{0}=0$. 
We  perform metadynamics simulations with two different paths: (i) the linear path also used in the unrestrained simulations; (ii) a {\it bumpy} path following the average path density in the unrestrained A15$\to$bcc transformations in Fig.~\ref{fig:forward_metady}(a). A detailed description and an illustration of the predefined bumpy path are given in the supplementary information (see Fig.~S2). 
The average path densities extracted from simulations along the linear and bumpy path are shown in Figs.~\ref{fig:tight_bumpy_combined}(a) and (b), respectively.
In both cases, the trajectories sampling the A15$\leftrightarrow$bcc transformations follow the predefined path with little deviation. In comparison to the fast sampling of A15$\to$bcc transformations within 70~ns, it takes $t=320$~ns to sample the first bcc$\to$A15 transformation for the linear path, and $t=240$~ns for the bumpy path. We extend the sampling of the continued A15$\leftrightarrow$bcc transformation   
to $t=781$~ns for the linear path and $t=1554$~ns for the bumpy path to extract converged free energy profiles from the metadynamics simulations shown in Figs.~\ref{fig:tight_bumpy_combined}(c) and (d). 
In both cases, we obtained minimal hysteresis in the path CV during recrossing in metadynamics and a well-defined
free energy profiles, demonstrating that the selected path CV provides a good description of structural transformations (see also supplemental Fig.~S3).
The relative fast A15$\to$bcc transformation but extremely slow bcc$\to$A15 transformation is due to the energy difference between the two phases.
The lattice mismatch in the current cell geometry compresses A15 by 0.57\% (see section~\ref{subsec:comp_detail}), which leads to an energy gain and decrease in pressure when growing bcc, whereas the formation of A15 leads to an increase in pressure and energy. Consequently, the free energy profile associated with the A15$\leftrightarrow$bcc transformation exhibits substantial asymmetry spanning a range of 92~eV for both the linear and bumpy path, which reflects the lower free energy of bcc  compared to A15. In order to efficiently explore the vastly asymmetric free energy landscapes, the height of the deposited Gaussians in the bias potential had to be increased by a factor of 10 compared to the unrestraint simulations. This may reduce the resolution of the free energy profiles, in particular around the transition states. However, in our study, the convergence of the metadynamics simulations indicates that the accuracy of the computed free energy profiles is sufficient. By employing refinements in the sampling, such as well-tempered metadynamics~\cite{laioMetadynamicsMethodSimulate2008,valsson2016enhancing,bussi2020using}, the estimate of the free energy barriers could be improved further.  

While the overall energy scale is comparable for both path CVs, the details of free energy profiles along the linear and bumpy path are notably different.
In particular, for the transformation along the linear path, the free energy profile shows ten energy minima, whereas for the profile along the bumpy path, there are approximately five. 
The differences in the free energy profiles are attributed to a change in the transformation mechanism. Specifically, as shown in Figs.~\ref{fig:tight_bumpy_combined}(e) and (f), the amount of disorder observed during the structural transformation differs for the two paths. For the bumpy path, as it follows the path density of the unrestrained A15$\to$bcc transformation, the width of the disordered interface region remains flexible and changes as the interface moves.   
As a result, the transformation  proceeds again through the migration of a disordered interface with varying width, allowing approximately four crystalline layers of A15 to transform from one free energy minimum to the next in a single step.
The estimated free energy barrier for interface migration during the A15$\to$bcc transformation is approximately $1.3-1.5$~eV, demonstrating quantitative agreement with the barrier obtained for the A15$\to$bcc transformation without any restraint. 
Due to the free energy difference between bcc and A15, the barriers associated with interface migration during bcc$\to$A15 transformation are much larger than for A15$\to$bcc.  
In the current supercell setup, the potential energy difference between bcc and A15 is 86~meV/atom. Transitioning from one free energy minimum to the next along the  A15$\to$bcc transformation increases the number of bcc atoms by $256-320$, resulting in a potential energy gain of $22-27$~eV which is comparable to the free energy difference between subsequent minima.
Along the linear path, the transformation proceeds through a disordered interface region with restricted fluctuations, see Fig.~\ref{fig:tight_bumpy_combined}(e), where the amount of disorder is significantly lower than observed along the bumpy path.
The smaller amount of disorder observed along the linear path results in larger energy barriers and additional minima and saddles in the free energy profile compared to the bumpy path, see Figs.~\ref{fig:tight_bumpy_combined}(c) and (d), respectively.  
Due to the limited fluctuations at the disordered interface, only approximately two layers of crystalline A15 transform along [001] at a time, with a much larger barrier of $4-6$~eV to an intermediate local minimum, followed by the transformation of the remaining two layers with an additional $1.3-1.7$~eV barrier.
The larger barrier is a consequence of the restrained width of the disordered region at the migrating interface, which leads to an 
increase in pressure and reduced space for structural rearrangements making the  transformation rather costly (see supplemental Fig.~S4).

Our simulations demonstrate that the formation of a disordered region at the interface plays a pivotal role in the transformation mechanism, fluctuating in width and facilitating structural rearrangements during solid-solid transformations. 
Furthermore, any collective variable suitable to describe the transformation does not only have to distinguish between and drive the formation of the crystalline phases, but also needs to correctly capture the fluctuations in disorder at the interface.

\subsection{Interface migration along different crystallographic directions}
\label{subsec:difforient}

The formation of a disordered region at the moving interface is an important aspect of the transformation mechanism.  The properties of the interface between the crystalline phases depend, however, strongly on the orientation relationship and coherency between the phases.  
Here, we focus on three systems with orientation relationships similar to those observed in W thin films, namely $[001]_\text{bcc}\parallel[001]_\text{A15}$, $[110]_\text{bcc}\parallel[110]_\text{A15}$, and $[210]_\text{bcc}\parallel[210]_\text{A15}$. 
All three systems were  setup with  supercell geometries optimized for bcc, leading to a 0.57\% compression of A15 (see Sec.~\ref{subsec:comp_detail}). This results in a driving force characterized by the potential energy difference  between bcc and A15 of approximately 86 meV/atom.
Conducting enhanced sampling simulations for a number of large supercells with several interface orientations is extremely demanding computationally. Therefore, we perform high temperature MD simulations of the A15$\to$bcc transformation for an initial screening of the various systems to identify potentially interesting interface geometries.
  
\begin{figure}[htbp]
    \centering
    \subfigure{
    \includegraphics[width=0.465\textwidth]{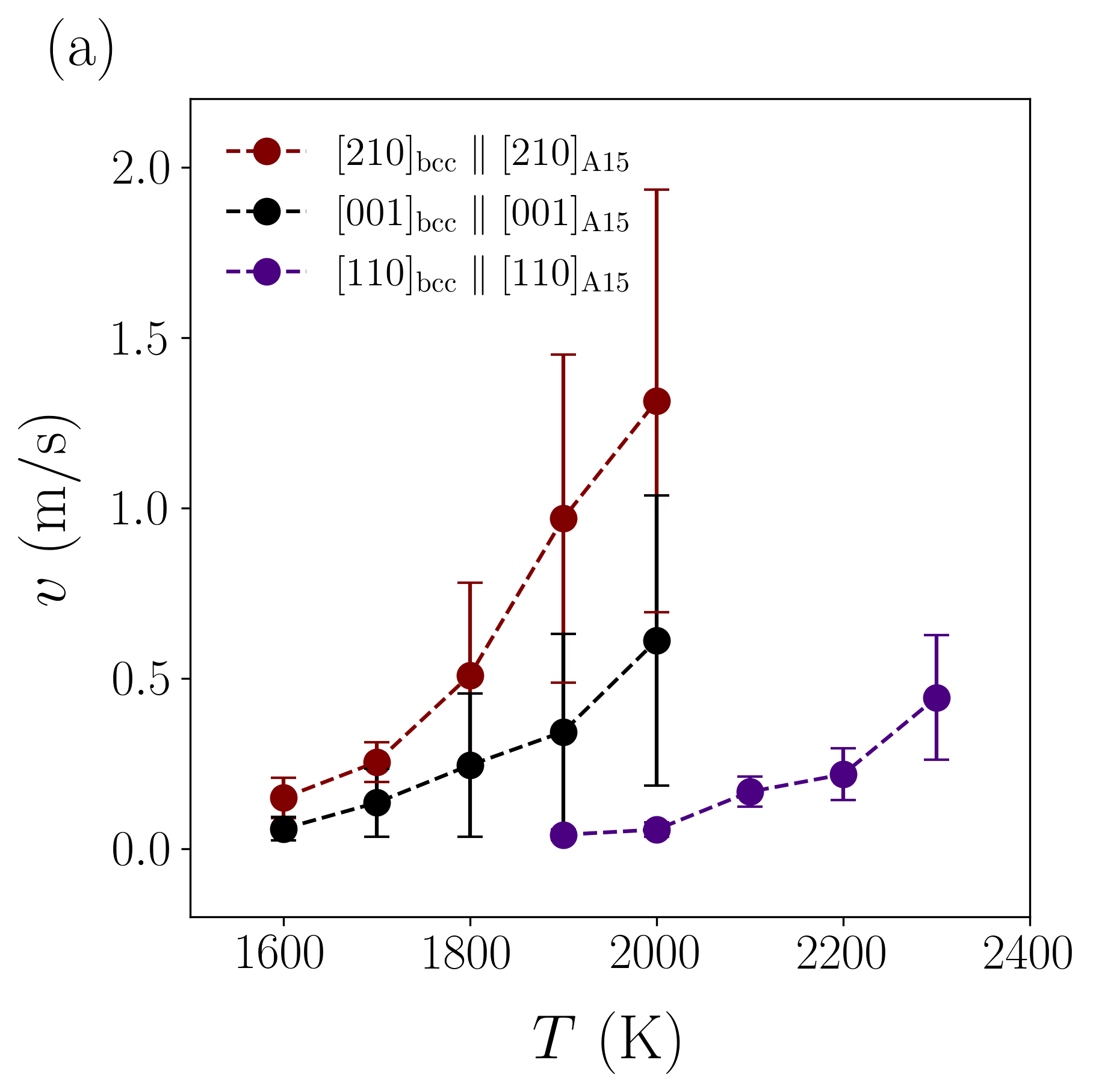}
    \label{fig:diffv}
    }
    \subfigure{
    \includegraphics[width=0.46\textwidth]{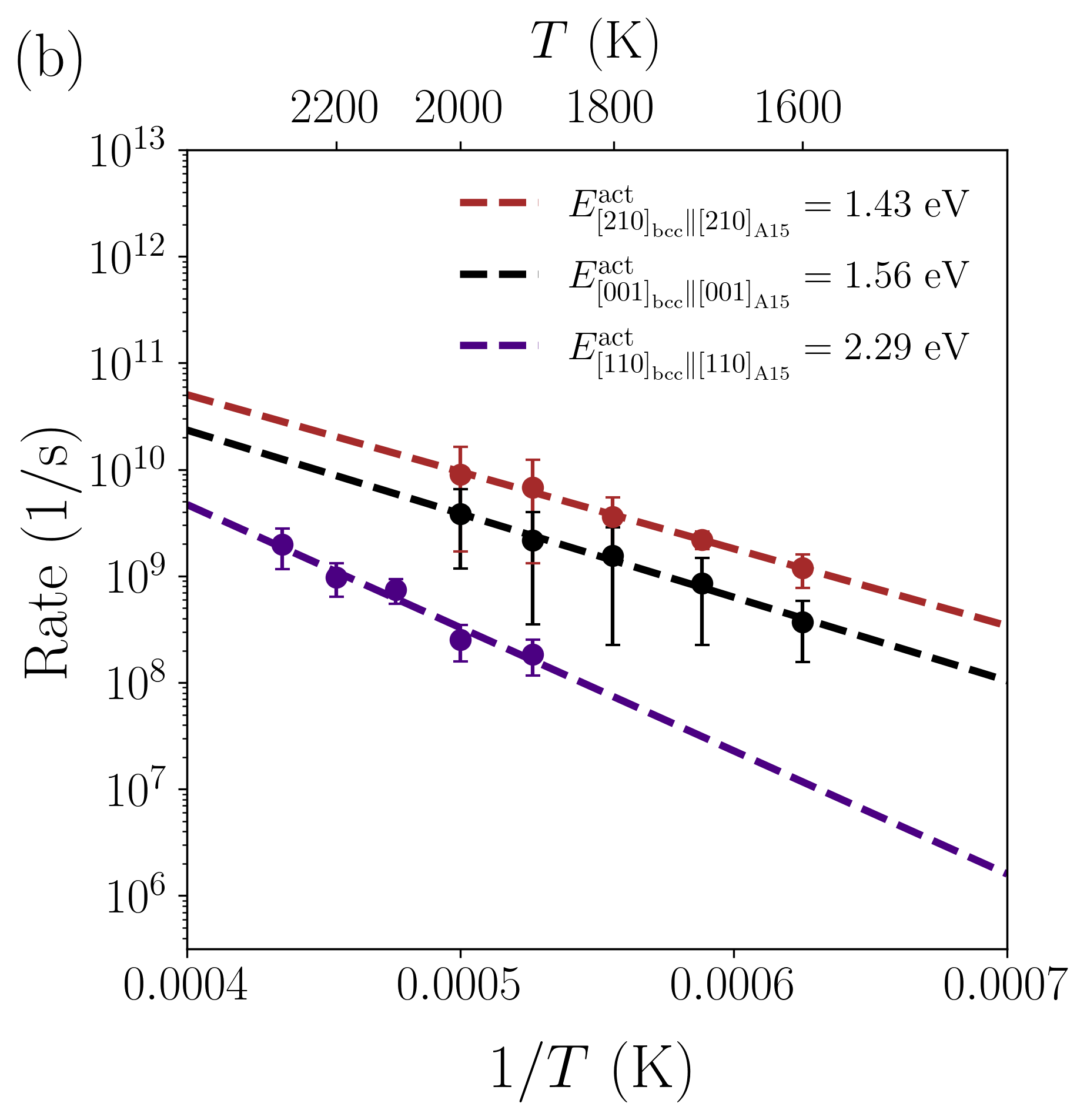}
    \label{fig:diffact}
    }
    \caption{ (a) Average interface velocity as a function of temperature, and (b) Arrhenius plot of interface migration rate vs. the inverse temperature for different orientation relationships. Each point in the average velocity is calculated from a minimum of 100 values in 10 MD runs at every temperature.}
    \label{fig:velocity_difforient}
\end{figure}

To be able to register any movement of the interface, simulations are run at very high temperatures between $T=1600-2300$~K.
For all systems, the A15$\to$ bcc transformation proceeds through the migration of a disordered interface where the disordered region dissolves preceding A15 layers, while bcc progressively grows layer-by-layer perpendicular to the interface. The average interface velocity at various temperatures for each orientation relationship 
is presented in Fig.~\ref{fig:velocity_difforient}~(a).
The interface velocity increases with increasing temperature and
varies significantly between distinct orientations 
with $v_{[210]_\text{bcc}\parallel[210]_\text{A15}}> v_{[001]_\text{bcc}\parallel[001]_\text{A15}}> v_{[110]_\text{bcc}\parallel[110]_\text{A15}}$. Notably, the $[110]_\text{bcc}\parallel[110]_\text{A15}$ interface appears to be extremely immobile compared to the other two orientations, with its velocity at $T=2000$~K being only 0.05~m/s, comparable to $[001]_\text{bcc}\parallel[001]_\text{A15}$ at a much lower temperature of $T=1600$~K. This is also reflected in the activation energies estimated from the Arrhenius relation between temperature and rate for interface migration shown in Fig.~\ref{fig:velocity_difforient}~(b),  with $E^\text{act}_{[110]_\text{bcc}\parallel[110]_\text{A15}}= 2.29~\text{eV} > E^\text{act}_{[001]_\text{bcc}\parallel[001]_\text{A15}} = 1.56~\text{eV} > E^\text{act}_{[210]_\text{bcc}\parallel[210]_\text{A15}}$ = 1.43~eV. 
The results for the $[001]_\text{bcc} \parallel [001]_\text{A15}$ interface are in quantitative agreement with the free energy barriers in our enhanced sampling simulations in Section~\ref{subsec:disorderinterface}, as well as with previous high-temperature MD simulations for slightly smaller (($8\times8)\text{bcc} \parallel(5\times5)\text{A15}$) and larger (($16\times16)\text{bcc} \parallel (10\times10)\text{A15}$) supercells reporting activation energies of 1.7~eV and 1.4~eV, respectively~\cite{barmak_TransformationTopologically_2017a}. 
The comparability in activation energies between different cell sizes indicates that the cell sizes used in our simulations do not introduce significant errors due to periodic boundary conditions frustrating the layerwise transformation. 
Our analysis reveals that
the variance in interface velocities and corresponding activation energies are predominantly governed by the planar densities of both the growing and shrinking phase parallel to the moving interface. Among the three orientations, bcc and A15 in (110) planes have the highest packing density (14.09~atom/nm$^2$ for bcc and an average of 8.26~atom/nm$^2$ for A15, see Tab.~\ref{tab:difforient}). 
Consequently, interface migration along this direction is largely hindered as the disordered region is frustrated  
in the densely packed (110) planes with limited free volume for atomic shuffling, resulting in a high activation energy.
The low planar density in both bcc and A15 (210) layers, on the other hand, easily accommodates  structural rearrangements in the disordered interface region as evidenced by the high interface velocity and low activation energy in ${[210]_\text{bcc}\parallel[210]_\text{A15}}$.  
The interface velocity is commonly connected with the interface mobility $M$ through the driving force $P$, $v=MP$~\cite{mendelevInterfaceMobilityDifferent2002}. 
Since all three systems are setup with identical lattice mismatch and contain the same number of bcc atoms per layer parallel to the interface, they share comparable initial driving forces given by the difference in potential energy and pressure.  
The variation in interface velocities can, therefore, mainly be attributed to the difference in interface mobilities following $M_{(210)}>M_{(001)}>M_{(110)}$, which is directly correlated with the planar densities. 

Consistent with the discussion in Section~\ref{subsec:disorderinterface}, 
the properties of the disordered region at the moving interface remains a key feature in the transformation mechanism for different orientation relationships. 
Specifically, the planar density of the crystalline phases perpendicular to the growth direction directly impacts the density of the disordered region and its ability to facility atomic rearrangements, which, in turn, determine the interface mobility.
While high-temperature MD simulations enable 
a fast screening of interface migration in  
different orientation relationships, the fast kinetics at elevated temperature might favour transformation mechanisms that are not representative at lower temperatures closer to experimental conditions, as discussed in the following Section. 

\subsection{Exploring low mobility interface migration towards realistic temperatures}
\label{subsec:lowTmechanism}
The  $[110]_\text{bcc}||[110]_\text{A15}$ interface exhibits by far the most sluggish migration and is seemingly immobile 
even at elevated temperatures that are far beyond experimental working conditions. 
This prompts an intriguing question regarding how interface migration proceeds at more realistic temperatures approaching experimental settings and whether 
the dominant transformation mechanisms differ from those observed in high-temperature MD simulations. 
Here, we employ enhanced sampling simulations to explore 
the migration of the low mobility $[110]_\text{bcc}||[110]_\text{A15}$ interface at a temperature of $T=1300$~K, much lower than temperatures applicable in straightforward MD.  
Specifically, we performed a series of 30 metadynamics simulations to sample the A15$\to$bcc transformation  along a linear path in the $Y^\text{bcc}$-$Y^\text{A15}$ space between two end points $\{Y^\text{bcc}:0.20,Y^\text{A15}:0.55\}$ and $\{Y^\text{bcc}:0.65,Y^\text{A15}:0.10\}$ (see supplementary Fig.~S5). 
The bcc$\to$A15 transformation is not sampled here since this requires the addition of a restraining potential due to the substantial asymmetry in the free energy profile, which leads to considerable hysteresis in unrestrained simulations.

\begin{figure}[htbp]
    \centering
    \includegraphics[width=0.99\textwidth]{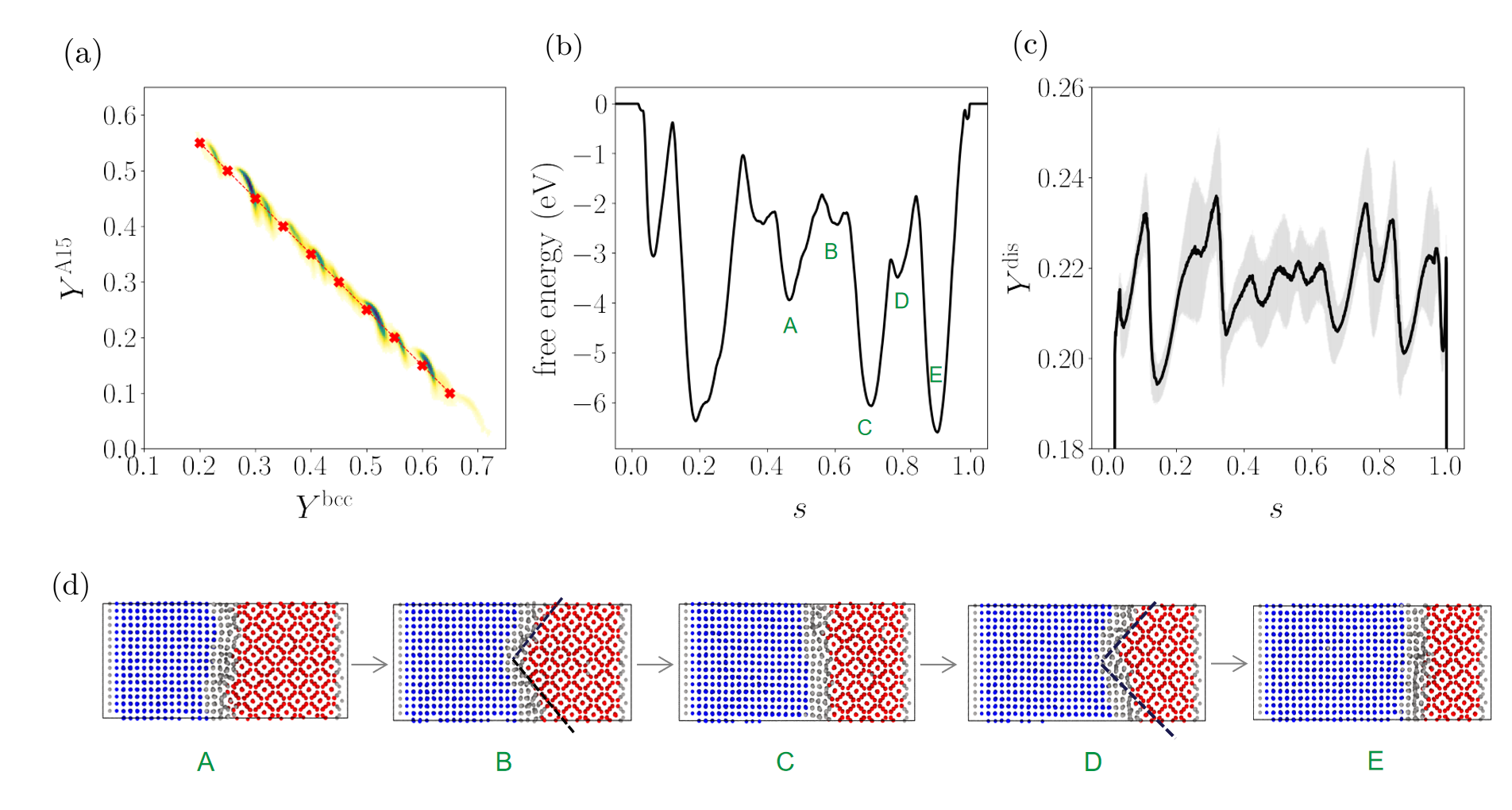}
    \caption{A15$\to$bcc transformations for $[110]_\text{bcc}\parallel[110]_\text{A15}$ interfaces. (a) Average path density together with the predefined linear path (red) in the $Y^\text{bcc}$-$Y^\text{A15}$ space; (b) average free energy profile; (c) amount of disorder at the interface projected along the path collective variable;  the black solid line represents the average amount of disorder, the gray shaded area represents the standard deviation. (d) Representative configurations for different states along the free energy profile. A15 is colored in red, bcc in blue, and disorder in gray, respectively. The dashed lines in the configurations indicate interfacial ledge formation in the [100] direction.}
    \label{fig:pb110}
\end{figure}

The A15$\to$bcc transition occurs within  $80-240$~ns across individual runs.   The average path density in Fig.~\ref{fig:pb110}(a) exhibits again a  zig-zag pattern with high (green/blue) and low (yellow) density regions, indicating that the sampled trajectories diverge from the predefined linear path. The average free energy profile of the transformation, shown in Fig.~\ref{fig:pb110}(b), 
exhibits large complexity, consisting of a number of energy minima with varying depths connected by rugged transition states.
The average amount of disorder at the interface projected along the path collective variable, Fig.~\ref{fig:pb110}(c), shows the same characteristics as the free energy, where the smallest amounts of disorder corresponds to free energy minima and the largest amounts of disorder to saddles. This indicates that an increase in width of the disordered region at the moving interface
is needed to facilitate the barrier crossing from one energy minimum to the next.
The rugged free energy profile as well as the average amount of disorder suggest that the transformation does not proceed through a  simple layer-by-layer mechanism.
Indeed, the interface migrates by combining interfacial ledge formation along [100] and subsequent flattening of the ledge to form complete layers of bcc and A15 along [110]. In Fig.~\ref{fig:pb110}(d), representative snapshots of each step in this complex transformation are illustrated. 
At the free energy minima, system configurations consist of full A15 and bcc layers with (110) planes, separated by a disordered interface of minimal width (configurations A, C, and E in Fig.~\ref{fig:pb110}(d)). 
Transitioning between energy minima involves an expansion of the disordered interface region and bcc growth occurs along the [100] direction rather than [110], reaching a local minimum at the saddle point (see configurations B and D in Fig.~\ref{fig:pb110}(d)). 
The interfacial ledge forms due to the difference in interface mobility in the [100] and [110] direction. (110) planes remain exceptionally immobile even at high temperatures, as demonstrated in the previous section. Consequently, interface migration along [100] exceeds that along [110], leading to the formation of interfacial ledges.
The periodic boundary conditions limit the further extension of the interfacial ledges   
and the transformation continues by flattening along [110] to form a complete layer,
reaching a local free energy minimum again with well-defined (110) planes of bcc and A15. 
In a macroscopic system, the propagation of interfacial ledges driven by differences in intrinsic mobilities along distinct orientations is anticipated to lead to faceting at the migrating interface. This, in turn, could give rise to diverse morphological characteristics in both the parent and product phase.
Despite being limited by the simulation cell geometry, preferred growth along the high mobility direction is clearly evident at this lower temperature, which has not been observed in the dynamical evolution of the interface in high-temperature MD simulations. 
At elevated temperatures, substantial thermal fluctuations enable the system to overcome the energy barrier associated with growth along the [110] direction. Conversely, at lower temperatures, growth takes an alternative path along [100] that is energetically more favourable.  
This indicates that the predominant mechanism governing the A15$\to$bcc transformation along [110] is strongly temperature-dependent.  
Therefore, extrapolating the transformation mechanism observed in high-temperature MD simulations to low-temperature conditions would lead to an incomplete understanding of the mechanism.
This highlights the necessity to explore structural transformations at varying thermal conditions. To attain a comprehensive understanding of the complete picture of different transformation mechanisms, high-temperature MD simulations need to be complemented with  low-temperature enhanced sampling.

\section{Conclusion}
Solid-solid transformations between structurally different crystalline phases are complex processes governed by the properties of the moving interface between the phases. Our results, employing extensive enhanced sampling simulations, reveal that a key aspect of the transformation mechanism is the formation of a disordered or amorphous region at the interface, which has also been observed experimentally in colloidal systems ~\cite{peng2015two,qi2015nonclassical,peng2017diffusive}, metals~\cite{wang1998thermodynamic,luo2005segregation,levitasCrystalcrystalPhaseTransformation2012,su2021solid}, and perovskites~\cite{levitas2014solid,song2022planar}.
Most notably, the disordered region facilitates the rearrangement of atoms during growth and varies in width as the interface migrates. For the A15$\to$bcc transformation in tungsten studied here, this fundamental mechanism is directly linked to the interface mobility, where orientation relationships between the two phases with low planar densities exhibit the largest mobility. For interfaces with high planar densities, the disordered region has a smaller free volume and becomes frustrated, limiting atomic rearrangements in the interface. Similarly, if the width of the disordered region is restrained, the shuffling of the atoms during the phase transformation is hindered which leads to a significant increase in the free energy barrier associated with interface migration.
Consequently, the incorporation of defects, such as vacancies, could increase the available space required for atomic shuffling and effectively enhance the interface mobility.

The interface mobility directly impacts the preferred growth direction.  The competition between different orientations is, however, temperature dependent. A direct consequence of this is that it is not always possible to extrapolate the growth behaviour observed in high-temperature MD simulations to lower temperatures. For the A15$\to$bcc transformation along the low mobility [110] direction, our enhanced sampling simulations revealed the formation of interfacial ledges that did not appear at elevated temperatures. In macroscopic systems, this growth behaviour would eventually give rise to the formation of distinct morphologies.  To be able to capture all relevant transformation mechanisms at meaningful temperatures, it is therefore essential to explore extended timescales with enhanced sampling methods.

Our findings offer an in-depth perspective on the fundamental atomic processes during interface migration and, specifically, the importance of the disordered interface region for the transformation mechanism. This could proof particularly valuable in promoting the control of microstructure evolution during structural phase transformations by manipulating the properties of interfaces, including structure, orientation, and defects, between different crystalline phases in metals and alloys.

\begin{acknowledgments}
The authors acknowledge the computation time by Interface-Dominated High Performance Materials (ZGH, Ruhr-Universit{\"a}t Bochum). 
This work is a project conducted during YL's PhD fellowship from the International Max Planck Research School for Interface Controlled Materials for Energy Conversion (IMPRS-SurMat). GDL acknowledges support from Conacyt-Mexico through fellowship Ref. No. 220644 and the Isaac Newton Trust Grant Ref No: 20.40(h). JR acknowledges financial support from the Deutsche Forschungsgemeinschaft (DFG) through the Heisenberg Programme project 428315600.
\end{acknowledgments}

\bibliography{W}

\providecommand{\noopsort}[1]{}\providecommand{\singleletter}[1]{#1}%
\begin{thebibliography}{83}%
\makeatletter
\providecommand \@ifxundefined [1]{%
 \@ifx{#1\undefined}
}%
\providecommand \@ifnum [1]{%
 \ifnum #1\expandafter \@firstoftwo
 \else \expandafter \@secondoftwo
 \fi
}%
\providecommand \@ifx [1]{%
 \ifx #1\expandafter \@firstoftwo
 \else \expandafter \@secondoftwo
 \fi
}%
\providecommand \natexlab [1]{#1}%
\providecommand \enquote  [1]{``#1''}%
\providecommand \bibnamefont  [1]{#1}%
\providecommand \bibfnamefont [1]{#1}%
\providecommand \citenamefont [1]{#1}%
\providecommand \href@noop [0]{\@secondoftwo}%
\providecommand \href [0]{\begingroup \@sanitize@url \@href}%
\providecommand \@href[1]{\@@startlink{#1}\@@href}%
\providecommand \@@href[1]{\endgroup#1\@@endlink}%
\providecommand \@sanitize@url [0]{\catcode `\\12\catcode `\$12\catcode `\&12\catcode `\#12\catcode `\^12\catcode `\_12\catcode `\%12\relax}%
\providecommand \@@startlink[1]{}%
\providecommand \@@endlink[0]{}%
\providecommand \url  [0]{\begingroup\@sanitize@url \@url }%
\providecommand \@url [1]{\endgroup\@href {#1}{\urlprefix }}%
\providecommand \urlprefix  [0]{URL }%
\providecommand \Eprint [0]{\href }%
\providecommand \doibase [0]{https://doi.org/}%
\providecommand \selectlanguage [0]{\@gobble}%
\providecommand \bibinfo  [0]{\@secondoftwo}%
\providecommand \bibfield  [0]{\@secondoftwo}%
\providecommand \translation [1]{[#1]}%
\providecommand \BibitemOpen [0]{}%
\providecommand \bibitemStop [0]{}%
\providecommand \bibitemNoStop [0]{.\EOS\space}%
\providecommand \EOS [0]{\spacefactor3000\relax}%
\providecommand \BibitemShut  [1]{\csname bibitem#1\endcsname}%
\let\auto@bib@innerbib\@empty
\bibitem [{\citenamefont {Kirby}\ \emph {et~al.}(1991)\citenamefont {Kirby}, \citenamefont {Durham},\ and\ \citenamefont {Stern}}]{kirbyMantlePhaseChanges1991}%
  \BibitemOpen
  \bibfield  {author} {\bibinfo {author} {\bibfnamefont {S.~H.}\ \bibnamefont {Kirby}}, \bibinfo {author} {\bibfnamefont {W.~B.}\ \bibnamefont {Durham}},\ and\ \bibinfo {author} {\bibfnamefont {L.~A.}\ \bibnamefont {Stern}},\ }\bibfield  {title} {\bibinfo {title} {Mantle phase changes and deep-earthquake faulting in subducting lithosphere},\ }\href {https://doi.org/10.1126/science.252.5003.216} {\bibfield  {journal} {\bibinfo  {journal} {Science}\ }\textbf {\bibinfo {volume} {252}},\ \bibinfo {pages} {216} (\bibinfo {year} {1991})}\BibitemShut {NoStop}%
\bibitem [{\citenamefont {Porter}\ and\ \citenamefont {Easterling}(2009)}]{porter2009phase}%
  \BibitemOpen
  \bibfield  {author} {\bibinfo {author} {\bibfnamefont {D.~A.}\ \bibnamefont {Porter}}\ and\ \bibinfo {author} {\bibfnamefont {K.~E.}\ \bibnamefont {Easterling}},\ }\href@noop {} {\emph {\bibinfo {title} {Phase transformations in metals and alloys (revised reprint)}}}\ (\bibinfo  {publisher} {CRC press},\ \bibinfo {year} {2009})\BibitemShut {NoStop}%
\bibitem [{\citenamefont {Smith}(1986)}]{smith1986principles}%
  \BibitemOpen
  \bibfield  {author} {\bibinfo {author} {\bibfnamefont {W.~F.}\ \bibnamefont {Smith}},\ }\href@noop {} {\emph {\bibinfo {title} {Principles of materials science and engineering}}}\ (\bibinfo  {publisher} {McGraw Hill Book Co., New York, NY},\ \bibinfo {year} {1986})\BibitemShut {NoStop}%
\bibitem [{\citenamefont {Manoharan}(2015)}]{manoharan2015colloidal}%
  \BibitemOpen
  \bibfield  {author} {\bibinfo {author} {\bibfnamefont {V.~N.}\ \bibnamefont {Manoharan}},\ }\bibfield  {title} {\bibinfo {title} {Colloidal matter: Packing, geometry, and entropy},\ }\href {https://doi.org/10.1126/science.1253751} {\bibfield  {journal} {\bibinfo  {journal} {Science}\ }\textbf {\bibinfo {volume} {349}},\ \bibinfo {pages} {1253751} (\bibinfo {year} {2015})}\BibitemShut {NoStop}%
\bibitem [{\citenamefont {Molodov}\ \emph {et~al.}(1999)\citenamefont {Molodov}, \citenamefont {Gottstein}, \citenamefont {Heringhaus},\ and\ \citenamefont {Shvindlerman}}]{molodovMagneticallyForcedMotion1999}%
  \BibitemOpen
  \bibfield  {author} {\bibinfo {author} {\bibfnamefont {D.~A.}\ \bibnamefont {Molodov}}, \bibinfo {author} {\bibfnamefont {G.}~\bibnamefont {Gottstein}}, \bibinfo {author} {\bibfnamefont {F.}~\bibnamefont {Heringhaus}},\ and\ \bibinfo {author} {\bibfnamefont {L.~S.}\ \bibnamefont {Shvindlerman}},\ }\bibfield  {title} {\bibinfo {title} {Magnetically forced motion of specific planar boundaries in {{Bi-bicrystals}}},\ }in\ \href {https://doi.org/https://doi.org/10.4028/www.scientific.net/MSF.294-296.127} {\emph {\bibinfo {booktitle} {Materials Science Forum}}},\ Vol.\ \bibinfo {volume} {294}\ (\bibinfo  {publisher} {{Trans Tech Publ}},\ \bibinfo {year} {1999})\ pp.\ \bibinfo {pages} {127--130}\BibitemShut {NoStop}%
\bibitem [{\citenamefont {Straumal}\ and\ \citenamefont {Baretzky}(2004)}]{straumal2004grain}%
  \BibitemOpen
  \bibfield  {author} {\bibinfo {author} {\bibfnamefont {B.}~\bibnamefont {Straumal}}\ and\ \bibinfo {author} {\bibfnamefont {B.}~\bibnamefont {Baretzky}},\ }\bibfield  {title} {\bibinfo {title} {Grain boundary phase transitions and their influence on properties of polycrystals},\ }\href {https://doi.org/https://doi.org/10.1023/B:INTS.0000028645.30358.f5} {\bibfield  {journal} {\bibinfo  {journal} {Interface Sci.}\ }\textbf {\bibinfo {volume} {12}},\ \bibinfo {pages} {147} (\bibinfo {year} {2004})}\BibitemShut {NoStop}%
\bibitem [{\citenamefont {Meiners}\ \emph {et~al.}(2020)\citenamefont {Meiners}, \citenamefont {Frolov}, \citenamefont {Rudd}, \citenamefont {Dehm},\ and\ \citenamefont {Liebscher}}]{meiners2020observations}%
  \BibitemOpen
  \bibfield  {author} {\bibinfo {author} {\bibfnamefont {T.}~\bibnamefont {Meiners}}, \bibinfo {author} {\bibfnamefont {T.}~\bibnamefont {Frolov}}, \bibinfo {author} {\bibfnamefont {R.~E.}\ \bibnamefont {Rudd}}, \bibinfo {author} {\bibfnamefont {G.}~\bibnamefont {Dehm}},\ and\ \bibinfo {author} {\bibfnamefont {C.~H.}\ \bibnamefont {Liebscher}},\ }\bibfield  {title} {\bibinfo {title} {Observations of grain-boundary phase transformations in an elemental metal},\ }\href {https://doi.org/https://doi.org/10.1038/s41586-020-2082-6} {\bibfield  {journal} {\bibinfo  {journal} {Nature}\ }\textbf {\bibinfo {volume} {579}},\ \bibinfo {pages} {375} (\bibinfo {year} {2020})}\BibitemShut {NoStop}%
\bibitem [{\citenamefont {Niu}\ \emph {et~al.}(2016)\citenamefont {Niu}, \citenamefont {Shu}, \citenamefont {Zhang}, \citenamefont {Gao}, \citenamefont {Jin}, \citenamefont {Zhou},\ and\ \citenamefont {Lu}}]{niuAtomisticInsightsShearcoupled2016}%
  \BibitemOpen
  \bibfield  {author} {\bibinfo {author} {\bibfnamefont {L.-L.}\ \bibnamefont {Niu}}, \bibinfo {author} {\bibfnamefont {X.}~\bibnamefont {Shu}}, \bibinfo {author} {\bibfnamefont {Y.}~\bibnamefont {Zhang}}, \bibinfo {author} {\bibfnamefont {F.}~\bibnamefont {Gao}}, \bibinfo {author} {\bibfnamefont {S.}~\bibnamefont {Jin}}, \bibinfo {author} {\bibfnamefont {H.-B.}\ \bibnamefont {Zhou}},\ and\ \bibinfo {author} {\bibfnamefont {G.-H.}\ \bibnamefont {Lu}},\ }\bibfield  {title} {\bibinfo {title} {Atomistic insights into shear-coupled grain boundary migration in bcc tungsten},\ }\href {https://doi.org/10.1016/j.msea.2016.09.029} {\bibfield  {journal} {\bibinfo  {journal} {Mater. Sci. Eng. A}\ }\textbf {\bibinfo {volume} {677}},\ \bibinfo {pages} {20} (\bibinfo {year} {2016})}\BibitemShut {NoStop}%
\bibitem [{\citenamefont {Olmsted}\ \emph {et~al.}(2007)\citenamefont {Olmsted}, \citenamefont {Foiles},\ and\ \citenamefont {Holm}}]{olmsted2007grain}%
  \BibitemOpen
  \bibfield  {author} {\bibinfo {author} {\bibfnamefont {D.~L.}\ \bibnamefont {Olmsted}}, \bibinfo {author} {\bibfnamefont {S.~M.}\ \bibnamefont {Foiles}},\ and\ \bibinfo {author} {\bibfnamefont {E.~A.}\ \bibnamefont {Holm}},\ }\bibfield  {title} {\bibinfo {title} {Grain boundary interface roughening transition and its effect on grain boundary mobility for non-faceting boundaries},\ }\href {https://doi.org/https://doi.org/10.1016/j.scriptamat.2007.07.045} {\bibfield  {journal} {\bibinfo  {journal} {Scr. Mater.}\ }\textbf {\bibinfo {volume} {57}},\ \bibinfo {pages} {1161} (\bibinfo {year} {2007})}\BibitemShut {NoStop}%
\bibitem [{\citenamefont {Olmsted}\ \emph {et~al.}(2009)\citenamefont {Olmsted}, \citenamefont {Holm},\ and\ \citenamefont {Foiles}}]{olmstedSurveyComputedGrain2009}%
  \BibitemOpen
  \bibfield  {author} {\bibinfo {author} {\bibfnamefont {D.~L.}\ \bibnamefont {Olmsted}}, \bibinfo {author} {\bibfnamefont {E.~A.}\ \bibnamefont {Holm}},\ and\ \bibinfo {author} {\bibfnamefont {S.~M.}\ \bibnamefont {Foiles}},\ }\bibfield  {title} {\bibinfo {title} {Survey of computed grain boundary properties in face-centered cubic metals\textemdash{{II}}: {{Grain}} boundary mobility},\ }\href {https://doi.org/10.1016/j.actamat.2009.04.015} {\bibfield  {journal} {\bibinfo  {journal} {Acta Mater.}\ }\textbf {\bibinfo {volume} {57}},\ \bibinfo {pages} {3704} (\bibinfo {year} {2009})}\BibitemShut {NoStop}%
\bibitem [{\citenamefont {Rahman}\ \emph {et~al.}(2014)\citenamefont {Rahman}, \citenamefont {Zurob},\ and\ \citenamefont {Hoyt}}]{rahman2014comprehensive}%
  \BibitemOpen
  \bibfield  {author} {\bibinfo {author} {\bibfnamefont {M.}~\bibnamefont {Rahman}}, \bibinfo {author} {\bibfnamefont {H.~S.}\ \bibnamefont {Zurob}},\ and\ \bibinfo {author} {\bibfnamefont {J.~J.}\ \bibnamefont {Hoyt}},\ }\bibfield  {title} {\bibinfo {title} {A comprehensive molecular dynamics study of low-angle grain boundary mobility in a pure aluminum system},\ }\href {https://doi.org/https://doi.org/10.1016/j.actamat.2014.03.063} {\bibfield  {journal} {\bibinfo  {journal} {Acta Mater.}\ }\textbf {\bibinfo {volume} {74}},\ \bibinfo {pages} {39} (\bibinfo {year} {2014})}\BibitemShut {NoStop}%
\bibitem [{\citenamefont {Race}\ \emph {et~al.}(2014)\citenamefont {Race}, \citenamefont {von Pezold},\ and\ \citenamefont {Neugebauer}}]{race2014role}%
  \BibitemOpen
  \bibfield  {author} {\bibinfo {author} {\bibfnamefont {C.~P.}\ \bibnamefont {Race}}, \bibinfo {author} {\bibfnamefont {J.}~\bibnamefont {von Pezold}},\ and\ \bibinfo {author} {\bibfnamefont {J.}~\bibnamefont {Neugebauer}},\ }\bibfield  {title} {\bibinfo {title} {Role of the mesoscale in migration kinetics of flat grain boundaries},\ }\href {https://doi.org/https://doi.org/10.1103/PhysRevB.89.214110} {\bibfield  {journal} {\bibinfo  {journal} {Phys. Rev. B.}\ }\textbf {\bibinfo {volume} {89}},\ \bibinfo {pages} {214110} (\bibinfo {year} {2014})}\BibitemShut {NoStop}%
\bibitem [{\citenamefont {Hoyt}(2014)}]{hoytAtomisticSimulationsGrain2014}%
  \BibitemOpen
  \bibfield  {author} {\bibinfo {author} {\bibfnamefont {J.~J.}\ \bibnamefont {Hoyt}},\ }\bibfield  {title} {\bibinfo {title} {Atomistic simulations of grain and interphase boundary mobility},\ }\bibfield  {journal} {\bibinfo  {journal} {Modelling Simul. Mater. Sci. Eng.}\ }\textbf {\bibinfo {volume} {22}},\ \href {https://doi.org/10.1088/0965-0393/22/3/033001} {10.1088/0965-0393/22/3/033001} (\bibinfo {year} {2014})\BibitemShut {NoStop}%
\bibitem [{\citenamefont {Korneva}\ \emph {et~al.}(2020)\citenamefont {Korneva}, \citenamefont {Starikov}, \citenamefont {Zhilyaev}, \citenamefont {Akhatov},\ and\ \citenamefont {Zhilyaev}}]{korneva2020atomistic}%
  \BibitemOpen
  \bibfield  {author} {\bibinfo {author} {\bibfnamefont {M.~A.}\ \bibnamefont {Korneva}}, \bibinfo {author} {\bibfnamefont {S.~V.}\ \bibnamefont {Starikov}}, \bibinfo {author} {\bibfnamefont {A.~P.}\ \bibnamefont {Zhilyaev}}, \bibinfo {author} {\bibfnamefont {I.~S.}\ \bibnamefont {Akhatov}},\ and\ \bibinfo {author} {\bibfnamefont {P.~A.}\ \bibnamefont {Zhilyaev}},\ }\bibfield  {title} {\bibinfo {title} {Atomistic modeling of grain boundary migration in nickel},\ }\href {https://doi.org/https://doi.org/10.1002/adem.202000115} {\bibfield  {journal} {\bibinfo  {journal} {Adv. Eng. Mater.}\ }\textbf {\bibinfo {volume} {22}},\ \bibinfo {pages} {2000115} (\bibinfo {year} {2020})}\BibitemShut {NoStop}%
\bibitem [{\citenamefont {Wei}\ \emph {et~al.}(2021)\citenamefont {Wei}, \citenamefont {Feng}, \citenamefont {Ishikawa}, \citenamefont {Yokoi}, \citenamefont {Matsunaga}, \citenamefont {Shibata},\ and\ \citenamefont {Ikuhara}}]{wei2021direct}%
  \BibitemOpen
  \bibfield  {author} {\bibinfo {author} {\bibfnamefont {J.}~\bibnamefont {Wei}}, \bibinfo {author} {\bibfnamefont {B.}~\bibnamefont {Feng}}, \bibinfo {author} {\bibfnamefont {R.}~\bibnamefont {Ishikawa}}, \bibinfo {author} {\bibfnamefont {T.}~\bibnamefont {Yokoi}}, \bibinfo {author} {\bibfnamefont {K.}~\bibnamefont {Matsunaga}}, \bibinfo {author} {\bibfnamefont {N.}~\bibnamefont {Shibata}},\ and\ \bibinfo {author} {\bibfnamefont {Y.}~\bibnamefont {Ikuhara}},\ }\bibfield  {title} {\bibinfo {title} {Direct imaging of atomistic grain boundary migration},\ }\href {https://doi.org/https://doi.org/10.1038/s41563-020-00879-z} {\bibfield  {journal} {\bibinfo  {journal} {Nat. Mater}\ }\textbf {\bibinfo {volume} {20}},\ \bibinfo {pages} {951} (\bibinfo {year} {2021})}\BibitemShut {NoStop}%
\bibitem [{\citenamefont {Wang}\ \emph {et~al.}(2022)\citenamefont {Wang}, \citenamefont {Zhang}, \citenamefont {Zeng}, \citenamefont {Zhou}, \citenamefont {He}, \citenamefont {Liu}, \citenamefont {Chen}, \citenamefont {Han}, \citenamefont {Srolovitz}, \citenamefont {Teng} \emph {et~al.}}]{wang2022tracking}%
  \BibitemOpen
  \bibfield  {author} {\bibinfo {author} {\bibfnamefont {L.}~\bibnamefont {Wang}}, \bibinfo {author} {\bibfnamefont {Y.}~\bibnamefont {Zhang}}, \bibinfo {author} {\bibfnamefont {Z.}~\bibnamefont {Zeng}}, \bibinfo {author} {\bibfnamefont {H.}~\bibnamefont {Zhou}}, \bibinfo {author} {\bibfnamefont {J.}~\bibnamefont {He}}, \bibinfo {author} {\bibfnamefont {P.}~\bibnamefont {Liu}}, \bibinfo {author} {\bibfnamefont {M.}~\bibnamefont {Chen}}, \bibinfo {author} {\bibfnamefont {J.}~\bibnamefont {Han}}, \bibinfo {author} {\bibfnamefont {D.~J.}\ \bibnamefont {Srolovitz}}, \bibinfo {author} {\bibfnamefont {J.}~\bibnamefont {Teng}}, \emph {et~al.},\ }\bibfield  {title} {\bibinfo {title} {Tracking the sliding of grain boundaries at the atomic scale},\ }\href {https://doi.org/DOI: 10.1126/science.abm2612} {\bibfield  {journal} {\bibinfo  {journal} {Science}\ }\textbf {\bibinfo {volume} {375}},\ \bibinfo {pages} {1261} (\bibinfo {year} {2022})}\BibitemShut {NoStop}%
\bibitem [{\citenamefont {Pitsch}(1959)}]{pitsch1959martensite}%
  \BibitemOpen
  \bibfield  {author} {\bibinfo {author} {\bibfnamefont {W.}~\bibnamefont {Pitsch}},\ }\bibfield  {title} {\bibinfo {title} {The martensite transformation in thin foils of iron-nitrogen alloys},\ }\href {https://doi.org/https://doi.org/10.1080/14786435908238253} {\bibfield  {journal} {\bibinfo  {journal} {Philos. Mag.}\ }\textbf {\bibinfo {volume} {4}},\ \bibinfo {pages} {577} (\bibinfo {year} {1959})}\BibitemShut {NoStop}%
\bibitem [{\citenamefont {Furuhara}\ \emph {et~al.}(1990)\citenamefont {Furuhara}, \citenamefont {Lee}, \citenamefont {Menon},\ and\ \citenamefont {Aaronson}}]{furuhara1990interphase}%
  \BibitemOpen
  \bibfield  {author} {\bibinfo {author} {\bibfnamefont {T.}~\bibnamefont {Furuhara}}, \bibinfo {author} {\bibfnamefont {H.}~\bibnamefont {Lee}}, \bibinfo {author} {\bibfnamefont {E.}~\bibnamefont {Menon}},\ and\ \bibinfo {author} {\bibfnamefont {H.}~\bibnamefont {Aaronson}},\ }\bibfield  {title} {\bibinfo {title} {Interphase boundary structures associated with diffusional phase transformations in {Ti}-base alloys},\ }\href {https://doi.org/https://doi.org/10.1007/BF02672578} {\bibfield  {journal} {\bibinfo  {journal} {Metall Mater Trans A Phys Metall Mater Sci .}\ }\textbf {\bibinfo {volume} {21}},\ \bibinfo {pages} {1627} (\bibinfo {year} {1990})}\BibitemShut {NoStop}%
\bibitem [{\citenamefont {Bos}\ \emph {et~al.}(2006)\citenamefont {Bos}, \citenamefont {Sietsma},\ and\ \citenamefont {Thijsse}}]{bos2006molecular}%
  \BibitemOpen
  \bibfield  {author} {\bibinfo {author} {\bibfnamefont {C.}~\bibnamefont {Bos}}, \bibinfo {author} {\bibfnamefont {J.}~\bibnamefont {Sietsma}},\ and\ \bibinfo {author} {\bibfnamefont {B.~J.}\ \bibnamefont {Thijsse}},\ }\bibfield  {title} {\bibinfo {title} {Molecular dynamics simulation of interface dynamics during the fcc-bcc transformation of a martensitic nature},\ }\href {https://doi.org/https://doi.org/10.1103/PhysRevB.73.104117} {\bibfield  {journal} {\bibinfo  {journal} {Phys. Rev. B}\ }\textbf {\bibinfo {volume} {73}},\ \bibinfo {pages} {104117} (\bibinfo {year} {2006})}\BibitemShut {NoStop}%
\bibitem [{\citenamefont {Fukino}\ and\ \citenamefont {Tsurekawa}(2008)}]{fukino2008situ}%
  \BibitemOpen
  \bibfield  {author} {\bibinfo {author} {\bibfnamefont {T.}~\bibnamefont {Fukino}}\ and\ \bibinfo {author} {\bibfnamefont {S.}~\bibnamefont {Tsurekawa}},\ }\bibfield  {title} {\bibinfo {title} {In-situ {SEM/EBSD} observation of $\alpha$/ $\gamma$ phase transformation in {Fe-Ni} alloy},\ }\href {https://doi.org/https://doi.org/10.2320/matertrans.MAW200824} {\bibfield  {journal} {\bibinfo  {journal} {Mater. Trans.}\ }\textbf {\bibinfo {volume} {49}},\ \bibinfo {pages} {2770} (\bibinfo {year} {2008})}\BibitemShut {NoStop}%
\bibitem [{\citenamefont {Song}\ and\ \citenamefont {Hoyt}(2012)}]{song2012molecular}%
  \BibitemOpen
  \bibfield  {author} {\bibinfo {author} {\bibfnamefont {H.}~\bibnamefont {Song}}\ and\ \bibinfo {author} {\bibfnamefont {J.}~\bibnamefont {Hoyt}},\ }\bibfield  {title} {\bibinfo {title} {A molecular dynamics simulation study of the velocities, mobility and activation energy of an austenite--ferrite interface in pure {Fe}},\ }\href {https://doi.org/https://doi.org/10.1016/j.actamat.2012.04.023} {\bibfield  {journal} {\bibinfo  {journal} {Acta Mater.}\ }\textbf {\bibinfo {volume} {60}},\ \bibinfo {pages} {4328} (\bibinfo {year} {2012})}\BibitemShut {NoStop}%
\bibitem [{\citenamefont {Wang}\ and\ \citenamefont {Urbassek}(2013)}]{wang2013molecular}%
  \BibitemOpen
  \bibfield  {author} {\bibinfo {author} {\bibfnamefont {B.}~\bibnamefont {Wang}}\ and\ \bibinfo {author} {\bibfnamefont {H.~M.}\ \bibnamefont {Urbassek}},\ }\bibfield  {title} {\bibinfo {title} {Molecular dynamics study of the $\alpha$--$\gamma$ phase transition in {Fe} induced by shear deformation},\ }\href {https://doi.org/https://doi.org/10.1016/j.actamat.2013.05.045} {\bibfield  {journal} {\bibinfo  {journal} {Acta Mater.}\ }\textbf {\bibinfo {volume} {61}},\ \bibinfo {pages} {5979} (\bibinfo {year} {2013})}\BibitemShut {NoStop}%
\bibitem [{\citenamefont {Yang}\ \emph {et~al.}(2014)\citenamefont {Yang}, \citenamefont {Sun}, \citenamefont {Wu}, \citenamefont {Ma},\ and\ \citenamefont {Zhang}}]{yang2014dissecting}%
  \BibitemOpen
  \bibfield  {author} {\bibinfo {author} {\bibfnamefont {X.-S.}\ \bibnamefont {Yang}}, \bibinfo {author} {\bibfnamefont {S.}~\bibnamefont {Sun}}, \bibinfo {author} {\bibfnamefont {X.-L.}\ \bibnamefont {Wu}}, \bibinfo {author} {\bibfnamefont {E.}~\bibnamefont {Ma}},\ and\ \bibinfo {author} {\bibfnamefont {T.-Y.}\ \bibnamefont {Zhang}},\ }\bibfield  {title} {\bibinfo {title} {Dissecting the mechanism of martensitic transformation via atomic-scale observations},\ }\href {https://doi.org/https://doi.org/10.1038/srep06141} {\bibfield  {journal} {\bibinfo  {journal} {Sci. Rep.}\ }\textbf {\bibinfo {volume} {4}},\ \bibinfo {pages} {6141} (\bibinfo {year} {2014})}\BibitemShut {NoStop}%
\bibitem [{\citenamefont {Duncan}\ \emph {et~al.}(2016)\citenamefont {Duncan}, \citenamefont {Harjunmaa}, \citenamefont {Terrell}, \citenamefont {Drautz}, \citenamefont {Henkelman},\ and\ \citenamefont {Rogal}}]{duncan2016collective}%
  \BibitemOpen
  \bibfield  {author} {\bibinfo {author} {\bibfnamefont {J.}~\bibnamefont {Duncan}}, \bibinfo {author} {\bibfnamefont {A.}~\bibnamefont {Harjunmaa}}, \bibinfo {author} {\bibfnamefont {R.}~\bibnamefont {Terrell}}, \bibinfo {author} {\bibfnamefont {R.}~\bibnamefont {Drautz}}, \bibinfo {author} {\bibfnamefont {G.}~\bibnamefont {Henkelman}},\ and\ \bibinfo {author} {\bibfnamefont {J.}~\bibnamefont {Rogal}},\ }\bibfield  {title} {\bibinfo {title} {Collective atomic displacements during complex phase boundary migration in solid-solid phase transformations},\ }\href {https://doi.org/10.1103/PhysRevLett.116.035701} {\bibfield  {journal} {\bibinfo  {journal} {Phys. Rev. Lett.}\ }\textbf {\bibinfo {volume} {116}},\ \bibinfo {pages} {035701} (\bibinfo {year} {2016})}\BibitemShut {NoStop}%
\bibitem [{\citenamefont {Ou}(2017)}]{ouMolecularDynamicsSimulations2017}%
  \BibitemOpen
  \bibfield  {author} {\bibinfo {author} {\bibfnamefont {X.}~\bibnamefont {Ou}},\ }\bibfield  {title} {\bibinfo {title} {Molecular dynamics simulations of fcc-to-bcc transformation in pure iron: a review},\ }\href {https://doi.org/https://doi.org/10.1080/02670836.2016.1204064} {\bibfield  {journal} {\bibinfo  {journal} {Mater. Sci. Technol.}\ }\textbf {\bibinfo {volume} {33}},\ \bibinfo {pages} {822} (\bibinfo {year} {2017})}\BibitemShut {NoStop}%
\bibitem [{\citenamefont {Meijer}\ \emph {et~al.}(2017)\citenamefont {Meijer}, \citenamefont {Pal}, \citenamefont {Ouhajji}, \citenamefont {Lekkerkerker}, \citenamefont {Philipse},\ and\ \citenamefont {Petukhov}}]{meijer2017observation}%
  \BibitemOpen
  \bibfield  {author} {\bibinfo {author} {\bibfnamefont {J.-M.}\ \bibnamefont {Meijer}}, \bibinfo {author} {\bibfnamefont {A.}~\bibnamefont {Pal}}, \bibinfo {author} {\bibfnamefont {S.}~\bibnamefont {Ouhajji}}, \bibinfo {author} {\bibfnamefont {H.~N.}\ \bibnamefont {Lekkerkerker}}, \bibinfo {author} {\bibfnamefont {A.~P.}\ \bibnamefont {Philipse}},\ and\ \bibinfo {author} {\bibfnamefont {A.~V.}\ \bibnamefont {Petukhov}},\ }\bibfield  {title} {\bibinfo {title} {Observation of solid--solid transitions in {3D} crystals of colloidal superballs},\ }\href {https://doi.org/https://doi.org/10.1038/ncomms14352} {\bibfield  {journal} {\bibinfo  {journal} {Nat. Commun.}\ }\textbf {\bibinfo {volume} {8}},\ \bibinfo {pages} {14352} (\bibinfo {year} {2017})}\BibitemShut {NoStop}%
\bibitem [{\citenamefont {Fei}\ \emph {et~al.}(2018)\citenamefont {Fei}, \citenamefont {Gan}, \citenamefont {Ng}, \citenamefont {Wang}, \citenamefont {Xu}, \citenamefont {Lu}, \citenamefont {Zhou}, \citenamefont {Leung}, \citenamefont {Mak},\ and\ \citenamefont {Wang}}]{fei2018observable}%
  \BibitemOpen
  \bibfield  {author} {\bibinfo {author} {\bibfnamefont {L.}~\bibnamefont {Fei}}, \bibinfo {author} {\bibfnamefont {X.}~\bibnamefont {Gan}}, \bibinfo {author} {\bibfnamefont {S.~M.}\ \bibnamefont {Ng}}, \bibinfo {author} {\bibfnamefont {H.}~\bibnamefont {Wang}}, \bibinfo {author} {\bibfnamefont {M.}~\bibnamefont {Xu}}, \bibinfo {author} {\bibfnamefont {W.}~\bibnamefont {Lu}}, \bibinfo {author} {\bibfnamefont {Y.}~\bibnamefont {Zhou}}, \bibinfo {author} {\bibfnamefont {C.~W.}\ \bibnamefont {Leung}}, \bibinfo {author} {\bibfnamefont {C.-L.}\ \bibnamefont {Mak}},\ and\ \bibinfo {author} {\bibfnamefont {Y.}~\bibnamefont {Wang}},\ }\bibfield  {title} {\bibinfo {title} {Observable two-step nucleation mechanism in solid-state formation of tungsten carbide},\ }\href {https://doi.org/https://doi.org/10.1021/acsnano.8b07864} {\bibfield  {journal} {\bibinfo  {journal} {ACS nano}\ }\textbf {\bibinfo {volume} {13}},\ \bibinfo {pages} {681} (\bibinfo {year} {2018})}\BibitemShut {NoStop}%
\bibitem [{\citenamefont {Immink}\ \emph {et~al.}(2020)\citenamefont {Immink}, \citenamefont {Bergman}, \citenamefont {Maris}, \citenamefont {Stenhammar},\ and\ \citenamefont {Schurtenberger}}]{immink2020crystal}%
  \BibitemOpen
  \bibfield  {author} {\bibinfo {author} {\bibfnamefont {J.~N.}\ \bibnamefont {Immink}}, \bibinfo {author} {\bibfnamefont {M.~J.}\ \bibnamefont {Bergman}}, \bibinfo {author} {\bibfnamefont {J.~E.}\ \bibnamefont {Maris}}, \bibinfo {author} {\bibfnamefont {J.}~\bibnamefont {Stenhammar}},\ and\ \bibinfo {author} {\bibfnamefont {P.}~\bibnamefont {Schurtenberger}},\ }\bibfield  {title} {\bibinfo {title} {Crystal-to-crystal transitions in binary mixtures of soft colloids},\ }\href {https://doi.org/https://doi.org/10.1021/acsnano.0c03966} {\bibfield  {journal} {\bibinfo  {journal} {ACS Nano}\ }\textbf {\bibinfo {volume} {14}},\ \bibinfo {pages} {14861} (\bibinfo {year} {2020})}\BibitemShut {NoStop}%
\bibitem [{\citenamefont {Chen}\ \emph {et~al.}(2021)\citenamefont {Chen}, \citenamefont {Huo},\ and\ \citenamefont {Yeddu}}]{chen2021molecular}%
  \BibitemOpen
  \bibfield  {author} {\bibinfo {author} {\bibfnamefont {J.}~\bibnamefont {Chen}}, \bibinfo {author} {\bibfnamefont {D.}~\bibnamefont {Huo}},\ and\ \bibinfo {author} {\bibfnamefont {H.~K.}\ \bibnamefont {Yeddu}},\ }\bibfield  {title} {\bibinfo {title} {Molecular dynamics study of phase transformations in {NiTi} shape memory alloy embedded with precipitates},\ }\href {https://doi.org/10.1088/2053-1591/ac2b57} {\bibfield  {journal} {\bibinfo  {journal} {Mater. Res. Express}\ }\textbf {\bibinfo {volume} {8}},\ \bibinfo {pages} {106508} (\bibinfo {year} {2021})}\BibitemShut {NoStop}%
\bibitem [{\citenamefont {Kuznetsov}\ \emph {et~al.}(2001)\citenamefont {Kuznetsov}, \citenamefont {Gornostyrev}, \citenamefont {Katsnelson},\ and\ \citenamefont {Trefilov}}]{kuznetsov2001effect}%
  \BibitemOpen
  \bibfield  {author} {\bibinfo {author} {\bibfnamefont {A.~R.}\ \bibnamefont {Kuznetsov}}, \bibinfo {author} {\bibfnamefont {Y.~N.}\ \bibnamefont {Gornostyrev}}, \bibinfo {author} {\bibfnamefont {M.~I.}\ \bibnamefont {Katsnelson}},\ and\ \bibinfo {author} {\bibfnamefont {A.~V.}\ \bibnamefont {Trefilov}},\ }\bibfield  {title} {\bibinfo {title} {Effect of the dislocations on the kinetics of a martensitic transition: {MD} simulation of bcc--hcp transformation in {Zr}},\ }\href {https://doi.org/https://doi.org/10.1016/S0921-5093(00)01677-4} {\bibfield  {journal} {\bibinfo  {journal} {Mater. Sci. and Eng. A}\ }\textbf {\bibinfo {volume} {309}},\ \bibinfo {pages} {168} (\bibinfo {year} {2001})}\BibitemShut {NoStop}%
\bibitem [{\citenamefont {Li}\ \emph {et~al.}(2021)\citenamefont {Li}, \citenamefont {Yue}, \citenamefont {Chen}, \citenamefont {Tong}, \citenamefont {Tanaka},\ and\ \citenamefont {Tan}}]{li2021revealing}%
  \BibitemOpen
  \bibfield  {author} {\bibinfo {author} {\bibfnamefont {M.}~\bibnamefont {Li}}, \bibinfo {author} {\bibfnamefont {Z.}~\bibnamefont {Yue}}, \bibinfo {author} {\bibfnamefont {Y.}~\bibnamefont {Chen}}, \bibinfo {author} {\bibfnamefont {H.}~\bibnamefont {Tong}}, \bibinfo {author} {\bibfnamefont {H.}~\bibnamefont {Tanaka}},\ and\ \bibinfo {author} {\bibfnamefont {P.}~\bibnamefont {Tan}},\ }\bibfield  {title} {\bibinfo {title} {Revealing thermally-activated nucleation pathways of diffusionless solid-to-solid transition},\ }\href {https://doi.org/https://doi.org/10.1038/s41467-021-24256-9} {\bibfield  {journal} {\bibinfo  {journal} {Nat. Commun.}\ }\textbf {\bibinfo {volume} {12}},\ \bibinfo {pages} {4042} (\bibinfo {year} {2021})}\BibitemShut {NoStop}%
\bibitem [{\citenamefont {Fu}\ \emph {et~al.}(2022)\citenamefont {Fu}, \citenamefont {Wang}, \citenamefont {Zhao}, \citenamefont {Zhang}, \citenamefont {Sun}, \citenamefont {Wang}, \citenamefont {Zhang}, \citenamefont {Gu}, \citenamefont {Zhang}, \citenamefont {Zhang} \emph {et~al.}}]{fu2022atomic}%
  \BibitemOpen
  \bibfield  {author} {\bibinfo {author} {\bibfnamefont {X.}~\bibnamefont {Fu}}, \bibinfo {author} {\bibfnamefont {X.-D.}\ \bibnamefont {Wang}}, \bibinfo {author} {\bibfnamefont {B.}~\bibnamefont {Zhao}}, \bibinfo {author} {\bibfnamefont {Q.}~\bibnamefont {Zhang}}, \bibinfo {author} {\bibfnamefont {S.}~\bibnamefont {Sun}}, \bibinfo {author} {\bibfnamefont {J.-J.}\ \bibnamefont {Wang}}, \bibinfo {author} {\bibfnamefont {W.}~\bibnamefont {Zhang}}, \bibinfo {author} {\bibfnamefont {L.}~\bibnamefont {Gu}}, \bibinfo {author} {\bibfnamefont {Y.}~\bibnamefont {Zhang}}, \bibinfo {author} {\bibfnamefont {W.-Z.}\ \bibnamefont {Zhang}}, \emph {et~al.},\ }\bibfield  {title} {\bibinfo {title} {Atomic-scale observation of non-classical nucleation-mediated phase transformation in a titanium alloy},\ }\href {https://doi.org/https://doi.org/10.1038/s41563-021-01144-7} {\bibfield  {journal} {\bibinfo  {journal} {Nat. Mater}\ }\textbf {\bibinfo {volume} {21}},\ \bibinfo {pages} {290} (\bibinfo {year} {2022})}\BibitemShut
  {NoStop}%
\bibitem [{\citenamefont {Zhang}\ \emph {et~al.}(2022)\citenamefont {Zhang}, \citenamefont {Zhang}, \citenamefont {Wang}, \citenamefont {Rogal}, \citenamefont {Li}, \citenamefont {Wei},\ and\ \citenamefont {Hickel}}]{zhang2022defect}%
  \BibitemOpen
  \bibfield  {author} {\bibinfo {author} {\bibfnamefont {X.}~\bibnamefont {Zhang}}, \bibinfo {author} {\bibfnamefont {J.}~\bibnamefont {Zhang}}, \bibinfo {author} {\bibfnamefont {H.}~\bibnamefont {Wang}}, \bibinfo {author} {\bibfnamefont {J.}~\bibnamefont {Rogal}}, \bibinfo {author} {\bibfnamefont {H.-Y.}\ \bibnamefont {Li}}, \bibinfo {author} {\bibfnamefont {S.-H.}\ \bibnamefont {Wei}},\ and\ \bibinfo {author} {\bibfnamefont {T.}~\bibnamefont {Hickel}},\ }\bibfield  {title} {\bibinfo {title} {Defect-characterized phase transition kinetics},\ }\bibfield  {journal} {\bibinfo  {journal} {Appl. Phys. Rev.}\ }\textbf {\bibinfo {volume} {9}},\ \href {https://doi.org/https://doi.org/10.1063/5.0117234} {https://doi.org/10.1063/5.0117234} (\bibinfo {year} {2022})\BibitemShut {NoStop}%
\bibitem [{\citenamefont {Jiang}\ \emph {et~al.}(2023)\citenamefont {Jiang}, \citenamefont {Duchamp}, \citenamefont {Ang}, \citenamefont {Yan}, \citenamefont {Tan},\ and\ \citenamefont {Mirsaidov}}]{jiang2023dynamics}%
  \BibitemOpen
  \bibfield  {author} {\bibinfo {author} {\bibfnamefont {Y.}~\bibnamefont {Jiang}}, \bibinfo {author} {\bibfnamefont {M.}~\bibnamefont {Duchamp}}, \bibinfo {author} {\bibfnamefont {S.~J.}\ \bibnamefont {Ang}}, \bibinfo {author} {\bibfnamefont {H.}~\bibnamefont {Yan}}, \bibinfo {author} {\bibfnamefont {T.~L.}\ \bibnamefont {Tan}},\ and\ \bibinfo {author} {\bibfnamefont {U.}~\bibnamefont {Mirsaidov}},\ }\bibfield  {title} {\bibinfo {title} {Dynamics of the fcc-to-bcc phase transition in single-crystalline {PdCu} alloy nanoparticles},\ }\href {https://doi.org/https://doi.org/10.1038/s41467-022-35325-y} {\bibfield  {journal} {\bibinfo  {journal} {Nat. Commun}\ }\textbf {\bibinfo {volume} {14}},\ \bibinfo {pages} {104} (\bibinfo {year} {2023})}\BibitemShut {NoStop}%
\bibitem [{\citenamefont {Yang}\ \emph {et~al.}(2015)\citenamefont {Yang}, \citenamefont {Sun},\ and\ \citenamefont {Zhang}}]{yang2015mechanism}%
  \BibitemOpen
  \bibfield  {author} {\bibinfo {author} {\bibfnamefont {X.~S.}\ \bibnamefont {Yang}}, \bibinfo {author} {\bibfnamefont {S.}~\bibnamefont {Sun}},\ and\ \bibinfo {author} {\bibfnamefont {T.~Y.}\ \bibnamefont {Zhang}},\ }\bibfield  {title} {\bibinfo {title} {The mechanism of bcc $\alpha^\prime$ nucleation in single hcp $\varepsilon$ laths in the fcc $\gamma\to$hcp $\varepsilon\to$ bcc $\alpha^\prime$ martensitic phase transformation},\ }\href {https://doi.org/https://doi.org/10.1016/j.actamat.2015.05.034} {\bibfield  {journal} {\bibinfo  {journal} {Acta Mater.}\ }\textbf {\bibinfo {volume} {95}},\ \bibinfo {pages} {264} (\bibinfo {year} {2015})}\BibitemShut {NoStop}%
\bibitem [{\citenamefont {Howe}\ and\ \citenamefont {Howe}(1997)}]{howeInterfacesMaterialsAtomic1997}%
  \BibitemOpen
  \bibfield  {author} {\bibinfo {author} {\bibfnamefont {J.~M.}\ \bibnamefont {Howe}}\ and\ \bibinfo {author} {\bibfnamefont {J.~M.}\ \bibnamefont {Howe}},\ }\href@noop {} {\emph {\bibinfo {title} {Interfaces in {{Materials}}: {{Atomic Structure}}, {{Thermodynamics}} and {{Kinetics}} of {{Solid-Vapor}}, {{Solid-Liquid}} and {{Solid-Solid Interfaces}}}}}\ (\bibinfo  {publisher} {{Wiley}},\ \bibinfo {year} {1997})\BibitemShut {NoStop}%
\bibitem [{\citenamefont {Peng}\ \emph {et~al.}(2015)\citenamefont {Peng}, \citenamefont {Wang}, \citenamefont {Wang}, \citenamefont {Alsayed}, \citenamefont {Zhang}, \citenamefont {Yodh},\ and\ \citenamefont {Han}}]{peng2015two}%
  \BibitemOpen
  \bibfield  {author} {\bibinfo {author} {\bibfnamefont {Y.}~\bibnamefont {Peng}}, \bibinfo {author} {\bibfnamefont {F.}~\bibnamefont {Wang}}, \bibinfo {author} {\bibfnamefont {Z.}~\bibnamefont {Wang}}, \bibinfo {author} {\bibfnamefont {A.~M.}\ \bibnamefont {Alsayed}}, \bibinfo {author} {\bibfnamefont {Z.}~\bibnamefont {Zhang}}, \bibinfo {author} {\bibfnamefont {A.~G.}\ \bibnamefont {Yodh}},\ and\ \bibinfo {author} {\bibfnamefont {Y.}~\bibnamefont {Han}},\ }\bibfield  {title} {\bibinfo {title} {Two-step nucleation mechanism in solid--solid phase transitions},\ }\href {https://doi.org/https://doi.org/10.1038/nmat4083} {\bibfield  {journal} {\bibinfo  {journal} {Nat. Mater.}\ }\textbf {\bibinfo {volume} {14}},\ \bibinfo {pages} {101} (\bibinfo {year} {2015})}\BibitemShut {NoStop}%
\bibitem [{\citenamefont {Peng}\ \emph {et~al.}(2017)\citenamefont {Peng}, \citenamefont {Li}, \citenamefont {Wang}, \citenamefont {Still}, \citenamefont {Yodh},\ and\ \citenamefont {Han}}]{peng2017diffusive}%
  \BibitemOpen
  \bibfield  {author} {\bibinfo {author} {\bibfnamefont {Y.}~\bibnamefont {Peng}}, \bibinfo {author} {\bibfnamefont {W.}~\bibnamefont {Li}}, \bibinfo {author} {\bibfnamefont {F.}~\bibnamefont {Wang}}, \bibinfo {author} {\bibfnamefont {T.}~\bibnamefont {Still}}, \bibinfo {author} {\bibfnamefont {A.~G.}\ \bibnamefont {Yodh}},\ and\ \bibinfo {author} {\bibfnamefont {Y.}~\bibnamefont {Han}},\ }\bibfield  {title} {\bibinfo {title} {Diffusive and martensitic nucleation kinetics in solid-solid transitions of colloidal crystals},\ }\href {https://doi.org/https://doi.org/10.1038/ncomms14978} {\bibfield  {journal} {\bibinfo  {journal} {Nat. Commun.}\ }\textbf {\bibinfo {volume} {8}},\ \bibinfo {pages} {14978} (\bibinfo {year} {2017})}\BibitemShut {NoStop}%
\bibitem [{\citenamefont {Levitas}\ and\ \citenamefont {Momeni}(2014)}]{levitas2014solid}%
  \BibitemOpen
  \bibfield  {author} {\bibinfo {author} {\bibfnamefont {V.~I.}\ \bibnamefont {Levitas}}\ and\ \bibinfo {author} {\bibfnamefont {K.}~\bibnamefont {Momeni}},\ }\bibfield  {title} {\bibinfo {title} {Solid--solid transformations via nanoscale intermediate interfacial phase: Multiple structures, scale and mechanics effects},\ }\href {https://doi.org/https://doi.org/10.1016/j.actamat.2013.11.051} {\bibfield  {journal} {\bibinfo  {journal} {Acta Mater.}\ }\textbf {\bibinfo {volume} {65}},\ \bibinfo {pages} {125} (\bibinfo {year} {2014})}\BibitemShut {NoStop}%
\bibitem [{\citenamefont {Song}\ \emph {et~al.}(2022)\citenamefont {Song}, \citenamefont {Ge}, \citenamefont {Mao}, \citenamefont {Wang}, \citenamefont {Tai}, \citenamefont {Zhang}, \citenamefont {Tang}, \citenamefont {Hao}, \citenamefont {Yao}, \citenamefont {Wang} \emph {et~al.}}]{song2022planar}%
  \BibitemOpen
  \bibfield  {author} {\bibinfo {author} {\bibfnamefont {Y.-H.}\ \bibnamefont {Song}}, \bibinfo {author} {\bibfnamefont {J.}~\bibnamefont {Ge}}, \bibinfo {author} {\bibfnamefont {L.-B.}\ \bibnamefont {Mao}}, \bibinfo {author} {\bibfnamefont {K.-H.}\ \bibnamefont {Wang}}, \bibinfo {author} {\bibfnamefont {X.-L.}\ \bibnamefont {Tai}}, \bibinfo {author} {\bibfnamefont {Q.}~\bibnamefont {Zhang}}, \bibinfo {author} {\bibfnamefont {L.}~\bibnamefont {Tang}}, \bibinfo {author} {\bibfnamefont {J.-M.}\ \bibnamefont {Hao}}, \bibinfo {author} {\bibfnamefont {J.-S.}\ \bibnamefont {Yao}}, \bibinfo {author} {\bibfnamefont {J.-J.}\ \bibnamefont {Wang}}, \emph {et~al.},\ }\bibfield  {title} {\bibinfo {title} {Planar defect--free pure red perovskite light-emitting diodes via metastable phase crystallization},\ }\href {https://doi.org/DOI: 10.1126/sciadv.abq2321} {\bibfield  {journal} {\bibinfo  {journal} {Sci. Adv.}\ }\textbf {\bibinfo {volume} {8}},\ \bibinfo {pages} {eabq2321} (\bibinfo {year} {2022})}\BibitemShut {NoStop}%
\bibitem [{\citenamefont {Wang}\ and\ \citenamefont {Chiang}(1998)}]{wang1998thermodynamic}%
  \BibitemOpen
  \bibfield  {author} {\bibinfo {author} {\bibfnamefont {H.}~\bibnamefont {Wang}}\ and\ \bibinfo {author} {\bibfnamefont {Y.-M.}\ \bibnamefont {Chiang}},\ }\bibfield  {title} {\bibinfo {title} {Thermodynamic stability of intergranular amorphous films in bismuth-doped zinc oxide},\ }\href {https://doi.org/https://doi.org/10.1111/j.1151-2916.1998.tb02299.x} {\bibfield  {journal} {\bibinfo  {journal} {J. Am. Ceram. Soc.}\ }\textbf {\bibinfo {volume} {81}},\ \bibinfo {pages} {89} (\bibinfo {year} {1998})}\BibitemShut {NoStop}%
\bibitem [{\citenamefont {Luo}\ \emph {et~al.}(2005)\citenamefont {Luo}, \citenamefont {Gupta}, \citenamefont {Yoon},\ and\ \citenamefont {Meyer}}]{luo2005segregation}%
  \BibitemOpen
  \bibfield  {author} {\bibinfo {author} {\bibfnamefont {J.}~\bibnamefont {Luo}}, \bibinfo {author} {\bibfnamefont {V.}~\bibnamefont {Gupta}}, \bibinfo {author} {\bibfnamefont {D.}~\bibnamefont {Yoon}},\ and\ \bibinfo {author} {\bibfnamefont {H.}~\bibnamefont {Meyer}},\ }\bibfield  {title} {\bibinfo {title} {Segregation-induced grain boundary premelting in nickel-doped tungsten},\ }\bibfield  {journal} {\bibinfo  {journal} {Appl. Phys. Lett.}\ }\textbf {\bibinfo {volume} {87}},\ \href {https://doi.org/https://doi.org/10.1063/1.2138796} {https://doi.org/10.1063/1.2138796} (\bibinfo {year} {2005})\BibitemShut {NoStop}%
\bibitem [{\citenamefont {Levitas}\ \emph {et~al.}(2012)\citenamefont {Levitas}, \citenamefont {Ren}, \citenamefont {Zeng}, \citenamefont {Zhang},\ and\ \citenamefont {Han}}]{levitasCrystalcrystalPhaseTransformation2012}%
  \BibitemOpen
  \bibfield  {author} {\bibinfo {author} {\bibfnamefont {V.~I.}\ \bibnamefont {Levitas}}, \bibinfo {author} {\bibfnamefont {Z.}~\bibnamefont {Ren}}, \bibinfo {author} {\bibfnamefont {Y.}~\bibnamefont {Zeng}}, \bibinfo {author} {\bibfnamefont {Z.}~\bibnamefont {Zhang}},\ and\ \bibinfo {author} {\bibfnamefont {G.}~\bibnamefont {Han}},\ }\bibfield  {title} {\bibinfo {title} {Crystal-crystal phase transformation via surface-induced virtual premelting},\ }\href {https://doi.org/10.1103/PhysRevB.85.220104} {\bibfield  {journal} {\bibinfo  {journal} {Phys. Rev. B}\ }\textbf {\bibinfo {volume} {85}},\ \bibinfo {pages} {220104} (\bibinfo {year} {2012})}\BibitemShut {NoStop}%
\bibitem [{\citenamefont {Su}\ \emph {et~al.}(2021)\citenamefont {Su}, \citenamefont {Wang}, \citenamefont {Su}, \citenamefont {Du}, \citenamefont {Ren}, \citenamefont {St{\aa}hl}, \citenamefont {Cao}, \citenamefont {Zhang},\ and\ \citenamefont {Jiang}}]{su2021solid}%
  \BibitemOpen
  \bibfield  {author} {\bibinfo {author} {\bibfnamefont {Y.}~\bibnamefont {Su}}, \bibinfo {author} {\bibfnamefont {X.-D.}\ \bibnamefont {Wang}}, \bibinfo {author} {\bibfnamefont {Q.}~\bibnamefont {Su}}, \bibinfo {author} {\bibfnamefont {G.}~\bibnamefont {Du}}, \bibinfo {author} {\bibfnamefont {Y.}~\bibnamefont {Ren}}, \bibinfo {author} {\bibfnamefont {K.}~\bibnamefont {St{\aa}hl}}, \bibinfo {author} {\bibfnamefont {Q.}~\bibnamefont {Cao}}, \bibinfo {author} {\bibfnamefont {D.}~\bibnamefont {Zhang}},\ and\ \bibinfo {author} {\bibfnamefont {J.-Z.}\ \bibnamefont {Jiang}},\ }\bibfield  {title} {\bibinfo {title} {Solid-solid phase transition via the liquid in a {Pd$_{43}$Cu$_{27}$Ni$_{10}$P$_{20}$} bulk metallic glass under conventional conditions},\ }\href {https://doi.org/https://doi.org/10.1016/j.jallcom.2020.157802} {\bibfield  {journal} {\bibinfo  {journal} {J. Alloys Compd.}\ }\textbf {\bibinfo {volume} {859}},\ \bibinfo {pages} {157802} (\bibinfo {year} {2021})}\BibitemShut {NoStop}%
\bibitem [{\citenamefont {Watson}\ and\ \citenamefont {Bennett}(1984)}]{watson1984transition}%
  \BibitemOpen
  \bibfield  {author} {\bibinfo {author} {\bibfnamefont {R.~E.}\ \bibnamefont {Watson}}\ and\ \bibinfo {author} {\bibfnamefont {L.~H.}\ \bibnamefont {Bennett}},\ }\bibfield  {title} {\bibinfo {title} {Transition-metal alloy formation. {The} occurrence of topologically close packed phases-{I}},\ }\href {https://doi.org/https://doi.org/10.1016/0001-6160(84)90058-0} {\bibfield  {journal} {\bibinfo  {journal} {Acta Metal.}\ }\textbf {\bibinfo {volume} {32}},\ \bibinfo {pages} {477} (\bibinfo {year} {1984})}\BibitemShut {NoStop}%
\bibitem [{\citenamefont {Long}\ \emph {et~al.}(2018)\citenamefont {Long}, \citenamefont {Liu}, \citenamefont {Mao}, \citenamefont {Wei}, \citenamefont {Zhang}, \citenamefont {Ma}, \citenamefont {Deng}, \citenamefont {Chen}, \citenamefont {Zhang},\ and\ \citenamefont {Han}}]{long2018minimum}%
  \BibitemOpen
  \bibfield  {author} {\bibinfo {author} {\bibfnamefont {H.}~\bibnamefont {Long}}, \bibinfo {author} {\bibfnamefont {Y.}~\bibnamefont {Liu}}, \bibinfo {author} {\bibfnamefont {S.}~\bibnamefont {Mao}}, \bibinfo {author} {\bibfnamefont {H.}~\bibnamefont {Wei}}, \bibinfo {author} {\bibfnamefont {J.}~\bibnamefont {Zhang}}, \bibinfo {author} {\bibfnamefont {S.}~\bibnamefont {Ma}}, \bibinfo {author} {\bibfnamefont {Q.}~\bibnamefont {Deng}}, \bibinfo {author} {\bibfnamefont {Y.}~\bibnamefont {Chen}}, \bibinfo {author} {\bibfnamefont {Z.}~\bibnamefont {Zhang}},\ and\ \bibinfo {author} {\bibfnamefont {X.}~\bibnamefont {Han}},\ }\bibfield  {title} {\bibinfo {title} {Minimum interface misfit criterion for the precipitation morphologies of {TCP} phases in a {Ni}-based single crystal superalloy},\ }\href {https://doi.org/https://doi.org/10.1016/j.intermet.2017.12.020} {\bibfield  {journal} {\bibinfo  {journal} {Intermetallics}\ }\textbf {\bibinfo {volume} {94}},\ \bibinfo {pages} {55} (\bibinfo {year} {2018})}\BibitemShut
  {NoStop}%
\bibitem [{\citenamefont {Seiser}\ \emph {et~al.}(2011)\citenamefont {Seiser}, \citenamefont {Drautz},\ and\ \citenamefont {Pettifor}}]{seiser2011tcp}%
  \BibitemOpen
  \bibfield  {author} {\bibinfo {author} {\bibfnamefont {B.}~\bibnamefont {Seiser}}, \bibinfo {author} {\bibfnamefont {R.}~\bibnamefont {Drautz}},\ and\ \bibinfo {author} {\bibfnamefont {D.}~\bibnamefont {Pettifor}},\ }\bibfield  {title} {\bibinfo {title} {{TCP} phase predictions in {Ni}-based superalloys: structure maps revisited},\ }\href {https://doi.org/https://doi.org/10.1016/j.actamat.2010.10.013} {\bibfield  {journal} {\bibinfo  {journal} {Acta Mater.}\ }\textbf {\bibinfo {volume} {59}},\ \bibinfo {pages} {749} (\bibinfo {year} {2011})}\BibitemShut {NoStop}%
\bibitem [{\citenamefont {Long}\ \emph {et~al.}(2020)\citenamefont {Long}, \citenamefont {Mao}, \citenamefont {Liu}, \citenamefont {Yang}, \citenamefont {Wei}, \citenamefont {Deng}, \citenamefont {Chen}, \citenamefont {Li}, \citenamefont {Zhang},\ and\ \citenamefont {Han}}]{long2020structural}%
  \BibitemOpen
  \bibfield  {author} {\bibinfo {author} {\bibfnamefont {H.}~\bibnamefont {Long}}, \bibinfo {author} {\bibfnamefont {S.}~\bibnamefont {Mao}}, \bibinfo {author} {\bibfnamefont {Y.}~\bibnamefont {Liu}}, \bibinfo {author} {\bibfnamefont {H.}~\bibnamefont {Yang}}, \bibinfo {author} {\bibfnamefont {H.}~\bibnamefont {Wei}}, \bibinfo {author} {\bibfnamefont {Q.}~\bibnamefont {Deng}}, \bibinfo {author} {\bibfnamefont {Y.}~\bibnamefont {Chen}}, \bibinfo {author} {\bibfnamefont {A.}~\bibnamefont {Li}}, \bibinfo {author} {\bibfnamefont {Z.}~\bibnamefont {Zhang}},\ and\ \bibinfo {author} {\bibfnamefont {X.}~\bibnamefont {Han}},\ }\bibfield  {title} {\bibinfo {title} {Structural evolution of topologically closed packed phase in a {Ni}-based single crystal superalloy},\ }\href {https://doi.org/https://doi.org/10.1016/j.actamat.2019.12.014} {\bibfield  {journal} {\bibinfo  {journal} {Acta Mater.}\ }\textbf {\bibinfo {volume} {185}},\ \bibinfo {pages} {233} (\bibinfo {year} {2020})}\BibitemShut {NoStop}%
\bibitem [{\citenamefont {Choi}\ \emph {et~al.}(2011)\citenamefont {Choi}, \citenamefont {Wang}, \citenamefont {Chung}, \citenamefont {Liu}, \citenamefont {Darbal}, \citenamefont {Wise}, \citenamefont {Nuhfer}, \citenamefont {Barmak}, \citenamefont {Warren}, \citenamefont {Coffey},\ and\ \citenamefont {Toney}}]{choi_PhaseGrain_2011a}%
  \BibitemOpen
  \bibfield  {author} {\bibinfo {author} {\bibfnamefont {D.}~\bibnamefont {Choi}}, \bibinfo {author} {\bibfnamefont {B.}~\bibnamefont {Wang}}, \bibinfo {author} {\bibfnamefont {S.}~\bibnamefont {Chung}}, \bibinfo {author} {\bibfnamefont {X.}~\bibnamefont {Liu}}, \bibinfo {author} {\bibfnamefont {A.}~\bibnamefont {Darbal}}, \bibinfo {author} {\bibfnamefont {A.}~\bibnamefont {Wise}}, \bibinfo {author} {\bibfnamefont {N.~T.}\ \bibnamefont {Nuhfer}}, \bibinfo {author} {\bibfnamefont {K.}~\bibnamefont {Barmak}}, \bibinfo {author} {\bibfnamefont {A.~P.}\ \bibnamefont {Warren}}, \bibinfo {author} {\bibfnamefont {K.~R.}\ \bibnamefont {Coffey}},\ and\ \bibinfo {author} {\bibfnamefont {M.~F.}\ \bibnamefont {Toney}},\ }\bibfield  {title} {\bibinfo {title} {Phase, grain structure, stress, and resistivity of sputter-deposited tungsten films},\ }\href {https://doi.org/10.1116/1.3622619} {\bibfield  {journal} {\bibinfo  {journal} {J. Vac. Sci. Technol. A}\ }\textbf {\bibinfo {volume} {29}},\ \bibinfo {pages} {051512}
  (\bibinfo {year} {2011})}\BibitemShut {NoStop}%
\bibitem [{\citenamefont {Choi}\ \emph {et~al.}(2012)\citenamefont {Choi}, \citenamefont {Kim}, \citenamefont {Naveh}, \citenamefont {Chung}, \citenamefont {Warren}, \citenamefont {Nuhfer}, \citenamefont {Toney}, \citenamefont {Coffey},\ and\ \citenamefont {Barmak}}]{choi_ElectronMean_2012}%
  \BibitemOpen
  \bibfield  {author} {\bibinfo {author} {\bibfnamefont {D.}~\bibnamefont {Choi}}, \bibinfo {author} {\bibfnamefont {C.~S.}\ \bibnamefont {Kim}}, \bibinfo {author} {\bibfnamefont {D.}~\bibnamefont {Naveh}}, \bibinfo {author} {\bibfnamefont {S.}~\bibnamefont {Chung}}, \bibinfo {author} {\bibfnamefont {A.~P.}\ \bibnamefont {Warren}}, \bibinfo {author} {\bibfnamefont {N.~T.}\ \bibnamefont {Nuhfer}}, \bibinfo {author} {\bibfnamefont {M.~F.}\ \bibnamefont {Toney}}, \bibinfo {author} {\bibfnamefont {K.~R.}\ \bibnamefont {Coffey}},\ and\ \bibinfo {author} {\bibfnamefont {K.}~\bibnamefont {Barmak}},\ }\bibfield  {title} {\bibinfo {title} {Electron mean free path of tungsten and the electrical resistivity of epitaxial (110) tungsten films},\ }\href {https://doi.org/10.1103/PhysRevB.86.045432} {\bibfield  {journal} {\bibinfo  {journal} {Phys. Rev. B}\ }\textbf {\bibinfo {volume} {86}},\ \bibinfo {pages} {045432} (\bibinfo {year} {2012})}\BibitemShut {NoStop}%
\bibitem [{\citenamefont {Pai}\ \emph {et~al.}(2012)\citenamefont {Pai}, \citenamefont {Liu}, \citenamefont {Li}, \citenamefont {Tseng}, \citenamefont {Ralph},\ and\ \citenamefont {Buhrman}}]{paiSpinTransferTorque2012}%
  \BibitemOpen
  \bibfield  {author} {\bibinfo {author} {\bibfnamefont {C.-F.}\ \bibnamefont {Pai}}, \bibinfo {author} {\bibfnamefont {L.}~\bibnamefont {Liu}}, \bibinfo {author} {\bibfnamefont {Y.}~\bibnamefont {Li}}, \bibinfo {author} {\bibfnamefont {H.~W.}\ \bibnamefont {Tseng}}, \bibinfo {author} {\bibfnamefont {D.~C.}\ \bibnamefont {Ralph}},\ and\ \bibinfo {author} {\bibfnamefont {R.~A.}\ \bibnamefont {Buhrman}},\ }\bibfield  {title} {\bibinfo {title} {Spin transfer torque devices utilizing the giant spin hall effect of tungsten},\ }\href {https://doi.org/10.1063/1.4753947} {\bibfield  {journal} {\bibinfo  {journal} {Appl. Phys. Lett.}\ }\textbf {\bibinfo {volume} {101}},\ \bibinfo {pages} {122404} (\bibinfo {year} {2012})}\BibitemShut {NoStop}%
\bibitem [{\citenamefont {O'Keefe}\ and\ \citenamefont {Grant}(1996)}]{okeefePhaseTransformationSputter1996}%
  \BibitemOpen
  \bibfield  {author} {\bibinfo {author} {\bibfnamefont {M.~J.}\ \bibnamefont {O'Keefe}}\ and\ \bibinfo {author} {\bibfnamefont {J.~T.}\ \bibnamefont {Grant}},\ }\bibfield  {title} {\bibinfo {title} {Phase transformation of sputter deposited tungsten thin films with {{A}}-15 structure},\ }\href {https://doi.org/10.1063/1.362584} {\bibfield  {journal} {\bibinfo  {journal} {J. Appl. Phys.}\ }\textbf {\bibinfo {volume} {79}},\ \bibinfo {pages} {9134} (\bibinfo {year} {1996})}\BibitemShut {NoStop}%
\bibitem [{\citenamefont {Narasimham}\ \emph {et~al.}(2014)\citenamefont {Narasimham}, \citenamefont {Medikonda}, \citenamefont {Matsubayashi}, \citenamefont {Khare}, \citenamefont {Chong}, \citenamefont {Matyi}, \citenamefont {Diebold},\ and\ \citenamefont {LaBella}}]{narasimhamFabrication520Nm2014}%
  \BibitemOpen
  \bibfield  {author} {\bibinfo {author} {\bibfnamefont {A.~J.}\ \bibnamefont {Narasimham}}, \bibinfo {author} {\bibfnamefont {M.}~\bibnamefont {Medikonda}}, \bibinfo {author} {\bibfnamefont {A.}~\bibnamefont {Matsubayashi}}, \bibinfo {author} {\bibfnamefont {P.}~\bibnamefont {Khare}}, \bibinfo {author} {\bibfnamefont {H.}~\bibnamefont {Chong}}, \bibinfo {author} {\bibfnamefont {R.~J.}\ \bibnamefont {Matyi}}, \bibinfo {author} {\bibfnamefont {A.}~\bibnamefont {Diebold}},\ and\ \bibinfo {author} {\bibfnamefont {V.~P.}\ \bibnamefont {LaBella}},\ }\bibfield  {title} {\bibinfo {title} {Fabrication of 5-20 nm thick {$\beta$}-{{W}} films},\ }\href {https://doi.org/10.1063/1.4903165} {\bibfield  {journal} {\bibinfo  {journal} {AIP Advances}\ }\textbf {\bibinfo {volume} {4}},\ \bibinfo {pages} {117139} (\bibinfo {year} {2014})}\BibitemShut {NoStop}%
\bibitem [{\citenamefont {V{\"u}llers}\ and\ \citenamefont {Spolenak}(2015)}]{vullers2015alpha}%
  \BibitemOpen
  \bibfield  {author} {\bibinfo {author} {\bibfnamefont {F.}~\bibnamefont {V{\"u}llers}}\ and\ \bibinfo {author} {\bibfnamefont {R.}~\bibnamefont {Spolenak}},\ }\bibfield  {title} {\bibinfo {title} {Alpha-vs. beta-{W} nanocrystalline thin films: A comprehensive study of sputter parameters and resulting materials' properties},\ }\href {https://doi.org/https://doi.org/10.1016/j.tsf.2015.01.030} {\bibfield  {journal} {\bibinfo  {journal} {Thin Solid Films}\ }\textbf {\bibinfo {volume} {577}},\ \bibinfo {pages} {26} (\bibinfo {year} {2015})}\BibitemShut {NoStop}%
\bibitem [{\citenamefont {Liu}\ and\ \citenamefont {Barmak}(2016)}]{liuTopologicallyClosepackedPhases2016}%
  \BibitemOpen
  \bibfield  {author} {\bibinfo {author} {\bibfnamefont {J.}~\bibnamefont {Liu}}\ and\ \bibinfo {author} {\bibfnamefont {K.}~\bibnamefont {Barmak}},\ }\bibfield  {title} {\bibinfo {title} {Topologically close-packed phases: {{Deposition}} and formation mechanism of metastable {$\beta$}-{{W}} in thin films},\ }\href {https://doi.org/10.1016/j.actamat.2015.11.049} {\bibfield  {journal} {\bibinfo  {journal} {Acta Mater.}\ }\textbf {\bibinfo {volume} {104}},\ \bibinfo {pages} {223} (\bibinfo {year} {2016})}\BibitemShut {NoStop}%
\bibitem [{\citenamefont {Zhu}\ \emph {et~al.}(2018)\citenamefont {Zhu}, \citenamefont {Xie},\ and\ \citenamefont {Zhang}}]{zhu2018phase}%
  \BibitemOpen
  \bibfield  {author} {\bibinfo {author} {\bibfnamefont {F.}~\bibnamefont {Zhu}}, \bibinfo {author} {\bibfnamefont {Z.}~\bibnamefont {Xie}},\ and\ \bibinfo {author} {\bibfnamefont {Z.}~\bibnamefont {Zhang}},\ }\bibfield  {title} {\bibinfo {title} {Phase control and {Young’s} modulus of tungsten thin film prepared by dual ion beam sputtering deposition},\ }\bibfield  {journal} {\bibinfo  {journal} {AIP Advances}\ }\textbf {\bibinfo {volume} {8}},\ \href {https://doi.org/https://doi.org/10.1063/1.5021009} {https://doi.org/10.1063/1.5021009} (\bibinfo {year} {2018})\BibitemShut {NoStop}%
\bibitem [{\citenamefont {Barmak}\ \emph {et~al.}(2017)\citenamefont {Barmak}, \citenamefont {Liu}, \citenamefont {Harlan}, \citenamefont {Xiao}, \citenamefont {Duncan},\ and\ \citenamefont {Henkelman}}]{barmak_TransformationTopologically_2017a}%
  \BibitemOpen
  \bibfield  {author} {\bibinfo {author} {\bibfnamefont {K.}~\bibnamefont {Barmak}}, \bibinfo {author} {\bibfnamefont {J.}~\bibnamefont {Liu}}, \bibinfo {author} {\bibfnamefont {L.}~\bibnamefont {Harlan}}, \bibinfo {author} {\bibfnamefont {P.}~\bibnamefont {Xiao}}, \bibinfo {author} {\bibfnamefont {J.}~\bibnamefont {Duncan}},\ and\ \bibinfo {author} {\bibfnamefont {G.}~\bibnamefont {Henkelman}},\ }\bibfield  {title} {\bibinfo {title} {Transformation of topologically close-packed {$\beta$}-{{W}} to body-centered cubic {$\alpha$}-{{W}}: {{Comparison}} of experiments and computations},\ }\href {https://doi.org/10.1063/1.4995261} {\bibfield  {journal} {\bibinfo  {journal} {J. Chem. Phys.}\ }\textbf {\bibinfo {volume} {147}},\ \bibinfo {pages} {152709} (\bibinfo {year} {2017})}\BibitemShut {NoStop}%
\bibitem [{\citenamefont {Wang}\ \emph {et~al.}(2019)\citenamefont {Wang}, \citenamefont {Zang}, \citenamefont {Gao}, \citenamefont {Liu}, \citenamefont {Wang}, \citenamefont {Gong}, \citenamefont {Zha}, \citenamefont {Chen}, \citenamefont {Huang}, \citenamefont {Javaid} \emph {et~al.}}]{wang2019demand}%
  \BibitemOpen
  \bibfield  {author} {\bibinfo {author} {\bibfnamefont {X.}~\bibnamefont {Wang}}, \bibinfo {author} {\bibfnamefont {R.}~\bibnamefont {Zang}}, \bibinfo {author} {\bibfnamefont {J.}~\bibnamefont {Gao}}, \bibinfo {author} {\bibfnamefont {C.}~\bibnamefont {Liu}}, \bibinfo {author} {\bibfnamefont {L.}~\bibnamefont {Wang}}, \bibinfo {author} {\bibfnamefont {W.}~\bibnamefont {Gong}}, \bibinfo {author} {\bibfnamefont {X.}~\bibnamefont {Zha}}, \bibinfo {author} {\bibfnamefont {X.}~\bibnamefont {Chen}}, \bibinfo {author} {\bibfnamefont {F.}~\bibnamefont {Huang}}, \bibinfo {author} {\bibfnamefont {K.}~\bibnamefont {Javaid}}, \emph {et~al.},\ }\bibfield  {title} {\bibinfo {title} {On-demand preparation of $\alpha$-phase-dominated tungsten films for highly qualified thermal reflectors},\ }\href {https://doi.org/https://doi.org/10.1002/admi.201900031} {\bibfield  {journal} {\bibinfo  {journal} {Adv. Mater. Interfaces}\ }\textbf {\bibinfo {volume} {6}},\ \bibinfo {pages} {1900031} (\bibinfo {year} {2019})}\BibitemShut
  {NoStop}%
\bibitem [{\citenamefont {Rogal}\ \emph {et~al.}(2019)\citenamefont {Rogal}, \citenamefont {Schneider},\ and\ \citenamefont {Tuckerman}}]{rogal2019neural}%
  \BibitemOpen
  \bibfield  {author} {\bibinfo {author} {\bibfnamefont {J.}~\bibnamefont {Rogal}}, \bibinfo {author} {\bibfnamefont {E.}~\bibnamefont {Schneider}},\ and\ \bibinfo {author} {\bibfnamefont {M.~E.}\ \bibnamefont {Tuckerman}},\ }\bibfield  {title} {\bibinfo {title} {Neural-network-based path collective variables for enhanced sampling of phase transformations},\ }\href {https://doi.org/https://doi.org/10.1103/PhysRevLett.123.245701} {\bibfield  {journal} {\bibinfo  {journal} {Phys. Rev. Lett.}\ }\textbf {\bibinfo {volume} {123}},\ \bibinfo {pages} {245701} (\bibinfo {year} {2019})}\BibitemShut {NoStop}%
\bibitem [{\citenamefont {Rogal}\ and\ \citenamefont {Tuckerman}(2021)}]{rogal2021pathways}%
  \BibitemOpen
  \bibfield  {author} {\bibinfo {author} {\bibfnamefont {J.}~\bibnamefont {Rogal}}\ and\ \bibinfo {author} {\bibfnamefont {M.~E.}\ \bibnamefont {Tuckerman}},\ }\bibfield  {title} {\bibinfo {title} {Pathways in classification space: Machine learning as a route to predicting kinetics of structural transitions in atomic crystals},\ }in\ \href {https://doi.org/https://doi.org/10.1557/s43577-022-00407-1} {\emph {\bibinfo {booktitle} {Multiscale Dynamics Simulations}}}\ (\bibinfo  {publisher} {Royal Society of Chemistry},\ \bibinfo {year} {2021})\ pp.\ \bibinfo {pages} {312--348}\BibitemShut {NoStop}%
\bibitem [{\citenamefont {Laio}\ and\ \citenamefont {Parrinello}(2002)}]{laioEscapingFreeenergyMinima2002}%
  \BibitemOpen
  \bibfield  {author} {\bibinfo {author} {\bibfnamefont {A.}~\bibnamefont {Laio}}\ and\ \bibinfo {author} {\bibfnamefont {M.}~\bibnamefont {Parrinello}},\ }\bibfield  {title} {\bibinfo {title} {Escaping free-energy minima},\ }\href {https://doi.org/10.1073/pnas.202427399} {\bibfield  {journal} {\bibinfo  {journal} {Proc. Natl. Acad. Sci. U.S.A.}\ }\textbf {\bibinfo {volume} {99}},\ \bibinfo {pages} {12562} (\bibinfo {year} {2002})}\BibitemShut {NoStop}%
\bibitem [{\citenamefont {Laio}\ \emph {et~al.}(2005)\citenamefont {Laio}, \citenamefont {{Rodriguez-Fortea}}, \citenamefont {Gervasio}, \citenamefont {Ceccarelli},\ and\ \citenamefont {Parrinello}}]{laioAssessingAccuracyMetadynamics2005}%
  \BibitemOpen
  \bibfield  {author} {\bibinfo {author} {\bibfnamefont {A.}~\bibnamefont {Laio}}, \bibinfo {author} {\bibfnamefont {A.}~\bibnamefont {{Rodriguez-Fortea}}}, \bibinfo {author} {\bibfnamefont {F.~L.}\ \bibnamefont {Gervasio}}, \bibinfo {author} {\bibfnamefont {M.}~\bibnamefont {Ceccarelli}},\ and\ \bibinfo {author} {\bibfnamefont {M.}~\bibnamefont {Parrinello}},\ }\bibfield  {title} {\bibinfo {title} {Assessing the accuracy of metadynamics},\ }\href {https://doi.org/10.1021/jp045424k} {\bibfield  {journal} {\bibinfo  {journal} {J. Phys. Chem. B}\ }\textbf {\bibinfo {volume} {109}},\ \bibinfo {pages} {6714} (\bibinfo {year} {2005})}\BibitemShut {NoStop}%
\bibitem [{\citenamefont {Peng}\ \emph {et~al.}(2023)\citenamefont {Peng}, \citenamefont {Li}, \citenamefont {Still}, \citenamefont {Yodh},\ and\ \citenamefont {Han}}]{peng2023situ}%
  \BibitemOpen
  \bibfield  {author} {\bibinfo {author} {\bibfnamefont {Y.}~\bibnamefont {Peng}}, \bibinfo {author} {\bibfnamefont {W.}~\bibnamefont {Li}}, \bibinfo {author} {\bibfnamefont {T.}~\bibnamefont {Still}}, \bibinfo {author} {\bibfnamefont {A.~G.}\ \bibnamefont {Yodh}},\ and\ \bibinfo {author} {\bibfnamefont {Y.}~\bibnamefont {Han}},\ }\bibfield  {title} {\bibinfo {title} {In situ observation of coalescence of nuclei in colloidal crystal-crystal transitions},\ }\href {https://doi.org/https://doi.org/10.1038/s41467-023-40627-w} {\bibfield  {journal} {\bibinfo  {journal} {Nat. Commun.}\ }\textbf {\bibinfo {volume} {14}},\ \bibinfo {pages} {4905} (\bibinfo {year} {2023})}\BibitemShut {NoStop}%
\bibitem [{\citenamefont {Behler}\ and\ \citenamefont {Parrinello}(2007)}]{behler2007generalized}%
  \BibitemOpen
  \bibfield  {author} {\bibinfo {author} {\bibfnamefont {J.}~\bibnamefont {Behler}}\ and\ \bibinfo {author} {\bibfnamefont {M.}~\bibnamefont {Parrinello}},\ }\bibfield  {title} {\bibinfo {title} {Generalized neural-network representation of high-dimensional potential-energy surfaces},\ }\href {https://doi.org/https://doi.org/10.1103/PhysRevLett.98.146401} {\bibfield  {journal} {\bibinfo  {journal} {Phys. Rev. Lett.}\ }\textbf {\bibinfo {volume} {98}},\ \bibinfo {pages} {146401} (\bibinfo {year} {2007})}\BibitemShut {NoStop}%
\bibitem [{\citenamefont {Bart\'ok}\ \emph {et~al.}(2013)\citenamefont {Bart\'ok}, \citenamefont {Kondor},\ and\ \citenamefont {Cs\'anyi}}]{Bartok2013}%
  \BibitemOpen
  \bibfield  {author} {\bibinfo {author} {\bibfnamefont {A.~P.}\ \bibnamefont {Bart\'ok}}, \bibinfo {author} {\bibfnamefont {R.}~\bibnamefont {Kondor}},\ and\ \bibinfo {author} {\bibfnamefont {G.}~\bibnamefont {Cs\'anyi}},\ }\bibfield  {title} {\bibinfo {title} {On representing chemical environments},\ }\href {https://doi.org/10.1103/PhysRevB.87.184115} {\bibfield  {journal} {\bibinfo  {journal} {Phys. Rev. B}\ }\textbf {\bibinfo {volume} {87}},\ \bibinfo {pages} {184115} (\bibinfo {year} {2013})}\BibitemShut {NoStop}%
\bibitem [{\citenamefont {Drautz}(2019)}]{drautz2019}%
  \BibitemOpen
  \bibfield  {author} {\bibinfo {author} {\bibfnamefont {R.}~\bibnamefont {Drautz}},\ }\bibfield  {title} {\bibinfo {title} {Atomic cluster expansion for accurate and transferable interatomic potentials},\ }\href {https://doi.org/10.1103/PhysRevB.99.014104} {\bibfield  {journal} {\bibinfo  {journal} {Phys. Rev. B}\ }\textbf {\bibinfo {volume} {99}},\ \bibinfo {pages} {014104} (\bibinfo {year} {2019})}\BibitemShut {NoStop}%
\bibitem [{\citenamefont {Geiger}\ and\ \citenamefont {Dellago}(2013)}]{geiger2013neural}%
  \BibitemOpen
  \bibfield  {author} {\bibinfo {author} {\bibfnamefont {P.}~\bibnamefont {Geiger}}\ and\ \bibinfo {author} {\bibfnamefont {C.}~\bibnamefont {Dellago}},\ }\bibfield  {title} {\bibinfo {title} {Neural networks for local structure detection in polymorphic systems},\ }\href {https://doi.org/https://doi.org/10.1063/1.4825111} {\bibfield  {journal} {\bibinfo  {journal} {J. Chem. Phys.}\ }\textbf {\bibinfo {volume} {139}},\ \bibinfo {pages} {164105} (\bibinfo {year} {2013})}\BibitemShut {NoStop}%
\bibitem [{\citenamefont {Behler}(2011)}]{behler2011atom}%
  \BibitemOpen
  \bibfield  {author} {\bibinfo {author} {\bibfnamefont {J.}~\bibnamefont {Behler}},\ }\bibfield  {title} {\bibinfo {title} {Atom-centered symmetry functions for constructing high-dimensional neural network potentials},\ }\href {https://doi.org/https://doi.org/10.1063/1.3553717} {\bibfield  {journal} {\bibinfo  {journal} {J. Chem. Phys.}\ }\textbf {\bibinfo {volume} {134}},\ \bibinfo {pages} {074106} (\bibinfo {year} {2011})}\BibitemShut {NoStop}%
\bibitem [{\citenamefont {Steinhardt}\ \emph {et~al.}(1983)\citenamefont {Steinhardt}, \citenamefont {Nelson},\ and\ \citenamefont {Ronchetti}}]{steinhardt1983bond}%
  \BibitemOpen
  \bibfield  {author} {\bibinfo {author} {\bibfnamefont {P.~J.}\ \bibnamefont {Steinhardt}}, \bibinfo {author} {\bibfnamefont {D.~R.}\ \bibnamefont {Nelson}},\ and\ \bibinfo {author} {\bibfnamefont {M.}~\bibnamefont {Ronchetti}},\ }\bibfield  {title} {\bibinfo {title} {Bond-orientational order in liquids and glasses},\ }\href {https://doi.org/https://doi.org/10.1103/PhysRevB.28.784} {\bibfield  {journal} {\bibinfo  {journal} {Phys. Rev. B}\ }\textbf {\bibinfo {volume} {28}},\ \bibinfo {pages} {784} (\bibinfo {year} {1983})}\BibitemShut {NoStop}%
\bibitem [{\citenamefont {Lechner}\ and\ \citenamefont {Dellago}(2008)}]{lechnerAccurateDeterminationCrystal2008}%
  \BibitemOpen
  \bibfield  {author} {\bibinfo {author} {\bibfnamefont {W.}~\bibnamefont {Lechner}}\ and\ \bibinfo {author} {\bibfnamefont {C.}~\bibnamefont {Dellago}},\ }\bibfield  {title} {\bibinfo {title} {Accurate determination of crystal structures based on averaged local bond order parameters},\ }\href {https://doi.org/10.1063/1.2977970} {\bibfield  {journal} {\bibinfo  {journal} {J. Chem. Phys.}\ }\textbf {\bibinfo {volume} {129}},\ \bibinfo {pages} {114707} (\bibinfo {year} {2008})}\BibitemShut {NoStop}%
\bibitem [{\citenamefont {Branduardi}\ \emph {et~al.}(2007)\citenamefont {Branduardi}, \citenamefont {Gervasio},\ and\ \citenamefont {Parrinello}}]{branduardi2007b}%
  \BibitemOpen
  \bibfield  {author} {\bibinfo {author} {\bibfnamefont {D.}~\bibnamefont {Branduardi}}, \bibinfo {author} {\bibfnamefont {F.~L.}\ \bibnamefont {Gervasio}},\ and\ \bibinfo {author} {\bibfnamefont {M.}~\bibnamefont {Parrinello}},\ }\bibfield  {title} {\bibinfo {title} {From {A} to {B} in free energy space},\ }\href {https://doi.org/https://doi.org/10.1063/1.2432340} {\bibfield  {journal} {\bibinfo  {journal} {J. Chem. Phys.}\ }\textbf {\bibinfo {volume} {126}},\ \bibinfo {pages} {054103} (\bibinfo {year} {2007})}\BibitemShut {NoStop}%
\bibitem [{\citenamefont {Cuendet}\ \emph {et~al.}(2018)\citenamefont {Cuendet}, \citenamefont {Margul}, \citenamefont {Schneider}, \citenamefont {Vogt-Maranto},\ and\ \citenamefont {Tuckerman}}]{cuendet2018endpoint}%
  \BibitemOpen
  \bibfield  {author} {\bibinfo {author} {\bibfnamefont {M.~A.}\ \bibnamefont {Cuendet}}, \bibinfo {author} {\bibfnamefont {D.~T.}\ \bibnamefont {Margul}}, \bibinfo {author} {\bibfnamefont {E.}~\bibnamefont {Schneider}}, \bibinfo {author} {\bibfnamefont {L.}~\bibnamefont {Vogt-Maranto}},\ and\ \bibinfo {author} {\bibfnamefont {M.~E.}\ \bibnamefont {Tuckerman}},\ }\bibfield  {title} {\bibinfo {title} {Endpoint-restricted adiabatic free energy dynamics approach for the exploration of biomolecular conformational equilibria},\ }\href {https://doi.org/https://doi.org/10.1063/1.5027479} {\bibfield  {journal} {\bibinfo  {journal} {J. Chem. Phys.}\ }\textbf {\bibinfo {volume} {149}},\ \bibinfo {pages} {072316} (\bibinfo {year} {2018})}\BibitemShut {NoStop}%
\bibitem [{\citenamefont {Iannuzzi}\ \emph {et~al.}(2003)\citenamefont {Iannuzzi}, \citenamefont {Laio},\ and\ \citenamefont {Parrinello}}]{iannuzzi2003efficient}%
  \BibitemOpen
  \bibfield  {author} {\bibinfo {author} {\bibfnamefont {M.}~\bibnamefont {Iannuzzi}}, \bibinfo {author} {\bibfnamefont {A.}~\bibnamefont {Laio}},\ and\ \bibinfo {author} {\bibfnamefont {M.}~\bibnamefont {Parrinello}},\ }\bibfield  {title} {\bibinfo {title} {Efficient exploration of reactive potential energy surfaces using {Car-Parrinello} molecular dynamics},\ }\href {https://doi.org/https://doi.org/10.1103/PhysRevLett.90.238302} {\bibfield  {journal} {\bibinfo  {journal} {Phys. Rev. Lett.}\ }\textbf {\bibinfo {volume} {90}},\ \bibinfo {pages} {238302} (\bibinfo {year} {2003})}\BibitemShut {NoStop}%
\bibitem [{\citenamefont {Laio}\ and\ \citenamefont {Gervasio}(2008)}]{laioMetadynamicsMethodSimulate2008}%
  \BibitemOpen
  \bibfield  {author} {\bibinfo {author} {\bibfnamefont {A.}~\bibnamefont {Laio}}\ and\ \bibinfo {author} {\bibfnamefont {F.~L.}\ \bibnamefont {Gervasio}},\ }\bibfield  {title} {\bibinfo {title} {Metadynamics: A method to simulate rare events and reconstruct the free energy in biophysics, chemistry and material science},\ }\href {https://doi.org/10.1088/0034-4885/71/12/126601} {\bibfield  {journal} {\bibinfo  {journal} {Rep. Prog. Phys.}\ }\textbf {\bibinfo {volume} {71}},\ \bibinfo {pages} {126601} (\bibinfo {year} {2008})}\BibitemShut {NoStop}%
\bibitem [{\citenamefont {Valsson}\ \emph {et~al.}(2016)\citenamefont {Valsson}, \citenamefont {Tiwary},\ and\ \citenamefont {Parrinello}}]{valsson2016enhancing}%
  \BibitemOpen
  \bibfield  {author} {\bibinfo {author} {\bibfnamefont {O.}~\bibnamefont {Valsson}}, \bibinfo {author} {\bibfnamefont {P.}~\bibnamefont {Tiwary}},\ and\ \bibinfo {author} {\bibfnamefont {M.}~\bibnamefont {Parrinello}},\ }\bibfield  {title} {\bibinfo {title} {Enhancing important fluctuations: Rare events and metadynamics from a conceptual viewpoint},\ }\href {https://doi.org/https://doi.org/10.1146/annurev-physchem-040215-112229} {\bibfield  {journal} {\bibinfo  {journal} {Annu. Rev. Phys. Chem.}\ }\textbf {\bibinfo {volume} {67}},\ \bibinfo {pages} {159} (\bibinfo {year} {2016})}\BibitemShut {NoStop}%
\bibitem [{\citenamefont {Bussi}\ and\ \citenamefont {Laio}(2020)}]{bussi2020using}%
  \BibitemOpen
  \bibfield  {author} {\bibinfo {author} {\bibfnamefont {G.}~\bibnamefont {Bussi}}\ and\ \bibinfo {author} {\bibfnamefont {A.}~\bibnamefont {Laio}},\ }\bibfield  {title} {\bibinfo {title} {Using metadynamics to explore complex free-energy landscapes},\ }\href {https://doi.org/https://doi.org/10.1038/s42254-020-0153-0} {\bibfield  {journal} {\bibinfo  {journal} {Nat. Rev. Phys.}\ }\textbf {\bibinfo {volume} {2}},\ \bibinfo {pages} {200} (\bibinfo {year} {2020})}\BibitemShut {NoStop}%
\bibitem [{\citenamefont {Thompson}\ \emph {et~al.}(2022)\citenamefont {Thompson}, \citenamefont {Aktulga}, \citenamefont {Berger}, \citenamefont {Bolintineanu}, \citenamefont {Brown}, \citenamefont {Crozier}, \citenamefont {in~'t Veld}, \citenamefont {Kohlmeyer}, \citenamefont {Moore}, \citenamefont {Nguyen}, \citenamefont {Shan}, \citenamefont {Stevens}, \citenamefont {Tranchida}, \citenamefont {Trott},\ and\ \citenamefont {Plimpton}}]{LAMMPS}%
  \BibitemOpen
  \bibfield  {author} {\bibinfo {author} {\bibfnamefont {A.~P.}\ \bibnamefont {Thompson}}, \bibinfo {author} {\bibfnamefont {H.~M.}\ \bibnamefont {Aktulga}}, \bibinfo {author} {\bibfnamefont {R.}~\bibnamefont {Berger}}, \bibinfo {author} {\bibfnamefont {D.~S.}\ \bibnamefont {Bolintineanu}}, \bibinfo {author} {\bibfnamefont {W.~M.}\ \bibnamefont {Brown}}, \bibinfo {author} {\bibfnamefont {P.~S.}\ \bibnamefont {Crozier}}, \bibinfo {author} {\bibfnamefont {P.~J.}\ \bibnamefont {in~'t Veld}}, \bibinfo {author} {\bibfnamefont {A.}~\bibnamefont {Kohlmeyer}}, \bibinfo {author} {\bibfnamefont {S.~G.}\ \bibnamefont {Moore}}, \bibinfo {author} {\bibfnamefont {T.~D.}\ \bibnamefont {Nguyen}}, \bibinfo {author} {\bibfnamefont {R.}~\bibnamefont {Shan}}, \bibinfo {author} {\bibfnamefont {M.~J.}\ \bibnamefont {Stevens}}, \bibinfo {author} {\bibfnamefont {J.}~\bibnamefont {Tranchida}}, \bibinfo {author} {\bibfnamefont {C.}~\bibnamefont {Trott}},\ and\ \bibinfo {author} {\bibfnamefont {S.~J.}\ \bibnamefont {Plimpton}},\
  }\bibfield  {title} {\bibinfo {title} {{LAMMPS} - a flexible simulation tool for particle-based materials modeling at the atomic, meso, and continuum scales},\ }\href {https://doi.org/10.1016/j.cpc.2021.108171} {\bibfield  {journal} {\bibinfo  {journal} {Comp. Phys. Comm.}\ }\textbf {\bibinfo {volume} {271}},\ \bibinfo {pages} {108171} (\bibinfo {year} {2022})}\BibitemShut {NoStop}%
\bibitem [{\citenamefont {Zhou}\ \emph {et~al.}(2004)\citenamefont {Zhou}, \citenamefont {Johnson},\ and\ \citenamefont {Wadley}}]{zhou2004}%
  \BibitemOpen
  \bibfield  {author} {\bibinfo {author} {\bibfnamefont {X.~W.}\ \bibnamefont {Zhou}}, \bibinfo {author} {\bibfnamefont {R.~A.}\ \bibnamefont {Johnson}},\ and\ \bibinfo {author} {\bibfnamefont {H.~N.~G.}\ \bibnamefont {Wadley}},\ }\bibfield  {title} {\bibinfo {title} {Misfit-energy-increasing dislocations in vapor-deposited {CoFe/NiFe} multilayers},\ }\href {https://doi.org/10.1103/PhysRevB.69.144113} {\bibfield  {journal} {\bibinfo  {journal} {Phys. Rev. B}\ }\textbf {\bibinfo {volume} {69}},\ \bibinfo {pages} {144113} (\bibinfo {year} {2004})}\BibitemShut {NoStop}%
\bibitem [{\citenamefont {Evans}\ and\ \citenamefont {Holian}(1985)}]{evans1985nose}%
  \BibitemOpen
  \bibfield  {author} {\bibinfo {author} {\bibfnamefont {D.~J.}\ \bibnamefont {Evans}}\ and\ \bibinfo {author} {\bibfnamefont {B.~L.}\ \bibnamefont {Holian}},\ }\bibfield  {title} {\bibinfo {title} {The nose--hoover thermostat},\ }\href {https://doi.org/https://doi.org/10.1063/1.449071} {\bibfield  {journal} {\bibinfo  {journal} {J. Chem. Phys.}\ }\textbf {\bibinfo {volume} {83}},\ \bibinfo {pages} {4069} (\bibinfo {year} {1985})}\BibitemShut {NoStop}%
\bibitem [{\citenamefont {Nos{\'e}}(1984)}]{noseUnifiedFormulationConstant1984}%
  \BibitemOpen
  \bibfield  {author} {\bibinfo {author} {\bibfnamefont {S.}~\bibnamefont {Nos{\'e}}},\ }\bibfield  {title} {\bibinfo {title} {A unified formulation of the constant temperature molecular dynamics methods},\ }\href {https://doi.org/https://doi.org/10.1063/1.447334} {\bibfield  {journal} {\bibinfo  {journal} {J. Chem. Phys.}\ }\textbf {\bibinfo {volume} {81}},\ \bibinfo {pages} {511} (\bibinfo {year} {1984})}\BibitemShut {NoStop}%
\bibitem [{\citenamefont {Hoover}(1985)}]{hooverCanonicalDynamicsEquilibrium1985}%
  \BibitemOpen
  \bibfield  {author} {\bibinfo {author} {\bibfnamefont {W.~G.}\ \bibnamefont {Hoover}},\ }\bibfield  {title} {\bibinfo {title} {Canonical dynamics: {{Equilibrium}} phase-space distributions},\ }\href {https://doi.org/10.1103/PhysRevA.31.1695} {\bibfield  {journal} {\bibinfo  {journal} {Phys. Rev. A.}\ }\textbf {\bibinfo {volume} {31}},\ \bibinfo {pages} {1695} (\bibinfo {year} {1985})}\BibitemShut {NoStop}%
\bibitem [{\citenamefont {Mendelev}\ \emph {et~al.}(2002)\citenamefont {Mendelev}, \citenamefont {Srolovitz}, \citenamefont {Shvindlerman},\ and\ \citenamefont {Gottstein}}]{mendelevInterfaceMobilityDifferent2002}%
  \BibitemOpen
  \bibfield  {author} {\bibinfo {author} {\bibfnamefont {M.~I.}\ \bibnamefont {Mendelev}}, \bibinfo {author} {\bibfnamefont {D.~J.}\ \bibnamefont {Srolovitz}}, \bibinfo {author} {\bibfnamefont {L.~S.}\ \bibnamefont {Shvindlerman}},\ and\ \bibinfo {author} {\bibfnamefont {G.}~\bibnamefont {Gottstein}},\ }\bibfield  {title} {\bibinfo {title} {Interface mobility under different driving forces},\ }\href {https://doi.org/10.1557/JMR.2002.0033} {\bibfield  {journal} {\bibinfo  {journal} {J. Mater. Res.}\ }\textbf {\bibinfo {volume} {17}},\ \bibinfo {pages} {234} (\bibinfo {year} {2002})}\BibitemShut {NoStop}%
\bibitem [{\citenamefont {Qi}\ \emph {et~al.}(2015)\citenamefont {Qi}, \citenamefont {Peng}, \citenamefont {Han}, \citenamefont {Bowles},\ and\ \citenamefont {Dijkstra}}]{qi2015nonclassical}%
  \BibitemOpen
  \bibfield  {author} {\bibinfo {author} {\bibfnamefont {W.}~\bibnamefont {Qi}}, \bibinfo {author} {\bibfnamefont {Y.}~\bibnamefont {Peng}}, \bibinfo {author} {\bibfnamefont {Y.}~\bibnamefont {Han}}, \bibinfo {author} {\bibfnamefont {R.~K.}\ \bibnamefont {Bowles}},\ and\ \bibinfo {author} {\bibfnamefont {M.}~\bibnamefont {Dijkstra}},\ }\bibfield  {title} {\bibinfo {title} {Nonclassical nucleation in a solid-solid transition of confined hard spheres},\ }\href {https://doi.org/10.1103/PhysRevLett.115.185701} {\bibfield  {journal} {\bibinfo  {journal} {Phys. Rev. Lett.}\ }\textbf {\bibinfo {volume} {115}},\ \bibinfo {pages} {185701} (\bibinfo {year} {2015})}\BibitemShut {NoStop}%
\end{thebibliography}%


\providecommand{\noopsort}[1]{}\providecommand{\singleletter}[1]{#1}%
\begin{thebibliography}{4}%
\makeatletter
\providecommand \@ifxundefined [1]{%
 \@ifx{#1\undefined}
}%
\providecommand \@ifnum [1]{%
 \ifnum #1\expandafter \@firstoftwo
 \else \expandafter \@secondoftwo
 \fi
}%
\providecommand \@ifx [1]{%
 \ifx #1\expandafter \@firstoftwo
 \else \expandafter \@secondoftwo
 \fi
}%
\providecommand \natexlab [1]{#1}%
\providecommand \enquote  [1]{``#1''}%
\providecommand \bibnamefont  [1]{#1}%
\providecommand \bibfnamefont [1]{#1}%
\providecommand \citenamefont [1]{#1}%
\providecommand \href@noop [0]{\@secondoftwo}%
\providecommand \href [0]{\begingroup \@sanitize@url \@href}%
\providecommand \@href[1]{\@@startlink{#1}\@@href}%
\providecommand \@@href[1]{\endgroup#1\@@endlink}%
\providecommand \@sanitize@url [0]{\catcode `\\12\catcode `\$12\catcode `\&12\catcode `\#12\catcode `\^12\catcode `\_12\catcode `\%12\relax}%
\providecommand \@@startlink[1]{}%
\providecommand \@@endlink[0]{}%
\providecommand \url  [0]{\begingroup\@sanitize@url \@url }%
\providecommand \@url [1]{\endgroup\@href {#1}{\urlprefix }}%
\providecommand \urlprefix  [0]{URL }%
\providecommand \Eprint [0]{\href }%
\providecommand \doibase [0]{https://doi.org/}%
\providecommand \selectlanguage [0]{\@gobble}%
\providecommand \bibinfo  [0]{\@secondoftwo}%
\providecommand \bibfield  [0]{\@secondoftwo}%
\providecommand \translation [1]{[#1]}%
\providecommand \BibitemOpen [0]{}%
\providecommand \bibitemStop [0]{}%
\providecommand \bibitemNoStop [0]{.\EOS\space}%
\providecommand \EOS [0]{\spacefactor3000\relax}%
\providecommand \BibitemShut  [1]{\csname bibitem#1\endcsname}%
\let\auto@bib@innerbib\@empty
\bibitem [{\citenamefont {Behler}\ and\ \citenamefont {Parrinello}(2007)}]{behler2007generalized}%
  \BibitemOpen
  \bibfield  {author} {\bibinfo {author} {\bibfnamefont {J.}~\bibnamefont {Behler}}\ and\ \bibinfo {author} {\bibfnamefont {M.}~\bibnamefont {Parrinello}},\ }\bibfield  {title} {\bibinfo {title} {Generalized neural-network representation of high-dimensional potential-energy surfaces},\ }\href {https://doi.org/https://doi.org/10.1103/PhysRevLett.98.146401} {\bibfield  {journal} {\bibinfo  {journal} {Phys. Rev. Lett.}\ }\textbf {\bibinfo {volume} {98}},\ \bibinfo {pages} {146401} (\bibinfo {year} {2007})}\BibitemShut {NoStop}%
\bibitem [{\citenamefont {Behler}(2011)}]{behler2011atom}%
  \BibitemOpen
  \bibfield  {author} {\bibinfo {author} {\bibfnamefont {J.}~\bibnamefont {Behler}},\ }\bibfield  {title} {\bibinfo {title} {Atom-centered symmetry functions for constructing high-dimensional neural network potentials},\ }\href {https://doi.org/https://doi.org/10.1063/1.3553717} {\bibfield  {journal} {\bibinfo  {journal} {J. Chem. Phys.}\ }\textbf {\bibinfo {volume} {134}},\ \bibinfo {pages} {074106} (\bibinfo {year} {2011})}\BibitemShut {NoStop}%
\bibitem [{\citenamefont {Steinhardt}\ \emph {et~al.}(1983)\citenamefont {Steinhardt}, \citenamefont {Nelson},\ and\ \citenamefont {Ronchetti}}]{steinhardt1983bond}%
  \BibitemOpen
  \bibfield  {author} {\bibinfo {author} {\bibfnamefont {P.~J.}\ \bibnamefont {Steinhardt}}, \bibinfo {author} {\bibfnamefont {D.~R.}\ \bibnamefont {Nelson}},\ and\ \bibinfo {author} {\bibfnamefont {M.}~\bibnamefont {Ronchetti}},\ }\bibfield  {title} {\bibinfo {title} {Bond-orientational order in liquids and glasses},\ }\href {https://doi.org/https://doi.org/10.1103/PhysRevB.28.784} {\bibfield  {journal} {\bibinfo  {journal} {Phys. Rev. B}\ }\textbf {\bibinfo {volume} {28}},\ \bibinfo {pages} {784} (\bibinfo {year} {1983})}\BibitemShut {NoStop}%
\bibitem [{\citenamefont {Rogal}\ \emph {et~al.}(2019)\citenamefont {Rogal}, \citenamefont {Schneider},\ and\ \citenamefont {Tuckerman}}]{rogal2019neural}%
  \BibitemOpen
  \bibfield  {author} {\bibinfo {author} {\bibfnamefont {J.}~\bibnamefont {Rogal}}, \bibinfo {author} {\bibfnamefont {E.}~\bibnamefont {Schneider}},\ and\ \bibinfo {author} {\bibfnamefont {M.~E.}\ \bibnamefont {Tuckerman}},\ }\bibfield  {title} {\bibinfo {title} {Neural-network-based path collective variables for enhanced sampling of phase transformations},\ }\href {https://doi.org/https://doi.org/10.1103/PhysRevLett.123.245701} {\bibfield  {journal} {\bibinfo  {journal} {Phys. Rev. Lett.}\ }\textbf {\bibinfo {volume} {123}},\ \bibinfo {pages} {245701} (\bibinfo {year} {2019})}\BibitemShut {NoStop}%
\end{thebibliography}%

\end{document}


\preprint{APS/123-QED}

\title[Supplementary information]{Supplementary Information\vspace{0.5cm}\\ 
\large {Structural transformations driven by local disorder at interfaces}}

\author{Yanyan Liang}
\email{yanyan.liang@rub.de}
\affiliation{ICAMS, Ruhr-Universit{\"a}t Bochum, 44801 Bochum, Germany}
\author{Grisell D\'{i}az Leines}%
\affiliation{European Bioinformatics Institute (EMBL- EBI), Wellcome Genome Campus, Hinxton, Cambridge, United Kingdom} %
\author{Ralf Drautz}
\affiliation{ICAMS, Ruhr-Universit{\"a}t Bochum, 44801 Bochum, Germany}
\author{Jutta Rogal}
\affiliation{Department of Chemistry, New York University, New York, NY 10003, USA}
\affiliation{Fachbereich Physik, Freie Universit{\"a}t Berlin, 14195 Berlin, Germany}

\maketitle
\section{Descriptors of local structural environment}

We used two classes of descriptor functions, the Behler-Parrinello symmetry functions~\cite{behler2007generalized,behler2011atom} and the Steinhardt bond order parameters~\cite{steinhardt1983bond}, to serve as structural fingerprints for various local atomic environments. The descriptors are invariant towards translation, rotation, and exchange of two atoms of the same type. Specifically, there are 3 Steinhardt order parameters with  $l=6,7,8$,
\begin{equation}
    G_{{q}_{l}}^{(i)}(\textbf{r})=\sqrt{\frac{4\pi}{(2l+1)}\sum_{m=-l}^{l}|q_{lm}^{(i)}(\textbf{r})|^2}\quad,
    \label{eqn:steinhardt}
\end{equation}
with
\begin{equation}
    q_{lm}^{(i)}(\textbf{r})=\frac{\sum_{j\neq i}Y_{lm}(\textbf{r}_{ij})f_{c}(\textbf{r}_{ij})}{\sum_{j\neq i}f_{c}(\textbf{r}_{ij})}
\end{equation}
where $Y_{lm}(\textbf{r}_{ij})$ represents the spherical harmonics, $\textbf{r}_{ij}$ denotes the distance between atoms $i$ and $j$, and $f_c(\textbf{r}_{ij})$ is a smooth cutoff function,  
\begin{equation}
    f_{c}(\mathbf{r}_{ij})=
   \begin{cases}
   1 &\ \text{if}\ |\mathbf{r}_{ij}|\leq r_\text{min} \\
   0.5\big( \cos \big[\frac{(|\mathbf{r}_{ij}|-r_{\text{min}})}{(r_{\text{max}}-r_{\text{min}})}\pi\big]+1 \big) & \ \text{if}\ r_{\text{min}}<|\mathbf{r}_{ij}|\leq r_{\text{max}} \\
   0 &\ \text{if}\ |\mathbf{r}_{ij}|> r_\text{max}
   \end{cases}
\label{eqn:cutofffunc}
\end{equation}
with $r_{\text{min}}=3.8~\mathrm{\AA}$ and $r_{\text{max}}=4.0~\text{\AA}$.
We utilized two types of Behler-Parrinello radial symmetry functions,
\begin{equation}
    G_{2}^{i}(\textbf{r})=\sum_{j\neq i}e^{-\eta(|\mathbf{r}_{ij}|-R_s)^2}f_c(\mathbf{r}_{ij})
    \label{eqn:symmG2}\quad,
\end{equation}
\begin{equation}
    G_{3}^{i}(\textbf{r})=\sum_{j\neq i}\cos(\kappa|\mathbf{r}_{ij}|)f_{c}(\mathbf{r}_{ij})
    \label{eqn:symmG3}\quad.
\end{equation}
The values of the cutoff function are adjusted  to
$r_\text{min}=6.2$~\AA~and $r_\text{max}=6.4$~\AA, accordingly.
The parameters $\eta$, $R_s$, and $\kappa$ are determined from the overlap of the histograms of corresponding values computed for different structures in Mo obtained from MD simulations, as tabulated Tab.~\ref{tab:parameter_symmetry}.

\begin{table}[ht]
\setlength{\tabcolsep}{16pt}
\centering
\caption{Various parameters for the symmetry functions. These parameters are determined from the overlap of histograms of corresponding values computed for different structures in Mo obtained from MD simulations~\cite{rogal2019neural}.}
\begin{tabular}{lcccc}
\toprule
$G_2$  & $R_s (\mathrm{\AA})$ & $\eta (\mathrm{\AA}^{-2}$) & $G_3$ & $\kappa (\mathrm{\AA}^{-1}$) \\
\midrule
1 & 2.8 & 20.0 & 9 & 3.5 \\
2 &3.2 & 20.0 & 10 & 4.5 \\
3 & 4.4 & 20.0 & 11 & 7.0 \\
4 & 4.8 & 20.0  & & \\
5 & 5.0 & 20.0 & & \\
6 & 5.3 & 20.0 & & \\
7 & 5.7 & 20.0  & & \\
8 & 6.0 & 20.0 & & \\
\bottomrule
\end{tabular}
\label{tab:parameter_symmetry}
\end{table}

\section{Validation of the trained classification neural network}
In order to ensure the trained model of the classification NN is robust for identifying structures at various thermal conditions and transferable to our W system, we generate a large test dataset of 125000 atomic environments by performing MD simulations at a constant pressure ($P=0$~bar). 
Specifically, we include the data computed for configurations extracted from MD runs at $T=300, 450, 600, 800, 1000, 1200, 1300, 1500, 1800$, and 2200~K for various bulk structures and $T=4000, 5000, 6000$~K for liquid (25000 each for liquid and bulk structures). In Tab.~\ref{tab:NNaccuracy}, the accuracy score obtained for each test set is presented. We obtain a minimum accuracy of 98.9\%, validating the transferability and accuracy of this NN model in the application of the structural identification in the W system.
\begin{table}[htbp]
    \centering
    \setlength{\tabcolsep}{16pt}
    \caption{The accuracy obtained for the test set of various phases in W. Each test set contains 25000 data  randomly extracted from MD configurations at different temperatures.}
   \begin{tabular}{cccccc}
   \toprule
    Phase & bcc & A15 & fcc & hcp & liquid \\
   \midrule
    Accuracy (\%) & 99.8 & 99.6 & 99.8& 99.5& 98.9 \\
  \bottomrule
 \end{tabular}
\label{tab:NNaccuracy}
\end{table}
\begin{figure}[h]
    \centering
    \includegraphics[scale=0.5]
    {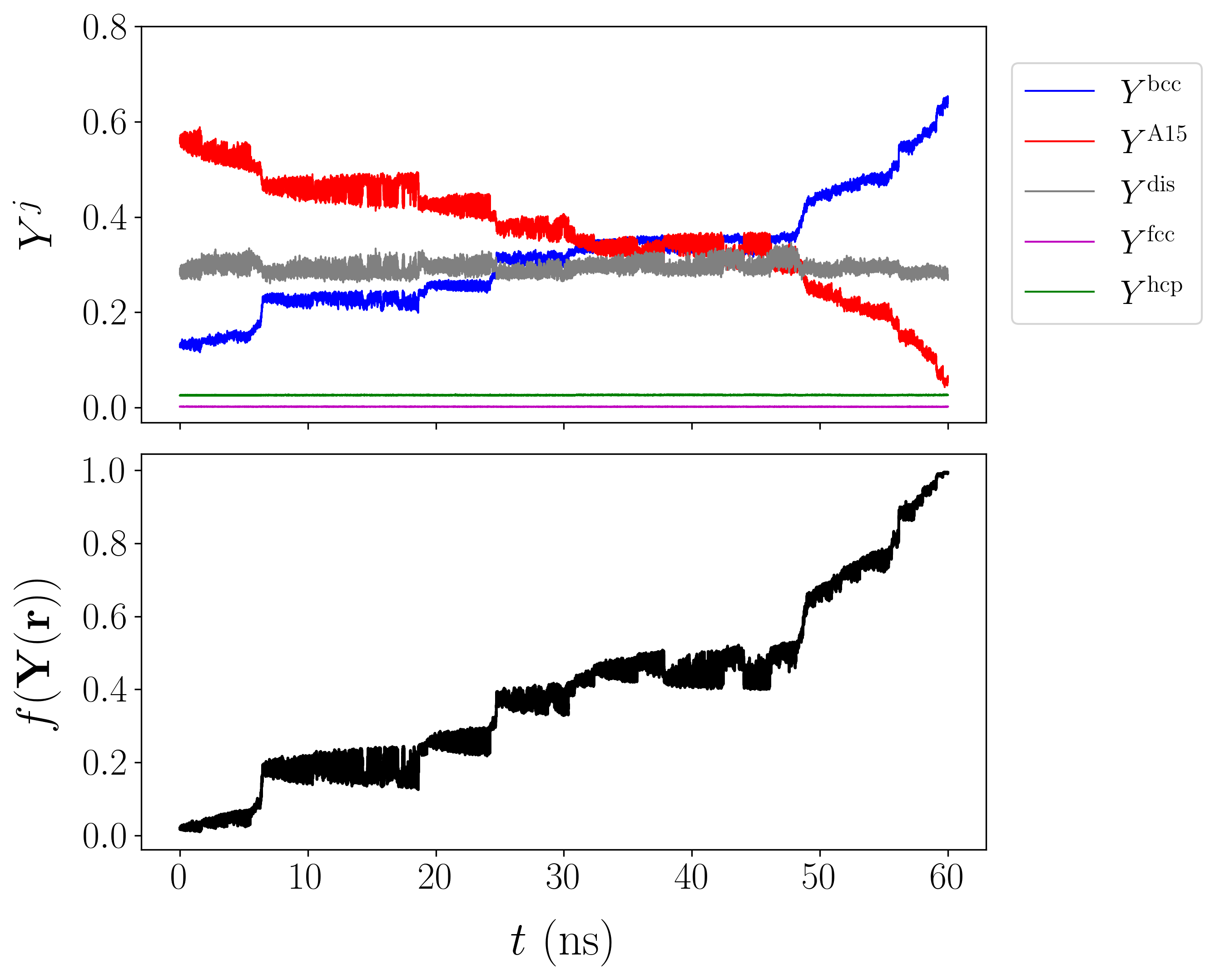}
    \caption{Evolution of the global classifier $Y^{j}$ for various phase of interest (top) and the path collective variable (bottom panel) extracted from a representative metadynamics run sampling the A15$\to$bcc transformation. The A15 fraction gradually decreases as the bcc phase grows. During this process, fcc and hcp phase are negligible, while disordered phase of 26-33\% with respect to the current system size is always present. Along the interface migration during the A15$\to$bcc transformation, the path collective variable $f(\bf{Y(r)})$ gradually increases from 0 to 1.}
    \label{fig:forward_classfier_pathCV}
\end{figure}
\begin{figure}[htbp]
    \centering
    \begin{subfloat}[]
     {\includegraphics[scale=0.45]{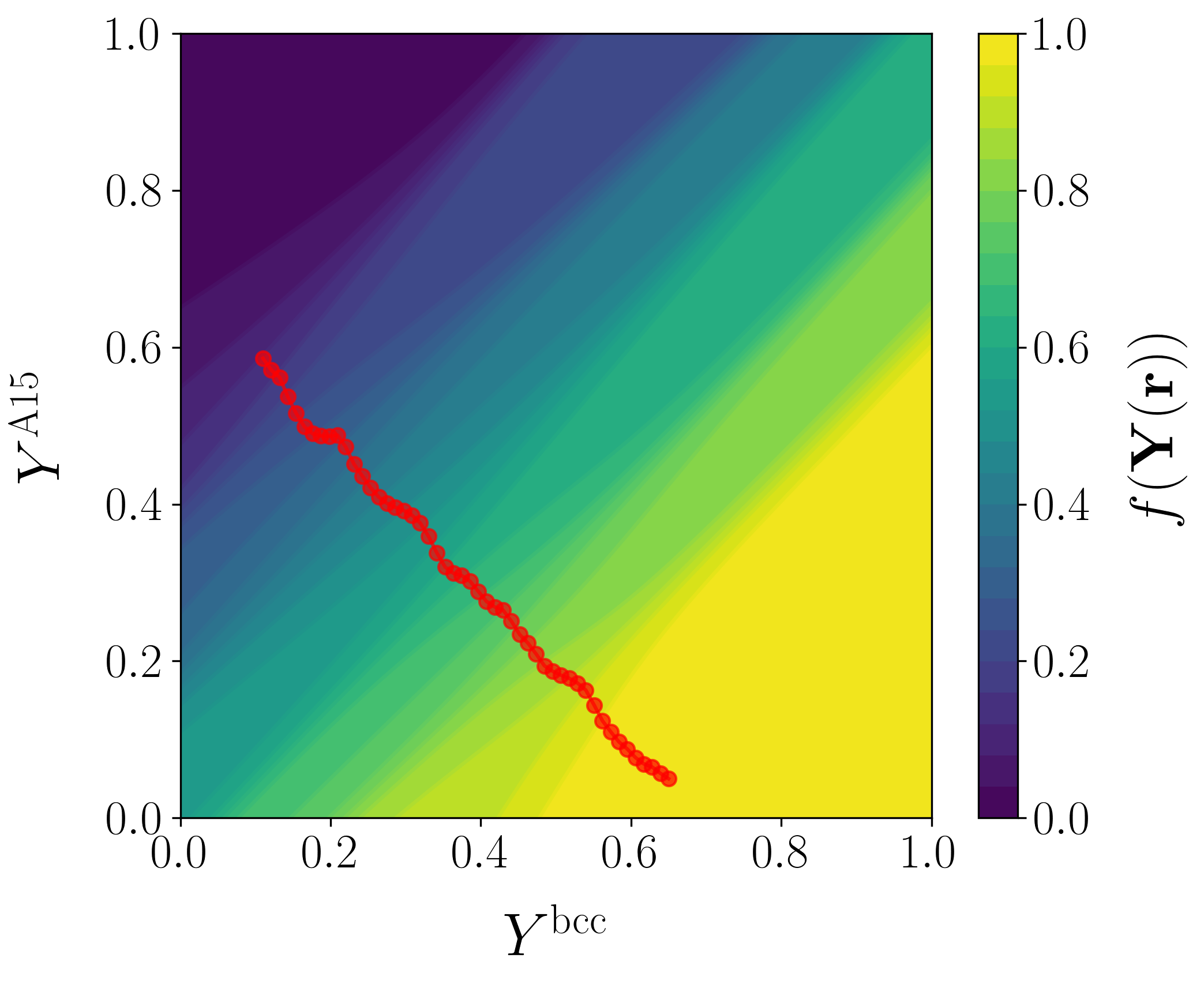}}
    \end{subfloat}
    \quad
    \begin{subfloat}[]
     {\includegraphics[scale=0.45]{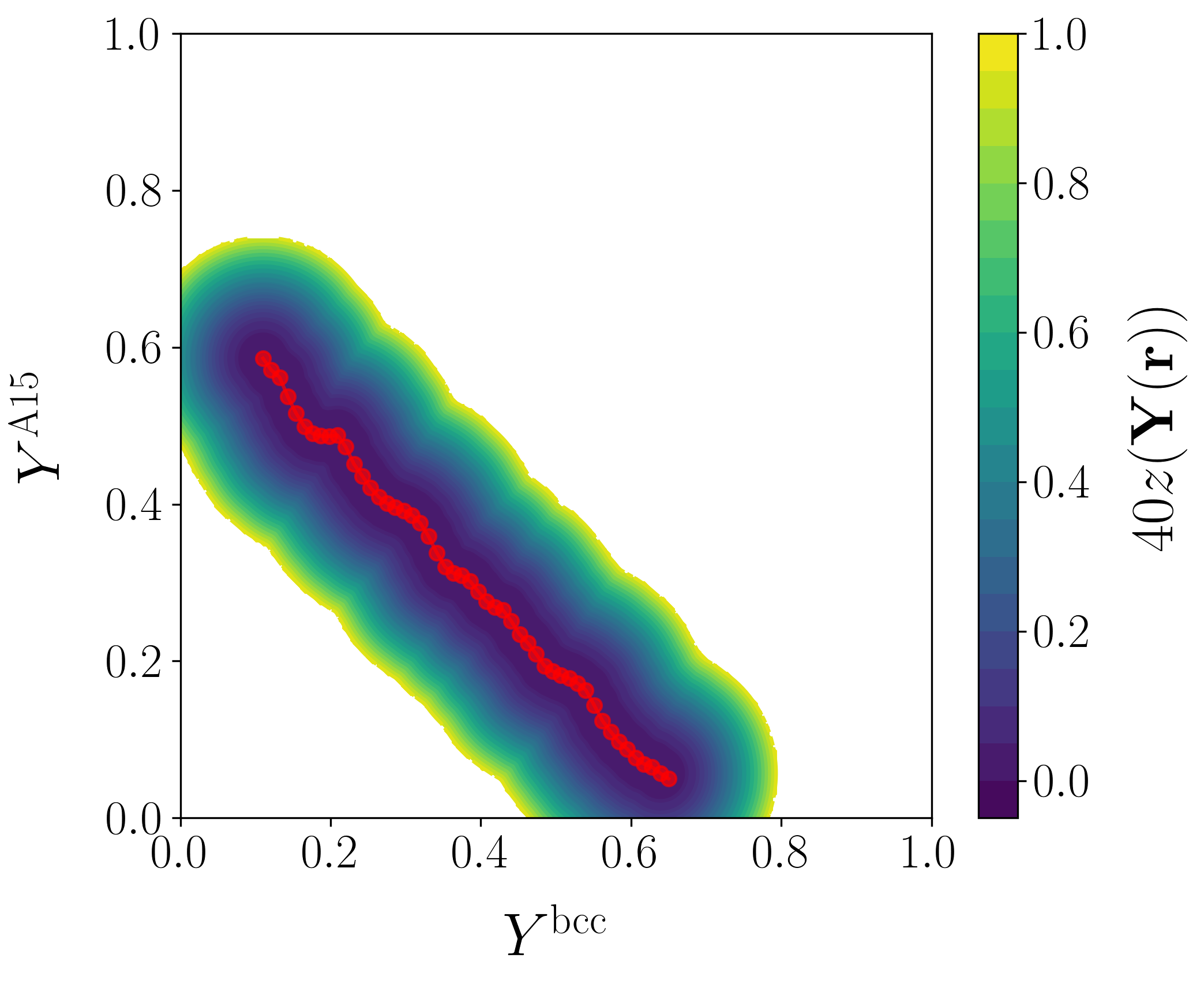}}
    \end{subfloat}
    \caption{Illustration of (a) the 'bumpy' path collective variable $f(\mathbf{Y}(\mathbf{r}))$ and (b) the distance function $z(\mathbf{Y}(\mathbf{r}))$ (multiplied by 40) together with the path (red points) in the $Y^\text{bcc}$-$Y^\text{A15}$ space.  The path is  based on the average path density extracted from 60 metadynamics sampling A15$\to$bcc transformations without any restraint. The bumpy path is constructed by connecting 50 nodal points that are displaced equidistantly.
    }
    \label{fig:bumpypathCV}
\end{figure}
\begin{figure}[htbp]
  \begin{subfigure}[b]{0.495\textwidth}
    \centering
    \includegraphics[width=\textwidth]{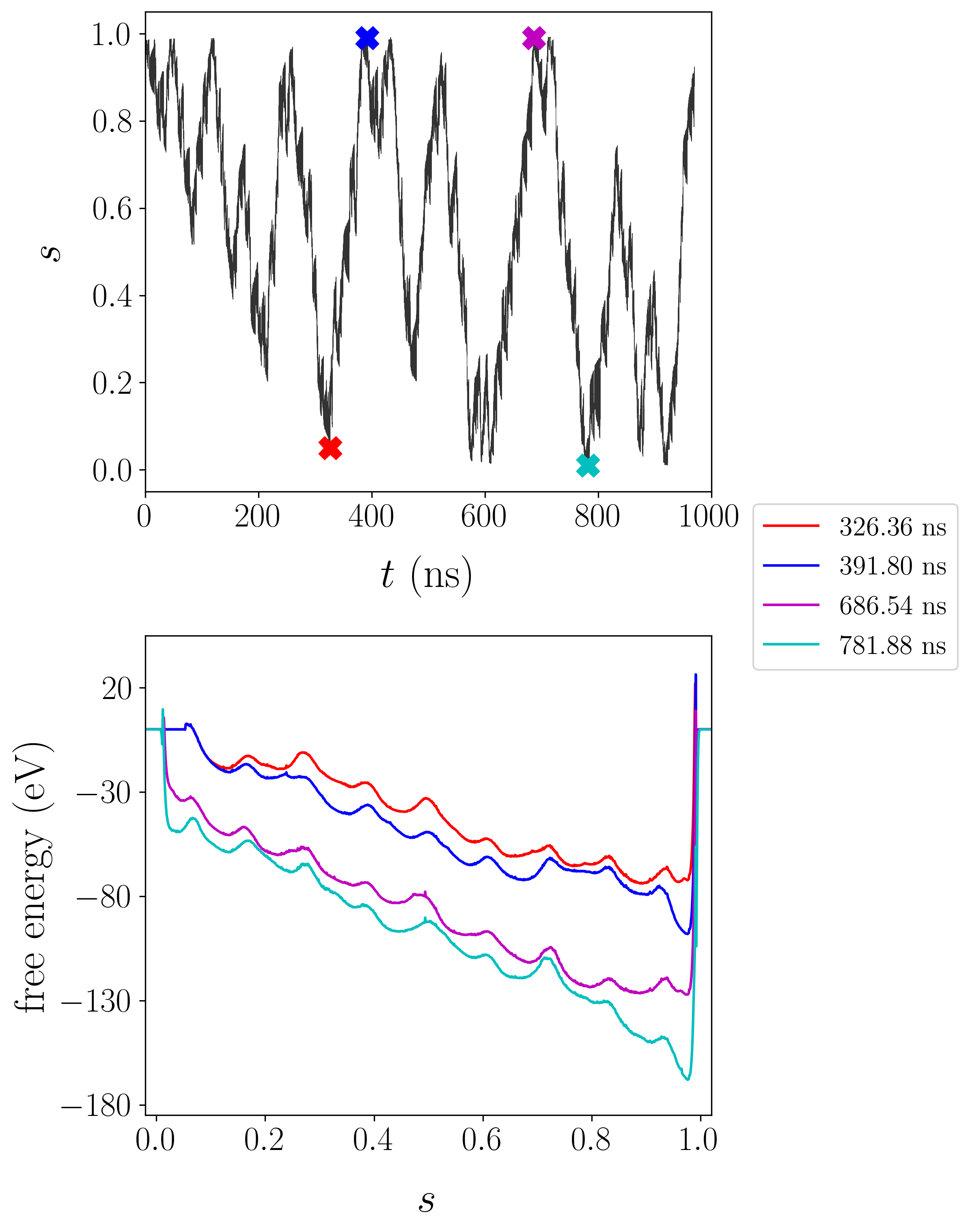}
    \caption{along a linear path}
    \label{fig:tightstraight}
  \end{subfigure}
  \begin{subfigure}[b]{0.495\textwidth}
    \centering
    \includegraphics[width=\textwidth]{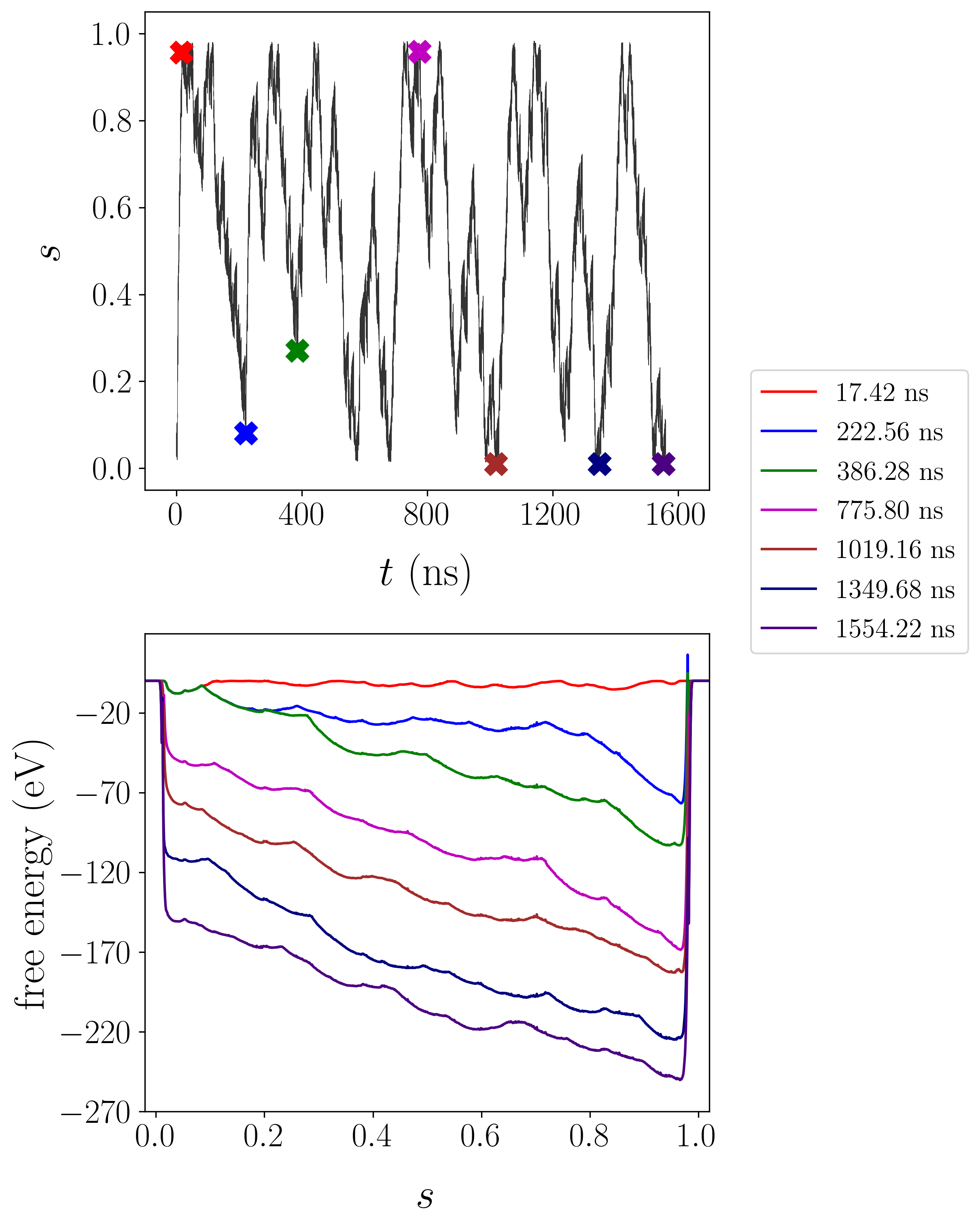}
    \caption{along a bumpy path}
    \label{fig:tightbumpy}
  \end{subfigure}
  \caption{Evolution of the path collective variable as a function of time (top) together with the extracted unbiased free energy profile (bottom) from metadynamics sampling of both the forward and backward A15$\leftrightarrow$bcc transformation with a tight restraining potential: (a) sampled along the linear path, and (b) sampled along the bumpy path. In both cases, there is minimal hysteresis in the path collective variable during recrossing. The free energy profiles exhibit qualitative convergence with a constant offset in energies but little change in profiles.
  }
  \label{fig:bothtight}
\end{figure}

\begin{figure}
    \centering
    \includegraphics[scale=0.5]{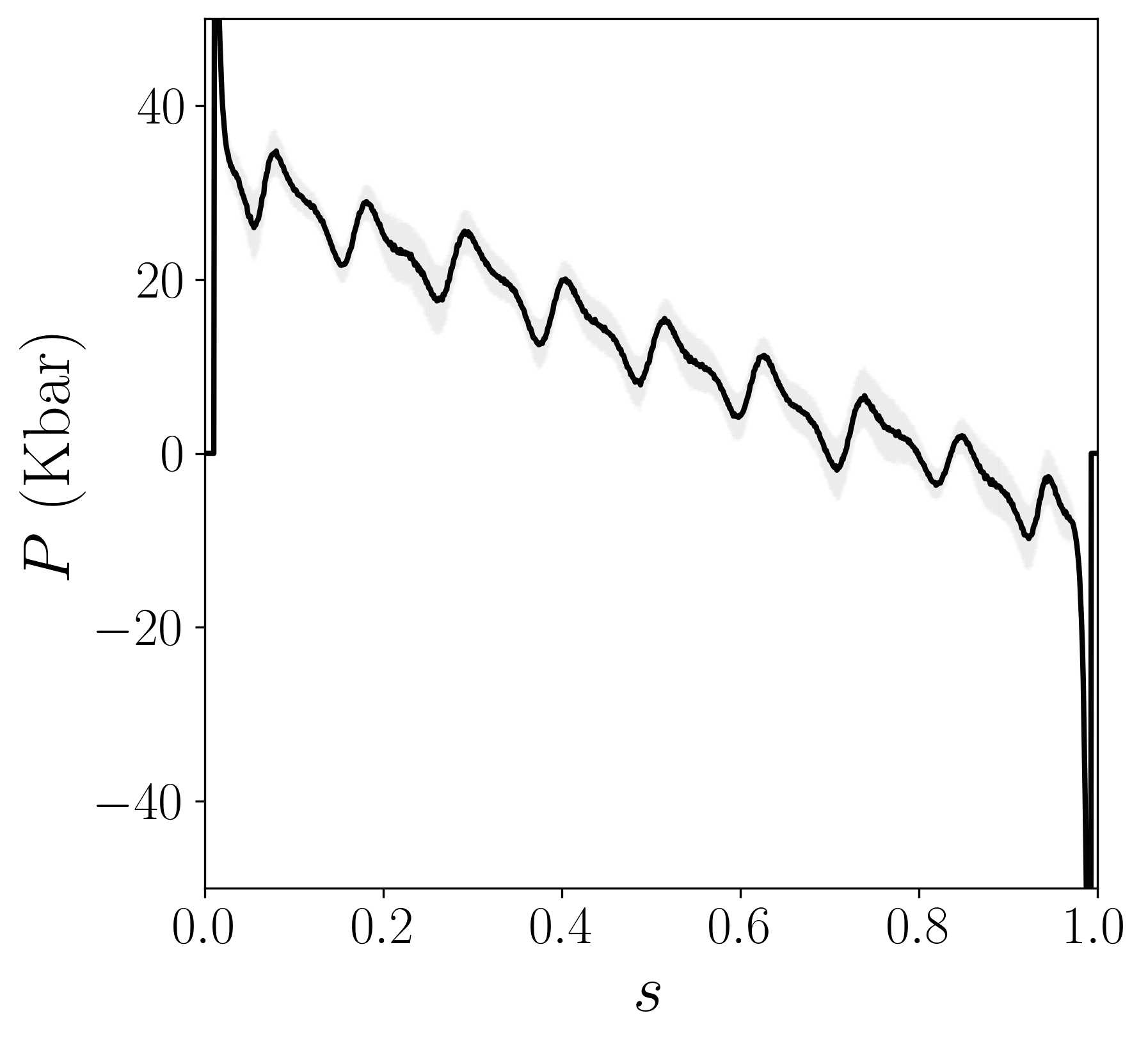}
    \caption{Average pressure as a function of the path collective variable extracted from metadynamics runs sampling the A15$\leftrightarrow$bcc transformation along a linear path with a tight restraining potential. In the current cell geometry, the lattice mismatch compresses the A15 phase by 0.57\%, and growing bcc leads to a decrease in pressure. 
    The restraining potential limits the width of the disordered region at the migrating interface leading to reduced space for atomic shuffling and thus a periodic increase in pressure along the path collective variable.
    }
    \label{fig:pressure_tight_straight}
\end{figure}

\begin{figure}[htbp]
    \centering
    \includegraphics[scale=0.48]{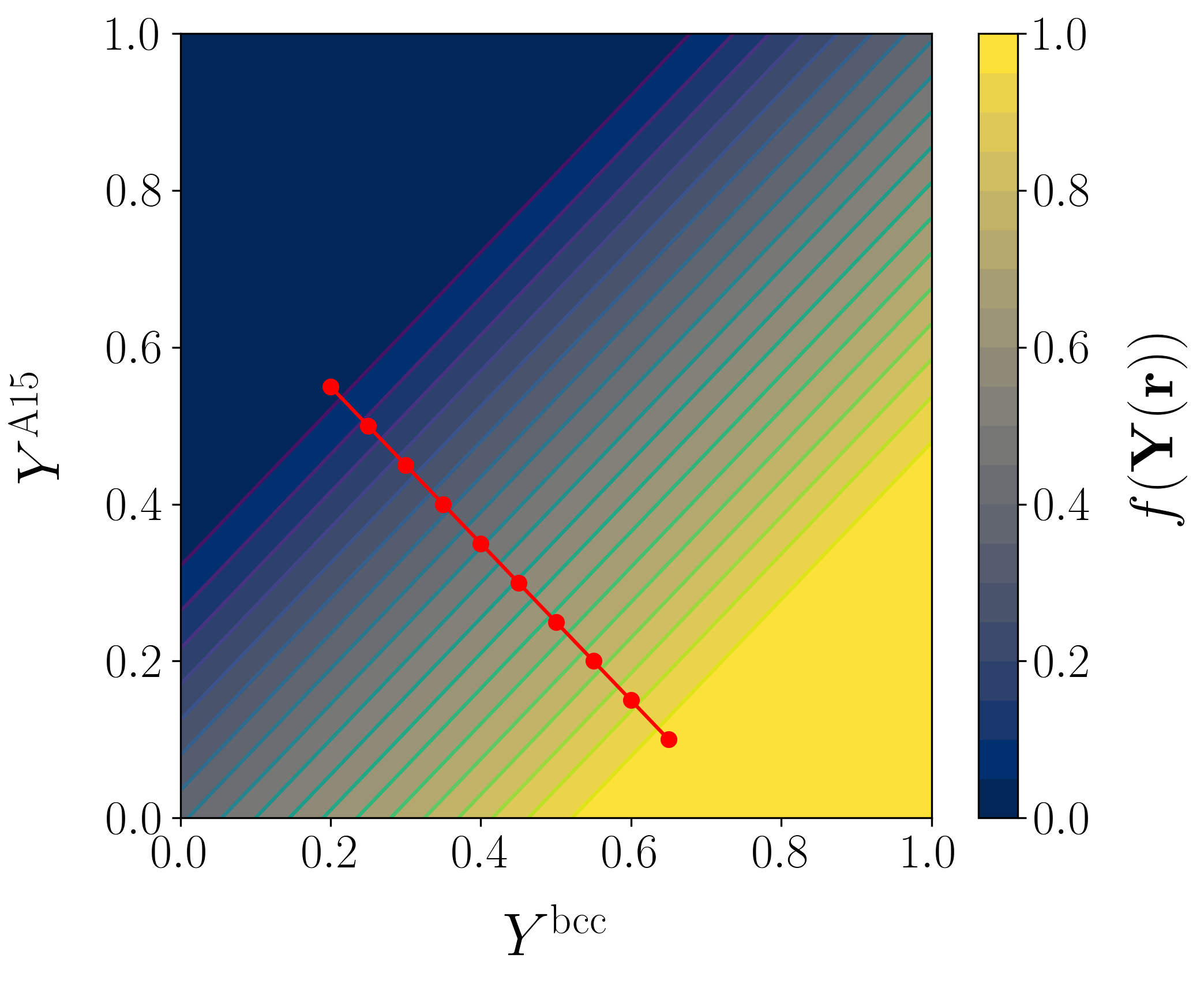}
    \caption{Illustration of the path collective variable $f(\mathbf{Y}(\mathbf{r}))$  with the path (red line) for the [110]$_\text{bcc}\parallel [110]_\text{A15}$ system in the $Y^{\text{bcc}}-Y^{\text{A15}}$ space. Ten equidistant points are distributed along this linear path. Two end points of the path are $\mathbf{Y}_{1}=\{Y^\text{bcc}:0.20,Y^\text{A15}:0.55\}$ and $\mathbf{Y}_{10}=\{Y^\text{bcc}:0.65,Y^\text{A15}:0.10\}$. }
    \label{fig:pathCV_110}
\end{figure}

\bibliography{W}